# Replacing Gas with Low-cost, Abundant Long-duration Pumped Hydro in Electricity Systems

Timothy Weber[1] (corresponding author: Timothy.Weber@anu.edu.au): Conceptualisation, Writing – Original Draft, Software, Data Curation, Visualisation, Method, Investigation, Formal Analysis

Cheng Cheng[1]: Software, Visualisation, Method, Writing – Review & Editing

Harry Thawley[1]: Software, Method, Writing – Review & Editing

Kylie Catchpole[1]: Writing – Review & Editing, Visualisation

Andrew Blakers[1]: Writing – Review & Editing, Data Curation

Bin Lu[1]: Writing – Review & Editing

Jennifer Zhao[1]: Writing – Review & Editing

Anna Nadolny[1]: Writing – Review & Editing

## Summary

Fossil gas is sometimes presented as an enabler of variable solar and wind generation beyond 2050, despite being a primary source of greenhouse gas emissions from methane leakage and combustion. We find that balancing solar and wind generation with pumped hydro energy storage eliminates the need for fossil gas without incurring a cost penalty. However, many existing long-term electricity system plans are biased to rely on fossil gas due to using temporal aggregation methods that either heavily constrain storage cycling behaviour or lose track of the state-of-charge, failing to consider the potential of low-cost long-duration off-river pumped hydro, and ignoring the broad suite of near-optimal energy transition pathways. We show that a temporal aggregation method based on 'segmentation' (fitted chronology) closely resembles the full-series optimisation, captures long-duration storage behaviour (48- and 160-hour durations), and finds a near-optimal 100% renewable electricity solution. We develop a new electricity system model to rapidly evaluate millions of other near-optimal solutions, stressing the importance of modelling pumped hydro sites with a low energy volume cost (<US$50 per kilowatt-hour), long economic lifetime (~75 years), and low real discount rate akin to other natural monopolies (≤3%). Almost every region of the world has access to sufficient 50–5000 gigawatt-hour off-river pumped hydro options that enable them to entirely decarbonise their future electricity systems.

[1]Affiliation: The Australian National University, Canberra ACT 2601

# Context & scale

Biases in models used to develop long-term electricity system plans have fuelled the notion that fossil gas ought to be expanded to support the renewable energy transition. Common methods of simplifying multiple decades of high-resolution data to reduce computational complexity fail to capture long-duration storage behaviour by either losing track of the state-of-charge or heavily constraining the charging/discharging behaviour. Coupled with out-dated assumptions regarding pumped hydro storage duration, costs, and availability, these electricity system models have no alternative but to rely upon gas for balancing solar and wind generation. An alternative to gas already exists in the form of pumped hydro, which remains the lowest-cost option for long-duration energy storage (overnight storage and longer) and is abundantly available around the world (over 800,000 potential sites). A combination of pumped hydro and batteries is capable of providing the same frequency control, voltage control, black start, and inertia-related services as gas generators.

We show that with suitable methods for simplifying the input data and updated pumped hydro assumptions, long-term planning models can find reliable 100% renewable electricity systems at a similar cost to gas-dependent systems. Using a new model with high computational speed, we show that there is a wide variety of "near-optimal" clean energy development pathways. The cost of these 100% renewable electricity pathways ought to be based on large-scale pumped hydro with a long economic life, low energy volume cost, and a low real discount rate equivalent to regulated natural monopolies such as transmission. Fossil gas is a primary cause of climate change, both from methane leakage and from carbon dioxide due to combustion. With sufficient policy and financial support from governments to develop large-scale long-duration pumped hydro systems, fossil gas can be entirely replaced within the electricity systems of the future.

Costs are in 2025 US dollars.

# Main

Long-term electricity system planning models are used to identify grid configurations that achieve the objectives of the energy planner. About 89% of global net generation capacity added in 2024 was either solar photovoltaics (PV) or wind, with PV capacity doubling roughly every three years since 2015 [1, 2, 3]. Models of energy systems with a high penetration of solar PV and wind must capture spatial [4] and temporal [5, 6, 7] variation in renewable generation. This means that long-term planning models now incorporate both capacity expansion and unit commitment problems into their formulation [8, 9, 10]. Extreme periods, characterised by high load and low renewable generation, drive up system costs and require multi-year planning horizons [11]. Running a high-resolution model using multiple decades of data requires an impractical amount of computing resources. All energy system optimisation models employ some form of spatial, technological, temporal, or sector coupling simplifications to reduce model size, and therefore computational complexity, to make the optimisation manageable [12, 13, 14].

Early techniques for temporal simplification involved performing the energy balance for an approximate load duration curve (LDC) or residual LDC to reduce temporal resolution by grouping periods of similar load or net-load together, at the cost of losing chronology [15, 16, 17, 18]. Some chronological information can be added to LDC methods by clustering according to "system states"[2] and then defining a transition matrix that accounts for state changes [19, 20]. To maintain full chronology, temporal resolution can instead be reduced through either down-sampling or segmentation into adjacent intervals of mutual similarity [21, 12, 22, 23]. Capacity expansion models are highly sensitive to temporal resolution [24].

Evaluating a sample of typical periods[3] based on simple heuristics, clustering algorithms, or optimisation-based approaches is now standard practice to reduce the number of intervals evaluated by electricity planning models [25, 26, 27, 21, 12, 23]. Methods that incorporate extreme periods into the set of typical periods have also been developed [21, 12, 28, 27]. A rolling horizon or myopic multi-step optimisation would reduce the time horizon evaluated by the optimiser [12], although horizons shorter than about 2 months would be heavily biased against building any long-duration energy storage [29]. Unfortunately, all temporal simplification techniques come at the cost of losing information about energy balance options at different timescales.

In § "Temporal Simplification and Energy Storage" we explain how some of these temporal simplification techniques may not adequately capture long-duration balancing behaviour. The use of these temporal simplification techniques and outdated pumped hydro assumptions in global energy plans is investigated in § "Problems Modelling Pumped Hydro in Existing Long-term Energy Plans". In § "Results", a PLEXOS model of Australia's National Electricity Market (NEM) is used to evaluate the bias towards gas and away from long-duration pumped hydro caused by these modelling decisions. The Australian NEM is a useful example since it is in a low- to mid-latitude region (where most of the global population resides), physically isolated which prevents sharing of electricity with its neighbours, supports an advanced economy, and is being rapidly decarbonised using solar photovoltaics (PV) and wind with a government target of 82% renewable electricity by 2030. We then use a new non-linear model to rapidly search the near-optimal solution space for millions of feasible grid configurations for the NEM. The conditions that support 100% renewable electricity systems are evaluated using a sample of these feasible solutions.

---

[2] The term "system states" is noted by [21] to be misleading as the complete state of a system is actually dependent on endogenous variables, such as the storage states-of-charge and transmission flows. More accurately, this method applies the transition matrix to an early version of the "typical period" method.
[3] A subset of these methods refers to typical periods as "representative periods" or "time slices". In PLEXOS, it is called a "sampled chronology".

## Temporal Simplification and Energy Storage

Balancing of variable generation can be provided by increased transmission interconnection to smooth out local weather and load, demand management to smooth out peak demand periods, energy storage, over-building of variable generators coupled with curtailment, load-following generators, and flexible generators such as conventional hydroelectricity or gas [30]. Long-duration energy storage typically refers to storage durations of at least 10 h [31], and can be further broken down into long multiday (24–100 h), seasonal (100–8760 h), and interannual (>8760 h) storage[4] [32]. Seasonal and interannual storage would be most relevant for managing winter *dunkelflaute* in mid- to high-latitude regions, though they may also support system reliability during the wet season for low-latitude countries.

Energy planning models that reduce temporal resolution will lose information about short-duration balancing requirements from flexible generators and intra-day storage [5, 33, 34, 35, 24]. Temporal aggregation based on typical periods will instead fail to capture the need for long-duration energy storage because cycle durations are highly constrained by the length of the typical period [36, 29, 37]. An alternative system state method, based upon the expected value of available storage when moving between states, overestimated storage volume by a factor of 2 at low storage costs [38].

Sampling longer contiguous periods may improve the modelling of inter-day storage, but will still substantially underestimate seasonal storage needs [25, 37]. To take these longer chronological periods into account, typical periods were linked in [39] (model M1) and [40, 41, 42, 43, 44, 45] while assuming storage system dispatch is identical within repetitions of each typical period. However, if the number of typical periods is low, long-term arbitrage may be highly constrained. That is, if there is a net change in the state-of-charge within a typical period, that change will be multiplied by each repetition of that typical period and very quickly push the profile either down to 0 GWh or up to the energy volume constraint [40]. This effect will still be present in models using multiple superimposed time grids and *a posteriori* clustering methods [45, 42, 46]. Modelling storage behaviour as a linear combination of typical days, rather than just a copy of a single typical day, can capture some additional long-duration storage value, but requires 128 typical days per year to converge to the full time-series benchmark [47].

Linear programming models using full time-series chronological data capture long-duration storage requirements the best [48, 29, 49, 50, 51], but come with the cost of either coarse temporal resolution [48, 29], time horizons of only 1 year [50, 48, 29, 49], or simplifying the network to a single spatial node [49, 51]. Modelling a multi-year time horizon in several optimisation steps with a length of 1-year or more may help to limit complexity, losing interannual storage behaviour while still testing robustness over multiple years of weather conditions.

A separate class of long-term planning models relies on business rules to perform the unit commitment, rather than a linear programming formulation [52, 53, 54, 55, 56, 57, 58, 59, 60]. We will refer to this class of models as the business rules-based long-term planning (BR-LTP) models.

---

[4] Seasonal and interannual storage is sometimes collectively referred to as "deep storage".

The business rules are simple heuristics that are sequentially executed to perform the energy balance while complying with the constraints of the system in every time interval. By using business rules for unit commitment, the number of decision variables is reduced by several orders of magnitude. The substantially smaller problem is capable of rapidly exploring the solution space using a metaheuristic optimiser with full chronological, high-resolution (≤1 hour) modelling over multiple decades of data. Since business rules are applied to the full time-series data and bound by the constraints of the system, BR-LTP solutions are guaranteed to be feasible and represent an upper-bound cost for solutions that could be obtained from a hypothetical linear programming optimisation with equivalent constraints and no temporal aggregation. We developed a new BR-LTP model to rapidly search for near-optimal configurations of the NEM in § "Results – Searching the Near-optimal Space for Solutions with Minimal Gas". The use of a BR-LTP model to rapidly evaluate millions of near-optimal solutions is a novel technique, since linear and mixed-integer linear programming models with a slower optimisation time have typically been relied upon in the literature [61].

## Problems Modelling Pumped Hydro in Existing Long-term Energy Plans

A review of global long-term energy plans published on behalf of governments and inter-governmental organisations is summarised in Supplementary Information A, Table S1. Within the seventeen energy plans we reviewed from around the world, one model specified using an approximate LDC method [62], one specified using segmentation into adjacent intervals of mutual similarity [63] (Detailed Long-term Model), and eight specified using linked typical periods to simplify the time-series data [63] (Single-Stage Long-Term Model) [64, 65, 66, 67, 68, 69, 70]. Temporal aggregation techniques served to reduce model complexity since higher temporal resolutions over long optimisation step time horizons would drive up computation time.

A high temporal resolution of 1 hour or less was used in twelve of the models [63] (Single-Stage Long-Term Model) [64, 73, 72, 32, 74, 66, 75, 76, 69, 70] [71]. The 3-hour resolution in [65] is incapable of fully capturing short-duration balancing from batteries and flexible generators. The fitted blocks used in four of the models may lose some detail for short-term balancing [63] (Detailed Long-Term Model) [67, 68, 62], though the variable length timesteps could mitigate this impact. The optimisation step time horizon in one of the models was just one week, which prevents the model from making decisions about seasonal and interannual energy storage [72]. While most of the energy plans that we reviewed maintained a high temporal resolution (≤ 1 hour), this typically came at the cost of using temporal aggregation techniques that could fail to capture energy storage behaviour at different timescales. Only three models did not require any temporal aggregation while maintaining a 1-hour temporal resolution and an optimisation step time horizon of at least 1 year [71, 32].

Pumped hydro is a mature and low-cost technology that already provides about 95% of grid-connected energy storage volume around the world [77, 78, 79].[5] It is sometimes incorrectly classified as "mid-duration" storage with limited potential due to its dependence on local geography near rivers [80]. However, off-river (closed-loop) pumped hydro is far more widely available because the only requirement is a local elevation difference. Geographic information system analysis was previously used to developed a global atlas of 818,000 off-river pumped hydro options with a storage potential of 86 million GWh outside protected areas and large urban centres [81, 82, 83]. Although not all sites are expected to be feasible due to site-specific geological, environmental, or social impacts, only a small fraction of these sites would be required to decarbonise global electricity systems. Energy volumes of 50–5000 GWh are highlighted by the atlases, which would be more than capable of providing long-duration storage services. Yet, only one of the reviewed energy plans considered a large pumped hydro build limit (144,381 GWh across the entire United States of America in this case), with highly restricted build limits in all other plans likely acting as a strong constraint [65]. New build pumped hydro was either limited to a handful of specific options or was not considered at all in five of the models [66, 75, 84, 68, 62].

---

[5] 190 GWh total grid-connected batteries in 2023 [148], 169 GWh grid-connected batteries added in 2024 [149], and estimated 9000 GWh pumped hydro capacity currently installed worldwide [79]. Other technologies provide a negligible contribution to grid-scale storage.

Long-duration energy storage with a seasonal pattern of behaviour has been found to require storage durations of between 100–825 hours when modelling decarbonised electricity systems [49, 29, 48]. Snowy 2.0 is a 2.2 GW, 350 GWh (160 h duration) pumped hydro system that is currently under construction in Australia and will be capable of providing these seasonal storage services [85]. None of the plans summarised in Table S1 considered new build pumped hydro systems with a storage duration longer than 48 h, with only two plans considering storage durations longer than 12 h [86, 32]. While some studies did model multiple sizes of pumped hydro system [87, 32, 65, 67], deep energy storage would be expensive to deploy due to the overbuild of expensive power capacity coupled to small energy volumes resulting from the short storage durations they considered. One energy plan did investigate the need for generic energy storage of arbitrarily long durations [32].

Through an analysis of the long-duration energy storage design space, the energy volume cost ($/kWh) was found to be the greatest driver of total system costs [49]. When apportioning project costs over energy volume, Snowy 2.0 will have a cost of about $29/kWh (March 2025 USD) and a lifetime of 150 years.[6] The pumped hydro capital costs ($/kWh) considered in [63, 65, 73, 32, 76, 68] were 140–4300 % of the Snowy 2.0 costs. The capital costs in eight of the reviewed energy plans could not be evaluated in $/kWh without information about the storage durations assumed for pumped hydro systems [64, 72, 88, 75, 62, 69, 70, 71]. The costs assumed in [67] appear to be on the order of Snowy 2.0 costs, though the model only considered short storage durations of up to 11 h. Pumped hydro developed for long-duration storage achieves economies of scale for large reservoirs, and the vast opportunity available for off-river locations makes it relatively straightforward to find systems with large heads (>> 300 m) compared to on-river systems. The head is one of the main drivers of cost, since doubling the head can double the energy capacity while keeping similar reservoir construction costs [81]. The global atlas shows that most regions of the world have access to a large number of premium sites with equivalent or better quality than Snowy 2.0 [89]. The combination of these outdated pumped hydro assumptions with temporal aggregation techniques that fail to capture long-duration storage behaviour likely provides long-term energy planning models with no alternative but to invest in fossil gas to balance solar and wind generation.

# Results

We modified the Australian Energy Market Operator's (AEMO) publicly available *2024 Integrated System Plan (ISP) Model* of the NEM (developed using PLEXOS) [86] to evaluate how common temporal aggregation techniques and outdated pumped hydro assumptions bias existing long-term energy plans away from long-duration energy storage and towards gas. The large majority of Australia's population lives near the east and southeast coast, connected to the NEM. Almost all new generation capacity that is being added to the NEM is either solar PV or wind. The example of the NEM is comparable to any large, interconnected grid in a low- to mid-latitude region (where most of the global population resides). Higher latitudes will have a larger requirement for long multiday and seasonal balancing because of winter *dunkelflaute*. Countries in the sunbelt region, particularly those close to the equator without a winter, would predominantly depend on overnight balancing of solar, though they would still require longer-duration balancing during the wet season when there is increased cloud cover [57].

---

[6] Snowy Hydro recently announced that the US$8 billion costs are currently being reassessed and may be revised upwards [139]. As the project is already 67% complete, we have assumed a final cost of US$10 billion.

The modified ISP model optimised the configuration of the NEM in FY2052 (the final year in the original model) over 10 reference weather years (July 2042 – July 2052 in the original model). All PLEXOS scenarios were modelled using four common temporal aggregation methods to evaluate their effects on the deployment of long-duration storage and gas. The temporal aggregation methods tested using the PLEXOS model included sampling 2 typical days per month ("Sampled (days)"), sampling 4 typical weeks per year ("Sampled (weeks)"), segmentation into adjacent intervals of mutual similarity by fitting an 8-block step function to the operational demand for each day ("Fitted"), and fitting an 8-block step function to the daily LDC which results in loss of chronological information within each day ("Partial"). Most of the long-term energy plans evaluated in § "Problems Modelling Pumped Hydro in Existing Long-term Energy Plans" used some variation of each of these temporal aggregation techniques.

A BR-LTP model based upon the FIRM framework [60] was developed to rapidly evaluate a broad range of near-optimal configurations of the NEM. Unlike PLEXOS, the FIRM framework uses simple business rules for unit commitment meaning that no temporal aggregation is required. Candidate solutions are quickly evaluated for reliability and cost according to the business rules, with a differential evolution algorithm iteratively mutating the population of candidate solutions to minimise total system cost. A sample of near-optimal configurations of the NEM was used to evaluate key drivers for low-cost 100% renewable electricity systems compared to gas-dependent systems.

A more detailed description of methods and results is provided in § "Methods" and Supplementary Information A and B. The raw data and PLEXOS models are available in Supplementary Information C. A GitHub link to the FIRM model is provided in § "Code Availability".

## The Bias towards Gas in Existing Long-term Energy Plan Formulations

The PLEXOS models were optimised using the original pumped hydro energy storage (PHES) capital cost, build limit, and storage duration assumptions from the *2024 ISP Model* ("Original PHES Assumptions" scenario) to establish a baseline. The models were then modified to create an "Improved PHES Assumptions" scenario as follows:

- Tasmanian pumped hydro capital cost assumptions in the *2024 ISP Model* were adopted for all nodes in the network. The global pumped hydro atlas describes 200 premium cost class sites (AAA or AA) in the southeast corner of the mainland of equal or better quality to the 91 Tasmanian sites based on the comparative cost model used by the atlas. These assumed costs are still higher than the Snowy 2.0 costs, so they remain a conservative estimate compared to the highest-quality options available.

- Pumped hydro maximum build limits were relaxed. The *2024 ISP Model* had a total build limit for the NEM of 423 GWh, while the global pumped hydro atlas has a maximum potential of 287,000 GWh across the relevant Australian states.
- A 160-hour duration pumped hydro option was added to some nodes. This is equivalent to the storage duration of Snowy 2.0.

A third scenario modified the Improved PHES Assumptions models to exclude large open-cycle gas turbine (OCGT) development ("100% Renewables" scenario).

Using these three scenarios, the influence of the four temporal aggregation methods on long-duration pumped hydro behaviour, and thereby investment, within the model was evaluated. We found that the Fitted temporal aggregation method performed best when modelling long-duration storage. By using the Fitted method with improved pumped hydro assumptions, we found that the 100% Renewables scenario was equivalent in cost to the gas-dependent solutions.

*Long-duration Storage Behaviour Within Each Temporal Aggregation Method*

The total state-of-charge profile of pumped hydro systems for each temporal aggregation method (Sampled (days), Sampled (weeks), Fitted, and Partial) over the modelling horizon was converted to the frequency domain, as illustrated for the 100% Renewables scenario in Figure 1. Each peak indicates a strong contribution from a particular cycle frequency to the overall state-of-charge profile. The Sampled (days) and Sampled (weeks) methods were dominated by daily cycling behaviour and failed to capture long-duration storage behaviour. The Partial method was able to consider seasonal cycle frequencies but did not have a noticeable contribution from short-duration cycling because of the loss of chronological information within each day. Only the Fitted method captured both daily and long-duration cycling behaviour for the pumped hydro systems.

*Figure 1. Normalised magnitude of frequency spectra of total energy stored in pumped hydro systems across 10-year model horizon in 100% Renewables scenario for each temporal aggregation method. Spectra are vertically and horizontally offset for readability.*

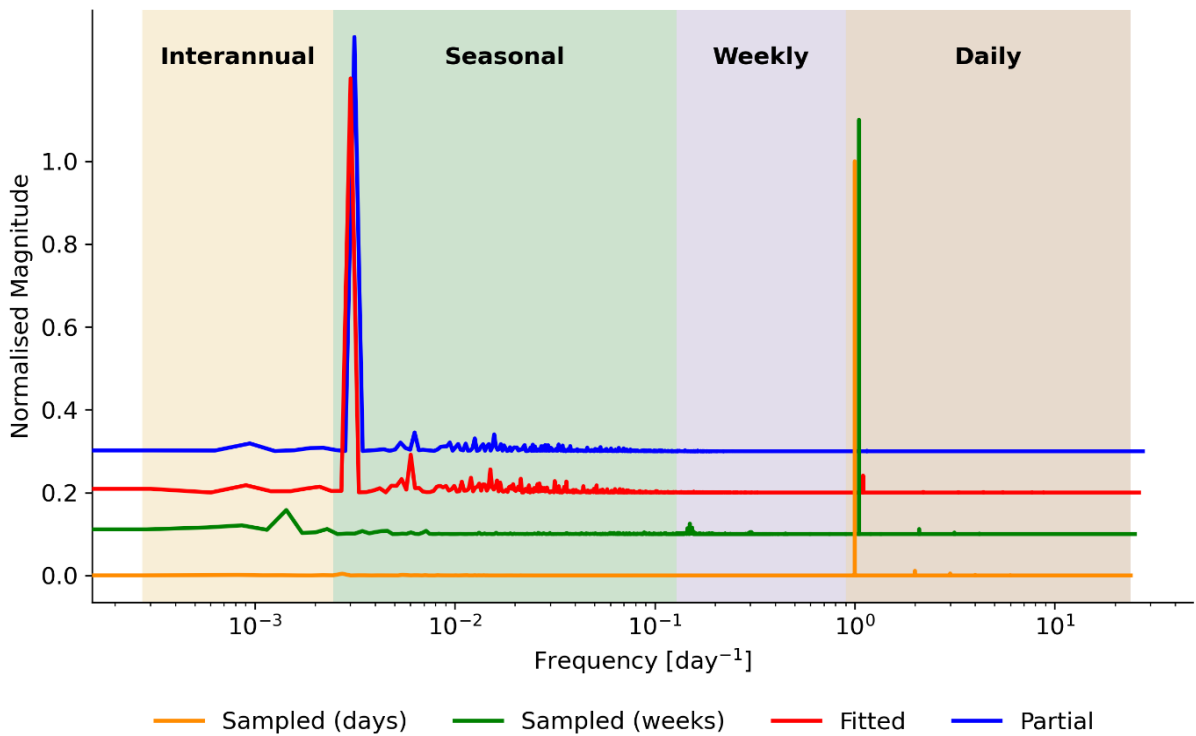

A two-week snapshot of each state-of-charge profile in the time domain is provided in Figure 2. The full series optimisation (FSO), absent of any temporal aggregation, was performed over a 1-year modelling horizon and is presented for comparison in Figure 2a. The full 10-year horizon was too complex for a single optimisation step using FSO, but is not necessary to highlight the main features.

*Figure 2. Two-week snapshot of pumped hydro state-of-charge during unit commitment in 100% Renewables scenario for: (a) FSO (1-year horizon), (b) Sampled (days), (c) Sampled (weeks), (d) Fitted, and (e) Partial temporal aggregation methods*

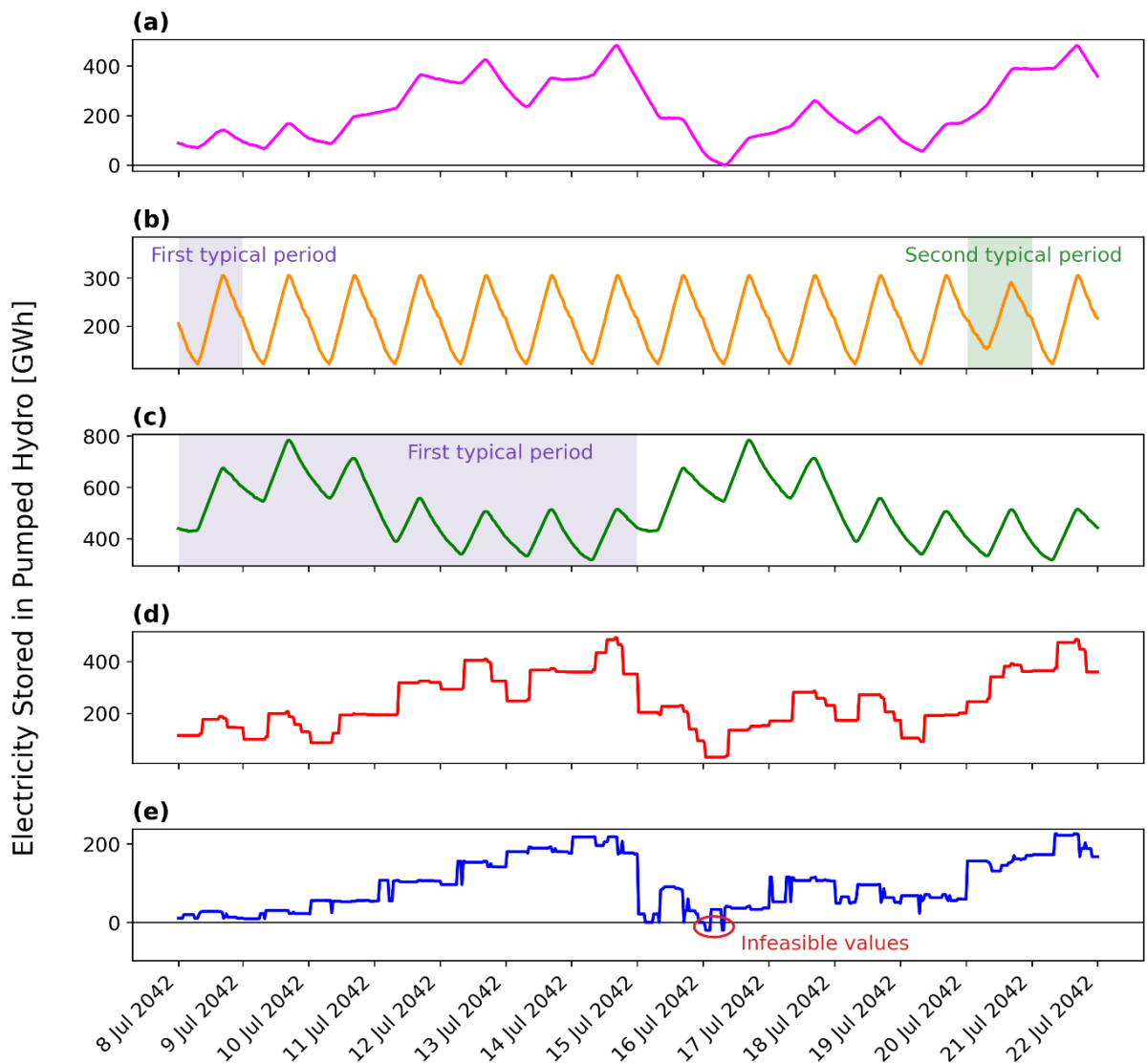

Within the PLEXOS sampled chronology methods (Figure 2b and Figure 2c), state-of-charge at the start and end of each typical period is constrained to be equal, providing no inter-period flexibility for any energy storage systems. Removing the constraint on initial and final state-of-charge within linked typical periods would allow some additional long-term flexibility in storage dispatch. However, dispatch profiles within each repetition of the typical period would still be identical, rapidly pushing the state-of-charge to the energy volume constraints over many repetitions of the typical period [40]. Cycle durations for typical period methods, such as Sampled (days) and Sampled (weeks), are either entirely restricted or highly biased towards being less than or equal to the length of the typical period itself, severely discriminating against the development of long-duration storage. Since clustering does not consider endogenous variables, such as transmission flows or stored energy, the typical period methods also produced state-of-charge profiles that lacked a resemblance to the FSO.

The state-of-charge profiles produced by the Fitted (Figure 2d) and Partial (Figure 2e) methods have a strong resemblance to the FSO method. In the absence of typical periods, storage systems can be flexibly charged and discharged according to a complex dispatch profile across the entire modelling horizon. However, the Partial method produces infeasible short-term dispatch for the pumped hydro due to the loss of chronological information within each daily LDC. Of the tested temporal aggregation methods, Fitted was the only method that produced a pumped hydro state-of-charge profile that resembled the FSO and complied with the energy volume constraints for all time intervals. Since the unit commitment performed by the Fitted method closely resembles the FSO, the method is also expected to produce a similar grid configuration through the optimisation.

*Energy Storage and Gas Investment Decisions Made by Each Temporal Aggregation Method*

We investigated how the difference in energy storage dispatch behaviour for each temporal aggregation method determined the investment decisions made by the model. The optimisation of each temporal aggregation method was repeated over a model horizon of just one year, allowing for a direct comparison with the FSO (which was too complex to optimise over the full 10-year horizon). A solution vector comprised of the annualised build cost, fixed operation and maintenance cost, variable operation and maintenance cost, and fuel cost of each asset in the system was developed for each optimisation. The L1-distance (Manhattan distance) between the solution vectors for each temporal aggregation method and the FSO was calculated, providing an asset-level comparison of each cost (refer § "Methods", Eq. (1)). A smaller L1-distance indicates a solution that is more similar to the FSO [6, 47]. The normalised L1-distances for the Improved PHES Assumptions scenario are presented in Figure 3, broken down by the contribution from the costs of each technology. The results for the 100% Renewables scenario are provided in Supplementary Information A, Figure S1b. Normalisation was performed with respect to the total system costs of the FSO solution.

The Fitted segmentation method produced solutions that were most similar to the FSO. The L1-distance between the solution vector for the Fitted method and the FSO was less than 3% of the total system costs of the FSO in both scenarios. The Sampled (days) and Sampled (weeks) methods produced solutions that were substantially different to the FSO, with an L1-distance of more than 59% and 69% of the total system costs of the FSO respectively. The large L1-distances were due to a difference in solar PV and wind investment at each location since clustering did not account for utility-scale generation, development of shorter duration 8- and 24-hour pumped hydro, and an increase in gas generation relative to the FSO. While typical period methods in PLEXOS may invest in a similar amount of gas capacity as the FSO, the reason for the investment is to balance short-term variability (on scales shorter than the typical period length) rather than seasonal variability.

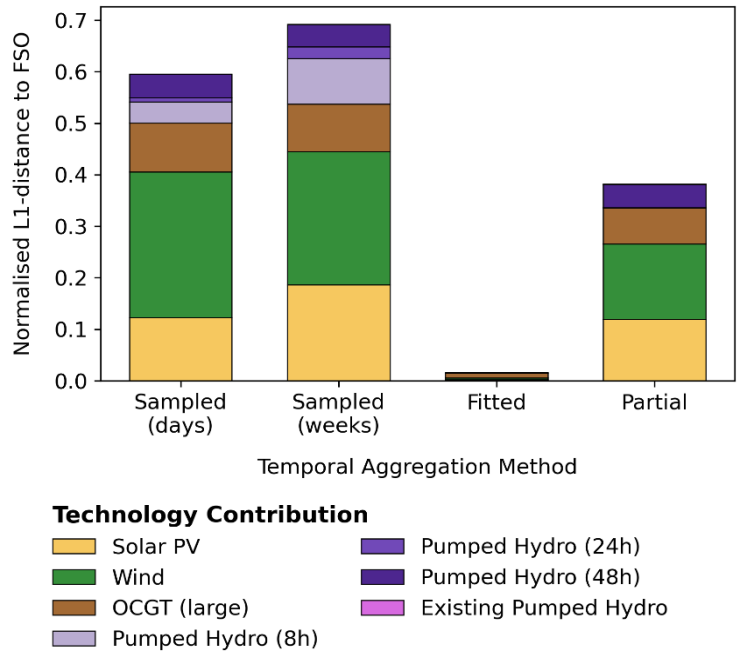

Figure 3. Similarity between solution vectors for each temporal aggregation method and the FSO. A small L1-distance indicates the solution is more similar to the FSO.

Figure 4 provides a summary of how each PLEXOS temporal aggregation method influences new build pumped hydro deployment, OCGT development, and total system costs for the least-cost solutions for each scenario, optimised over the 10 reference weather years. As shown in Figure 4a, total pumped hydro maximum build limits were not a binding constraint in the Original PHES Assumptions scenario because of the high assumed capital costs at most nodes. Once the capital costs were adjusted to reflect mainland Australia having equivalent low-cost pumped hydro options compared to Tasmania in the Improved PHES Assumptions scenario, new build pumped hydro capacity exceeded the original *2024 ISP Model* build limits when using the Fitted method. Excluding OCGT from the model in the 100% Renewables scenario resulted in all temporal aggregation methods building substantially more pumped hydro than the original *2024 ISP Model* build limits.

Without the option of developing gas turbines for balancing in the 100% Renewables scenario, the Fitted method invested in 1200 GWh of new build pumped hydro. The Fitted method invested in 840 GWh more pumped hydro than the Sampled (weeks) method, and 960 GWh more than the Sampled (days) method. This difference is the equivalent of 2.5–3 long-duration pumped hydro projects at the scale of Snowy 2.0, purely based upon the choice of temporal aggregation method. Part of this difference is driven by the heavy bias towards short-duration storage in the typical period methods, as evident in Figure 4b. Even in the absence of OCGT for balancing in the 100% Renewables scenario, the Sampled (days) and Sampled (weeks) methods developed new pumped hydro systems with an average storage duration of just 8 and 12 hours respectively. Both the Fitted and Partial methods invested in longer-duration pumped hydro storage, with an average duration of 75 and 87 hours respectively – much higher than the original maximum storage duration in the *2024 ISP Model*.

As shown in Figure 4c, replacing all OCGT capacity in the system required explicitly optimising for a 100% Renewables solution. Figure 4d shows that the cost of the 100% Renewables solutions was much more similar to the gas-dependent solutions when using the Fitted, Partial, and Sampled (weeks) methods. The solutions to these optimisations could be considered to exist in the same near-optimal space. However, as shown in Supplementary Information A, Figure S3, the Partial method produced solutions with an unserved energy of 0.16–0.94% of total demand when validated with a full time-series unit commitment optimisation, making the solutions far less reliable than other temporal aggregation methods.

*Figure 4. Influence of temporal aggregation method in each scenario on: (a) new build pumped hydro energy capacity, (b) average duration of new build pumped hydro, (c) new build OCGT capacity, and (d) total system cost over 10-year modelling horizon*

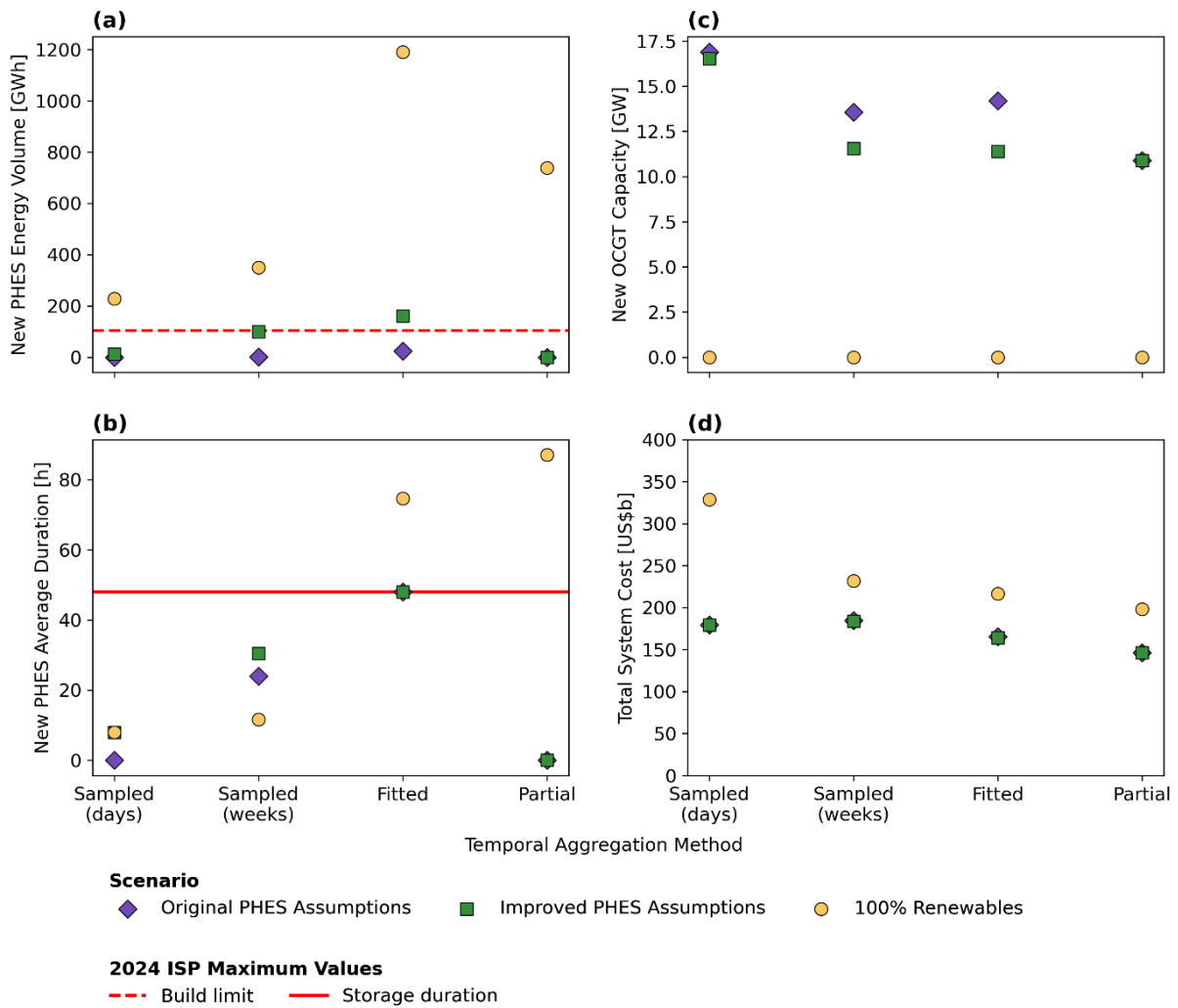

The Sampled (days) method produced a solution that was 83% more expensive in the 100% Renewables scenario compared to the Improved PHES Assumptions scenario. In the absence of gas for balancing, the typical days temporal aggregation technique opted to substantially over-build solar PV and wind capacity rather than develop large-scale energy storage systems. A modeller relying upon the Sampled (days) method for temporal aggregation would probably conclude that gas was an essential component of a reliable electricity system, and that a 100% renewable electricity network would be extremely expensive relative to the alternative. The failure to capture long-duration energy storage behaviour in the unit commitment component of the linear programming formulation will bias energy plans towards gas for balancing solar and wind generation.

## Searching the Near-optimal Space for Solutions with Minimal Gas

The similarity of costs between the gas-dependent and 100% renewable systems for three of the temporal aggregation methods indicates that there is a wide and relatively flat region in the solution space surrounding the global optimum. When forecasting fuel costs, technology costs, possible weather data traces, discount factors, and other parameters out to 2052, the small variation in total system costs within this flat near-optimal region is expected to be substantially smaller than the parametric and structural errors in the long-term energy planning model. Capacity expansion models ought to support planning and decision-making for the energy transition, they are not intended to predict the future. There is value in considering other near-optimal solutions as viable pathways, providing the energy planner with a range of reasonable energy transition options in the event of changes in costs, constraints (e.g., supply chain bottlenecks that slow the deployment of a particular technology), or objectives (e.g., increased pressure to develop a zero emissions grid).

To this end, the FIRM BR-LTP model was used to evaluate over 2 million solutions in two regions surrounding the global optima of the Improved PHES Assumptions and 100% Renewables scenarios. Such an exercise is impractical with a linear programming formulation, such as PLEXOS, because of the longer time required to evaluate each solution. The heuristic unit commitment of FIRM allows each near-optimal solution to be evaluated for reliability and approximate cost in a matter of seconds, but it does mean that BR-LTP models will over-build and over-use capacity relative to a linear programming model. To manage this over-build, flexible generation capacity was included in all FIRM scenarios. Using the Fitted method in PLEXOS, a second polishing step was performed for a sample of 20 solutions to replace this flexible capacity with the minimal OCGT (Low Pumped Hydro region) or solar PV and wind (High Pumped Hydro region) required for a reliable system.

*Solutions in the Near-optimal Regions*

The range of new build capacities in the near-optimal solutions explored by FIRM are summarised in Figure 5. The near-optimal region was based upon solutions that were within 20% of the build costs of the global optimum for the Fitted method, noting that the structural and parametric uncertainty within the model is expected to be in excess of this amount. This is roughly equivalent to considering a 20% contingency overhead for grid investment required by 2052. Scatter points on the boxplots indicate the total capacity of each technology in the polished representative solutions.

On average, the High Pumped Hydro representative solutions required an additional 18 GW of solar PV and 10 GW of wind capacity compared to the Low Pumped Hydro representative solutions. Over a 27-year timeframe (since electricity demand was based on FY2052 estimates), this amounts to an additional 670 MW of solar PV and 370 MW of wind capacity on average per year to displace all gas from the representative grids, assuming that sufficient pumped hydro capacity is developed concurrently. In 2024, Australia commissioned 5190 MW of solar PV and 2380 MW of wind [2]. Therefore, the additional solar PV and wind capacity required to support the long-duration pumped hydro to displace all gas generation represents a very modest fraction of current deployment rates. Eliminating gas from the electricity system is mostly contingent upon support for additional large-scale long-duration pumped hydro energy storage systems.

New build pumped hydro capacity in the Low Pumped Hydro near-optimal region explored by FIRM was in the range of 69 to 880 GWh. The High Pumped Hydro near-optimal region explored by FIRM contained candidate solutions with a range of 1200 to 2000 GWh of new build pumped hydro. Australia currently has a development pipeline of 192 GWh of additional pumped hydro [90].[7] In the absence of additional announcements for large-scale pumped hydro systems, Australia will continue down an energy transition pathway that is likely to depend upon gas for balancing.

*Figure 5. Boxplots depicting the range of capacities for solutions in the near-optimal space explored by FIRM. New build power capacity is depicted in (a), and new build energy volume in (b). Whiskers extend to the minimum and maximum values. A separate boxplot is provided for the Low Pumped Hydro and High Pumped Hydro regions.*

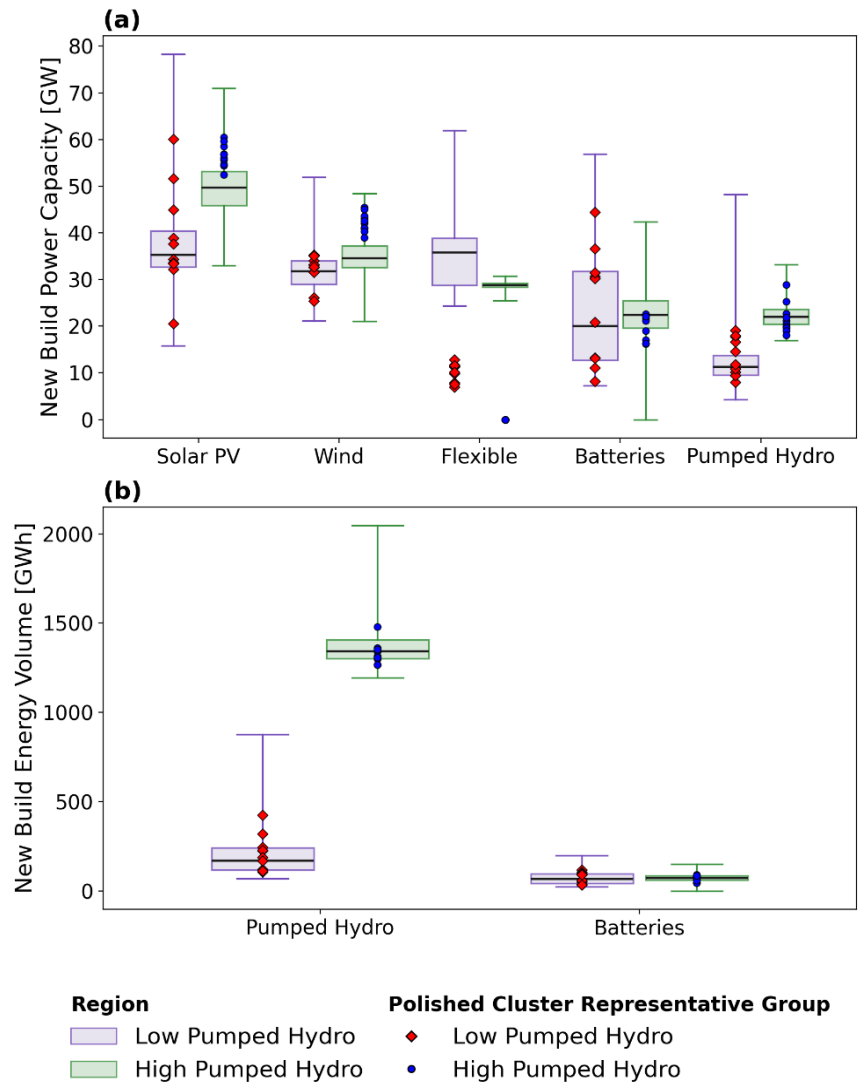

### Conditions that Determine Cost Competitiveness of Near-optimal Solutions

The total cost of the representative solutions was evaluated for a range of real (inflation-free) discount rates, gas turbine costs, pumped hydro costs, and gas fuel costs. The all-in levelised cost of energy (LCOE) for the system is used as the cost metric. Results for the sensitivity analysis are provided in Figure 6.

Real discount rate heavily impacts the levelised cost of capital-intensive energy technologies including solar PV, wind, pumped hydro, batteries and transmission. The appropriate real discount rate for a government project or a regulated monopoly is usually considerably below the rate used by a private company because of perceived lower risk and lower requirement for a return to shareholders. The Australian Government treasury bonds are currently on issue with coupon rates of up to 4.75%. Adjusted by an All Groups CPI inflation rate of 3.8%,[8] the current real bond rate is less than 1% [91, 92]. Natural monopolies, including transmission and distribution, have a rate of return that is regulated by the Australian Energy Regulator (AER) [93]. At the time of publishing the *2025 Inputs, Assumptions, and Scenarios Report*, the latest determination by the AER reflected a real discount rate of 3% for regulated monopolies [94].

---

[7] Snowy 2.0, Kidston, Lake Cethana, Shoalhaven expansion, and Borumba were excluded from this total, since they are already counted as committed or anticipated capacity within the modelling. Pioneer-Burdekin was also excluded because the project has since been cancelled.

*Figure 6. Sensitivity of all-in LCOE for representative near-optimal solutions to: (a) real discount rate, (b) gas fuel costs, (c) gas turbine capital costs, (d) pumped hydro energy volume capital costs, (e) pumped hydro power capacity capital costs, and (f) pumped hydro economic life. Baseline values assume 3% real discount rate and 75-year economic life for pumped hydro.*

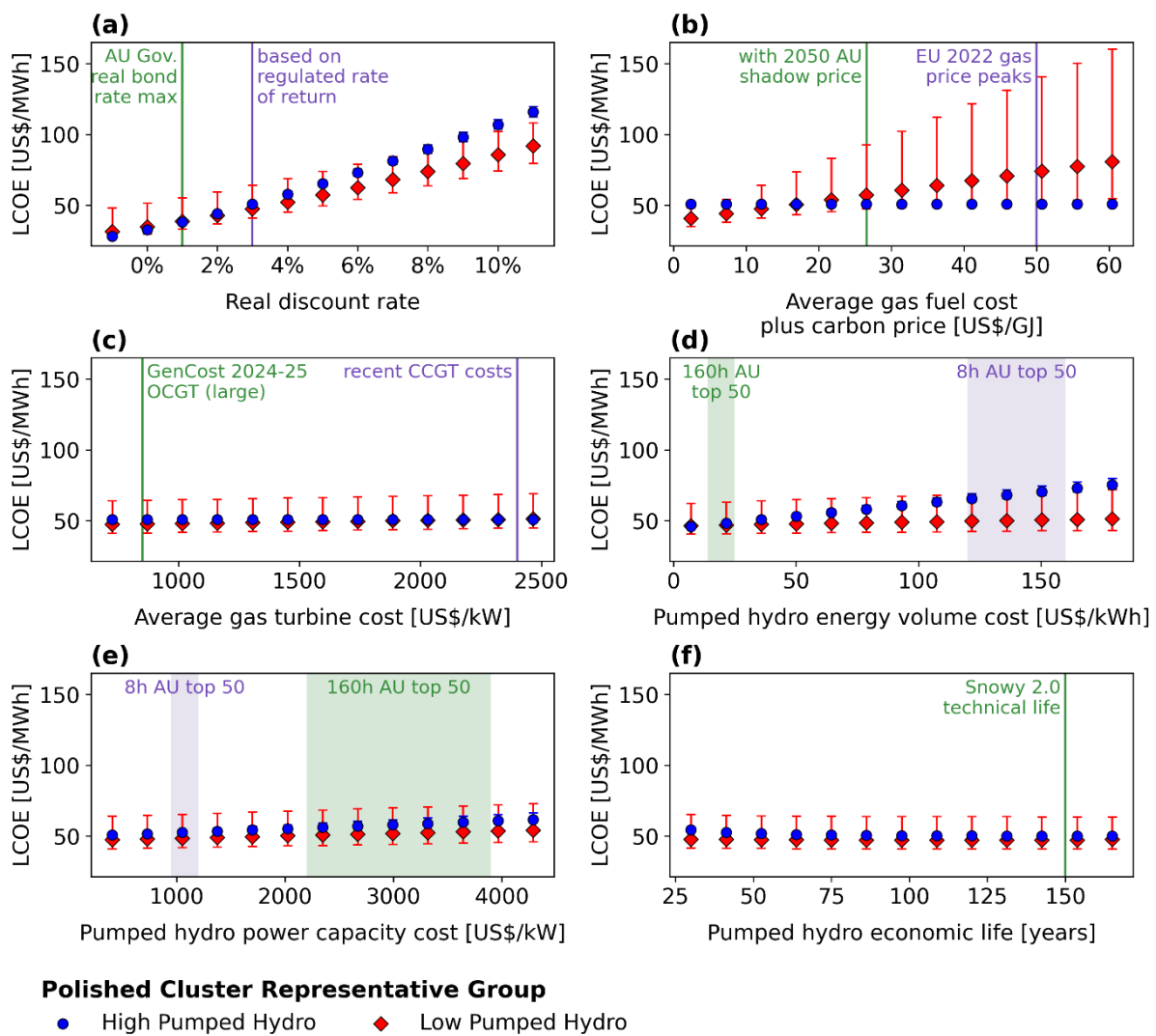

A serious shortcoming of many analyses is the selection of an overly large discount rate and short economic lifetime for large-scale pumped hydro systems. Our analysis suggests that just a few additional Snowy 2.0-sized systems are required to support a 100% renewable NEM, meaning that they constitute a natural monopoly akin to transmission. The discount rate pertaining to dozens of solar farms or gas turbines with lifetimes of 30 years could be quite inappropriate for a small number of large monopolistic assets with design lifetimes of 150 years. Similarly, the economic lifetime for a large-scale civil infrastructure asset that is government-owned or a regulated monopoly is expected to be longer than for private investment. The annualised build cost of an investment with a real discount rate of 3% and economic lifetime of 75 years (baseline for Figure 6) is less than half that of a real discount rate of 7% and economic lifetime of 40 years (baseline for Supplementary Information A, Figure S9).[9] For representative solutions evaluated in this analysis, increasing the real discount rate from 3% to 7% increased the all-in LCOE by up to 36%. Reducing the real discount rate from 3% to 1% decreased the LCOE by up to 50%.

---

[8] Australia's inflation rate in the 12 months to October 2025.

The gas fuel costs in this analysis were assumed to be US$12–13/GJ, as assumed for the original *2024 ISP Model*. Sensitivity to gas fuel price was highly variable since it was dependent on the contribution of gas to the generation mix in each representative solution. Gas prices are heavily dictated by unpredictable geopolitical and economic events. Fossil gas spot prices in Europe were regularly over US$50/GJ in 2022, primarily driven by the Russian invasion of Ukraine. The Australian Energy Market Commission (AEMC) estimated a value of emissions reduction of AU$75/t-$CO_2$-eq for 2025 rising to AU$420/t-$CO_2$-eq in 2050[10] [95]. By 2050, this would add a shadow price to fuel of about US$14.5/GJ for gas generators[11] – roughly equivalent to doubling the gas price for market bodies considering the emissions component of the National Electricity Objective. Doubling the gas price resulted in an increase in LCOE of between 2–28% for the Low Pumped Hydro representative solutions (average increase of 12%).

Production backlogs of seven to eight years are currently forecast for new gas turbines [96]. The high demand has more than doubled the cost of turbines in some markets, with new combined-cycle gas turbine (CCGT) facilities now costing approximately US$2400/kW [97]. The large OCGT generators for this analysis were assumed to be just US$747–780/kW, as assumed in the original *2024 ISP Model*. Low Pumped Hydro representative solutions had an increase in total costs of 4–9% when the gas turbine cost was increased to be roughly equivalent to recent CCGT costs.

The ANU pumped hydro cost model, including a 50% added overhead, was applied to all 1529 class A, AA, and AAA 15–500 GWh sites near the NEM transmission network, excluding national parks and large urban areas (refer Supplementary Information B for details). The range of capital costs across the top 50 sites for each duration was US$950–1200/kW or US$120–160/kWh (8-hours), US$1200–1800/kW or US$51–73/kWh (24-hours), US$1400–2300/kW or US$30–47/kWh (48-hours), and US$2200–3900/kW or US$14–25/kWh (160-hours). For comparison, Snowy 2.0 (2.2 GW, 350 GWh) will cost US$4500/kW or US$29/kWh.[12]

Although Snowy 2.0 has a large head (660 m) and utilises two existing reservoirs, it does include a very long tunnel (27 km) and is situated in a remote alpine National Park. The average tunnel length for the top 50 sites with 160-hour duration was 15 km, with a minimum value of 5 km. Short pressure tunnels will reduce the power capacity cost of the pumped hydro. Sites with large water-to-rock ratios (small dam walls impounding a large volume of water), large heads, and a large size

---

[9] Real discount rate 3%, economic lifetime of 75 years based on the useful life of civil works used by Snowy Hydro (government-owned corporation) [152]: annuity factor = $\frac{0.03}{1-(1/(1+0.03)^{75})} = 0.034$

Real discount rate 7%, economic lifetime of 40 years: annuity factor = $\frac{0.07}{1-(1/(1+0.07)^{40})} = 0.075$

[10] 2023 AUD. Roughly US$51/t-$CO_2$-eq in 2025 and US$287/t-$CO_2$-eq 2025 USD.

[11] Scope 1 emissions intensity of 373.55 kg/MWh for CCGT and 548.73 kg/MWh for large OCGT. Average heat rate of 7.25 GJ/MWh for CCGT and 10.93 GJ/MWh for large OCGT. Therefore, 51.5 kg/GJ for CCGT and 50.2 kg/GJ for OCGT. Emissions intensity and average heat rates from [88].

[12] Assuming a final cost of roughly US$10 billion.

(maximising economies of scale) will minimise the energy volume cost. The LCOE of the High Pumped Hydro representative solutions is more sensitive to energy volume cost than power capacity cost over the range of values estimated for the top 50 sites (refer Figure 6d and Figure 6e), meaning that a focus on small-scale short-duration systems (<48 hours) is likely to drive up total system costs.

# Discussion and Conclusion

Ignoring low-cost, reliable, low-gas solutions in long-term energy plans carries significant risk. Greenhouse gas emissions related to fossil gas are generated along the entire supply chain. Oil and gas emissions are likely 30% higher than the total reported in UNFCCC reports [98]. The International Energy Agency (IEA) estimated approximately 81 Mt of oil and gas methane emissions in 2024, with 77% produced through upstream processes (production, gathering and processing); 13% from transport (e.g., pipelines), storage and refining; and the remainder from abandoned oil and gas wells or satellite-detected large emissions [99]. This methane has a global warming potential 30 times higher than carbon dioxide over a 100-year time horizon and 83 times higher over a 20-year horizon [100]. These methane emissions are in addition to those that are actually produced from the combustion of oil and gas. There are no commercial-scale gas-fired power stations with carbon capture and storage operating anywhere in the world [101], meaning that there are currently no options for eliminating combustion emissions either.

A recent advisory opinion from the International Court of Justice stated that [102]:

> Failure of a State to take appropriate action to protect the climate system from GHG emissions — including through fossil fuel production, fossil fuel consumption, the granting of fossil fuel exploration licences or the provision of fossil fuel subsidies — may constitute an internationally wrongful act which is attributable to that State.

Ongoing approval to develop new natural gas-fired power stations while there is an alternative option for balancing solar and wind generation, in the form of long-duration pumped hydro coupled with short-duration batteries, may place a State at risk of breaching conventional and customary obligations under international law. These obligations, including the requirement to prevent significant harm to the environment, exist independent of the treaty obligations of any particular State. Aside from the emissions, gas-fired generators are exposed to financial risks associated with fuel price volatility and ongoing gas turbine shortages.

Unfortunately, existing long-term energy planning models rely upon techniques that are strongly biased towards gas and away from the alternative of long-duration energy storage. Eight of the long-term energy plans summarised in Supplementary Information A, Table S1 specified that they used a temporal aggregation method based upon typical periods. The PLEXOS sampled chronology methods evaluated for this analysis were incapable of modelling storage cycling behaviour longer than the typical period, which biased the optimisation towards investing in shorter 8- or 24-hour duration pumped hydro systems. Energy planners using some variation of the Sampled (days) method could incorrectly conclude that a 100% renewable electricity system is significantly more expensive than a gas-dependent system. Energy planners using a variation of the Partial method could instead erroneously conclude that 100% renewable electricity systems are less reliable than systems containing gas because the method will under-build storage volume since it cannot track the state-of-charge within the LDCs.

The only temporal aggregation method evaluated in this paper that reasonably captured long-duration energy storage behaviour was the Fitted segmentation method, which retained full chronology of the time-series data. The Fitted method also produced a solution that most closely resembled the FSO, with a normalised L1-distance between the solution vectors of just 2–3% of the total system costs of the FSO. The DLT model for the AEMO ISP was the only model summarised in Supplementary Information A, Table S1 that specified using a segmentation method for temporal aggregation.

Coupled with temporal aggregation methods that are not fit-for-purpose, outdated pumped hydro assumptions mean that existing energy plans are unlikely to consider any viable long-duration storage technologies that can compete with gas. Most regions of the world have orders of magnitude more off-river pumped hydro potential than would ever be required, as shown by the 818,000 sites with a storage potential of 86 million GWh on the global pumped hydro atlases. Existing energy plans greatly underestimate the maximum build limits for pumped hydro. High capital cost assumptions, potentially calibrated using existing on-river pumped hydro systems with low head and short storage duration, neglect the much better technical characteristics of modern off-river pumped hydro. By lowering pumped hydro capital costs across the NEM to be equivalent to Tasmanian sites in the *2024 ISP Model*, relaxing the build limits, and adding a long-duration 160-hour option, a 100% renewable electricity solution was found with approximately 1200 GWh of new build pumped hydro which had an average storage duration of 75 hours. This system had a very similar cost to the gas-dependent scenarios. The costs of a sample of 100% renewables systems were found to be sensitive to pumped hydro energy volume cost, making it essential to focus on large-scale long-duration sites with a low water-to-rock ratio and large head to minimise system costs.

Large-scale long-duration energy storage cannot be supplied by batteries at reasonable cost. The annualised build cost of 1200 GWh of batteries would be 6–12 times larger than that of the pumped hydro systems developed by the Fitted optimisation of the 100% Renewables scenario – equivalent to between US$17–20 billion extra per year.[13] The large difference in annualised cost is a combination of pumped hydro having a substantially lower cost of energy volume at large scales compared to batteries, as well as a much longer economic lifetime.

This is not to say that batteries are never competitive with pumped hydro, just that they are not cost-effective at very large energy volumes. Each scenario included 51 GW, 169 GWh of initial battery capacity which predominantly provided high-power intra-day storage that was complementary to the long-duration balancing. During winter *dunkelflaute* in mid- to high-latitude regions, long-duration pumped hydro can trickle-charge short-duration batteries during the day, making sure they are available to meet the evening demand peak [81]. A combination of

---

[13] The Commonwealth Scientific and Industrial Research Organisation (CSIRO) estimates a capital cost of US$190/kWh (AU$292/kWh) for 24-hour vanadium redox flow batteries [85] with a lifetime of up to 20 years [69]. We assumed pumped hydro capital costs derived from Tasmanian costs in the 2024 ISP, with US$2234/kW for 48-hour and US$6454/kW for 160-hour systems. Therefore,

Annualised battery build cost = $\frac{1200 \text{ [GWh]} \times 190 \text{ [\$/kWh]} \times 0.07}{1 - (1/(1+0.07)^{20})}$ = US$21 billion

Annualised pumped hydro build cost (Private, non-monopolistic: 7% real discount rate, 40-year economic life) = $\frac{(3.8 \text{ [GW]} \times 6454 \text{ [\$/kW]} + 12 \text{ [GW]} \times 2234 \text{ [\$/kW]}) \times 0.07}{1 - (1/(1+0.07)^{40})}$ = US$3.9 billion

Annualised pumped hydro build cost (Regulated monopoly: 3% real discount rate, 75-year economic life) = $\frac{(3.8 \text{ [GW]} \times 6454 \text{ [\$/kW]} + 12 \text{ [GW]} \times 2234 \text{ [\$/kW]}) \times 0.03}{1 - (1/(1+0.03)^{75})}$ = US$1.7 billion

pumped hydro and batteries can provide all of the frequency control, voltage control, black-start, and spinning reserve ancillary services that a gas generator can provide. In the NEM, 56% of frequency control ancillary services are already supplied by batteries, 12% by hydro, and less than 1% by gas [103]. Future work should investigate a range of 100% renewable electricity pathways to find those that offer the most resilient, secure power system and reliable supply of electricity.

Existing long-term energy plans are often focused on a single least-cost solution per scenario, because model complexity makes computation slow. Substantial parametric and structural uncertainty means that the solutions found through these optimisations are expected to contain a high degree of error, limiting the usefulness of a single global optimum. There is a large body of evidence demonstrating the existence of a broad near-optimal solution space consisting of a wide range of viable energy transition pathways [61]. As BR-LTP models mature further, they may offer a good quality option for rapidly exploring the near-optimal space. The philosophy of the BR-LTP class of models focuses on rapidly evaluating feasible solutions that achieve the normative objectives of the energy planner, not on finding the global least-cost solution (which is already distorted by temporal aggregation and uncertainty in the linear programming formulations). Energy planners could use similar techniques to generate a wide range of energy transition pathways that they can evaluate further using additional information that is not captured within the energy system model.

Energy planners have objectives beyond developing the lowest-cost electricity grid. In Australia, the legislated National Electricity Objective[14] must be used to guide activities by AEMO [104]. The National Electricity Objective requires consideration of price, quality, safety, reliability, security of supply, as well as the achievement of targets to reduce greenhouse gas emissions. The ISP is currently performed as a cost-minimisation (achieving the "price" objective) with constraints based upon jurisdictional renewable energy and greenhouse gas emission reduction targets, as well as a separate power system assessment to iteratively verify reliability and security. A near-optimal solution that minimised or eliminated all fossil fuels from the electricity system could still achieve the low-price objective, while also enhancing the ability to meet Australia's target of 62–70% reduction in emissions by 2035[15] and net zero greenhouse gas emissions by 2050 [105]. The same could be said for any country looking for a pathway to increase the ambition of their Nationally Determined Contribution (NDC) under the Paris Agreement.

Australia currently has a development pipeline of about 192 GWh of additional pumped hydro energy storage, in addition to about 416 GWh of anticipated and committed projects. Even if every single one of these projects is eventually commissioned, Australia will likely remain on a path that depends upon fossil fuel gas for decades to come. Under current policies, the *2025 World Energy Outlook* describes a future with 5400 GW of fossil fuel power stations generating 19,000 TWh of electricity in 2050 [106]. This is not the only option. Electricity grids that replaced gas with long-duration pumped hydro energy storage were found to have comparable cost to gas-based systems. Refocusing long-term energy planning tools on this opportunity to completely decarbonise the electricity system in all regions of the world is an important step in solving the global climate crisis.

---

[14] *National Electricity (South Australia) Act 1996* (SA) Schedule Part 1 Section 7. While this is an Act of Parliament for the state of South Australia, other states and territories that participate in the NEM have application statutes that adopt the law.
[15] Relative to 2005 baseline.

# Methods

The design of the modified *2024 ISP Model* in PLEXOS, the structure of the FIRM BR-LTP model, details on the L1-distance calculation, and process for searching the near-optimal solution space are provided in the sections below. § "Code Availability" provides directions for accessing the PLEXOS model files and FIRM BR-LTP code developed in accordance with these methods. All data required to reproduce the scenarios evaluated in this paper are provided in § "Data Availability". The detailed equations and algorithms that implement the FIRM BR-LTP formulation are provided in Supplementary Information A – FIRM Model Formulation.

## Modified *2024 ISP Model* in PLEXOS

The publicly available *2024 ISP Model* developed by AEMO [86] was downloaded and modified. The 2024 ISP solar and wind traces developed by AEMO were downloaded and added to the model [107, 108]. A detailed list of modifications with an explanation for each change is provided in Supplementary Information A, Table S3.

At a high level, the *2024 ISP Model* was modified from a full capacity expansion model into a simplified point-in-time model (i.e., all new build capacity is developed in the first year). The model finds the least-cost grid configuration to reliably meet FY2052 demand over 10 reference years of weather data (July 2042 – July 2052 from the *2024 ISP Model*). The removal of full capacity expansion behaviour allows the effects of temporal simplification on a single grid configuration to be evaluated without added complexity from construction lead times, retirement, or annual demand growth. Future work will investigate how these dynamics influence the trade-off between developing gas generators or energy storage for balancing solar and wind dominant grids.

Existing, anticipated and committed assets from the original *2024 ISP Model* were treated as initial capacity in the modified PLEXOS model. This includes solar PV (12 GW), onshore wind (15 GW), conventional hydro (6.4 GW), batteries (51 GW, 169 GWh), and pumped hydro (4.7 GW, 435 GWh). In each scenario, the system has 12 nodes (buses), 12 interconnections, 296–308 generators (66–78 for new build capacity), and 157–183 storage systems (60–88 for new build capacity).

New build capacities for solar PV and wind (onshore and offshore) in Renewable Energy Zones (REZs), large open-cycle gas turbines (OCGT), batteries (1–8-hour duration), and pumped hydro (8–160-hour duration) were optimised by the model, since these are overwhelmingly dominant in Australian and global electricity capacity construction. Large OCGTs have a lower capital cost than CCGTs, making them more competitive when operating with a low capacity factor (e.g., peaking and seasonal operation) despite their lower efficiency.

The assumed cost of the 160-hour pumped hydro was derived from the 48-hour Tasmanian pumped hydro cost, assuming the cost of power capacity ($/kW) and energy capacity ($/kWh) was in the same ratio as the class A cut-off of the global pumped hydro atlases. The estimated cost of US$6133/kW (US$38/kWh) was slightly higher than GHDs estimate of US$5390/kW (US$34/kWh) [90], which is higher than the rough cost of Snowy 2.0 at US$4500/kW or US$29/kWh.[16] Therefore, although the capital cost estimates for long-duration pumped hydro are improved relative to the original *2024 ISP Model,* they still remain quite conservative for this analysis to ensure robustness of the conclusions. Selecting the best quality sites for development is expected to make the 100% renewable energy grid configurations even cheaper.

---

[16] Assuming a cost increase to about US$10 billion.

The LT plan settings for each temporal simplification scenario are provided in Supplementary Information A, Table S2.

## FIRM Business Rules-based Long Term Planning Model

The original FIRM model was a bespoke model for Australia's NEM developed by [58]. It is a point-in-time model designed to find the least-cost grid configuration required to provide reliable electricity supply over decades of fully chronological high-resolution data without the need for temporal aggregation. FIRM belongs to the class of BR-LTP models defined in this paper.

For this paper, the FIRM framework was used to develop an entirely new codebase that allows for generic scenarios to be defined. Changing the network topology or adding generator and storage objects no longer requires changes to the codebase. An improved set of business rules was also incorporated for transmission, pre-charging of storage using flexible generators, and inter-storage transfers within the unit commitment algorithm. While this generic FIRM model has been developed as a point-in-time model, modifying the model to have full capacity expansion behaviour would just require additional decision variables to consider new build capacity in each year of the planning horizon.

The FIRM framework has the following basic characteristics:

- A least-cost configuration of the grid that can reliably meet electricity demand over a long planning horizon (≥ 10 years) of high-resolution data (≤ 1 hour) is optimised through a differential evolution. No additional temporal simplification is required for the input data.
- The capacity of each new build generator is a decision variable for the least-cost optimisation.
- Unit commitment is performed sequentially using business rules for each iteration of the differential evolution. No additional decision variables are required for the dispatch of generators and storage systems since unit commitment is not formulated as its own optimisation problem.
- The basic business rules for the unit commitment are:
    - Baseload, solar and wind generators are dispatched first according to exogenously defined availability traces.
    - Surplus generation at each node is transmitted to balance the remaining power deficits at each node.
    - Energy storage systems are then dispatched for each time interval, charging when there is surplus generation and discharging when there is a deficit. Power capacity and energy volume constrain storage system behaviour when necessary.
    - Flexible generation (e.g., gas or conventional hydro) is dispatched to fill the remaining deficit. Annual generation constraints and power capacity limit the dispatch of flexible generators as necessary.
- Unserved energy in excess of the reliability constraint is issued a very high cost through a penalty function. For example, the NEM has a reliability standard of 99.998% energy supplied under clause 3.9.3C(a) of the National Electricity Rules [109]. An additional fixed cost cut-off can be exogenously defined to apply a further penalty (allowing the unit commitment to be skipped for solutions that are likely to have costs exceeding the planning objectives).

- The levelised cost of energy for the total system is calculated. The objective function is the sum of the levelised cost of energy and any penalties.

The unit commitment problem for a population of new build capacity candidates can be run as an embarrassingly parallel process. Adding parallel processes up to the size of the population provides substantial performance improvements to the differential evolution but may demand a large amount of memory for high-resolution systems.

Under the FIRM formulation used for this analysis, storage systems are dispatched in order of shortest to longest storage duration. That is, short-duration batteries are expected to cycle more frequently than longer-duration pumped hydro. Flexible generators are dispatched in order of lowest to highest short-run marginal cost. Unlike PLEXOS, the FIRM model will independently size power and energy capacity for pumped hydro systems.

Pre-charging of the storage systems, including inter-storage transfers, is started upon reaching the end of a block of time intervals that contain unserved energy (i.e., a deficit block). It exploits symmetry relationships for charging/discharging behaviour when moving backwards through time. That is, discharging to balance a deficit at $t-1$ will increase the energy stored in that storage system at time $t-1$ relative to $t$ (and vice versa for charging). Pre-charging occurs according to the following steps (depicted in Figure 7):

1. Iterate backwards through $t$ within the deficit block to find the energy that each storage system must be pre-charged with to balance the unserved energy;
2. Upon reaching the start of the deficit block, iterate backwards through time to determine the charging/discharging powers required to achieve the pre-charging energies.
    a. Prioritise inter-storage transfers from storage systems with excess stored energy (trickle-chargers) to empty storage systems (pre-chargers). Trickle-chargers must retain enough reserves to make it through the deficit block after pre-charging has occurred.
    b. If there are no more trickle-chargers available, use flexible generators to pre-charge storage systems.
3. Once the sequence of charging/discharging powers is determined, iterate forwards through $t$ to enforce the energy volume constraints on the new charging/discharging powers. If the constraint is binding, this implies that the storage configuration cannot feasibly resolve the entire deficit block. Once the algorithm has reached the final $t$ in the original deficit block, return to performing the normal energy balance of residual demand.

The software that implements the FIRM BR-LTP formulation is provided in § "Code Availability". The FIRM BR-LTP has been built using Python 3.12. The unit commitment problem relies upon just-in-time compilation from Numba [110]. Capacity expansion is optimised through the Scipy differential evolution [111]. The Pandas [112, 113] and Numpy [114] packages are other software dependencies. The full formulation for the FIRM BR-LTP model is provided in Supplementary Information A – FIRM Model Formulation. The detailed raw inputs of the FIRM BR-LTP are provided in Supplementary Information C.

*Figure 7. Simplified Depiction of Storage Pre-charging Business Rules for (a) pre-charger, and (b) trickle-charger[17]*

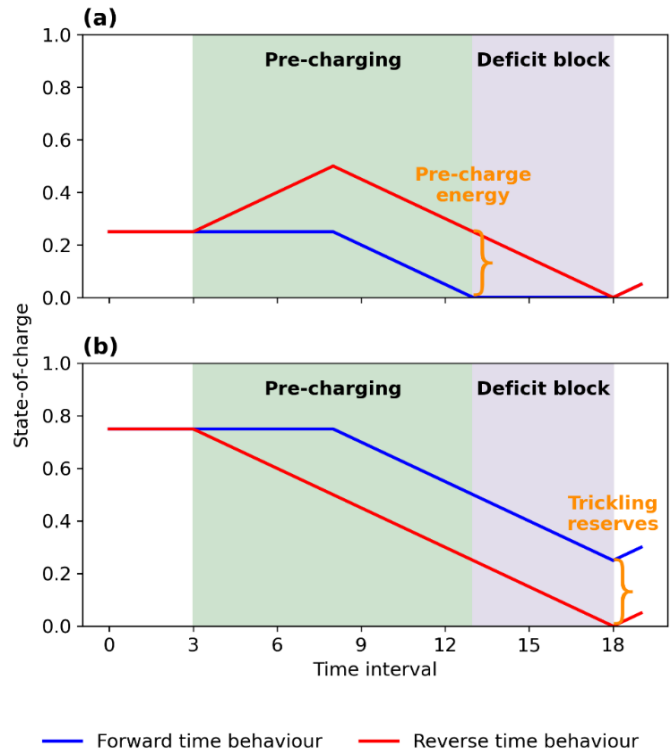

## Calculation of the L1-distance between Solution Vectors

Rather than simply comparing total system costs, the normalised L1-distance metric (adapted from [47] and [6]) is used to validate how close a solution is to the benchmark FSO (refer Eq. (1)).

$$L1 = \frac{|\vec{a} \cdot (\vec{z}^m - \vec{z}^{FSO})|}{SC^{\text{FSO}}} \tag{1}$$

The normalised L1-distance is calculated from:

- the cost vector ($\vec{a}$) which defines annualised build, fixed operation and maintenance, variable operation and maintenance, and fuel cost assumptions.
- the decision vector ($\vec{z}$) representing the new build capacity, total capacity, and total generation of each asset in the optimal solution for temporal aggregation method $m$ and the FSO. Each variable in $\vec{z}$ has a corresponding cost in $\vec{a}$.
- normalisation is performed with respect to the total system costs of the FSO ($SC^{\text{FSO}}$).

In the absence of normalisation, the L1-distance between the solution vectors has units of [$] and refers to the Manhattan distance between the $m$ and FSO vectors of costs for developing, operating and maintaining each separate asset in the system. Two solutions may have identical total system costs, but have a very large difference in the investment and dispatch decisions for each asset in the system leading to a large L1-distance.[18]

---

[17] The trickling reserve is actually the minimum state-of-charge within the deficit block and may not occur in the final deficit interval. Similarly, the maximum state-of-charge within the deficit block of a pre-charger constrains the amount of pre-charge energy.

[18] Strictly speaking, the complete L1-distance between one solution vector and the benchmark would be based upon vectors of all decision variables (including the hundreds of thousands of unit commitment variables). Such a metric would encode information about whether assets in the test system are

## Searching the Near-Optimal Space

The solution space for the NEM model has over 100 dimensions, making it impossible to map the entire space at any useful resolution using either the PLEXOS formulation or FIRM model. Exogenously defined constraints on flexible capacity and pumped hydro build limits were chosen to restrict the optimiser to exploring specific regions of interest in the solution space.

An initial optimisation equivalent to the Improved PHES Assumptions scenario was performed with the FIRM model. The lowest-cost solution from this initial optimisation was used to define the maximum build limits of flexible generation capacity (priced the same as large OCGT) for the subsequent near-optimal searches. A second optimisation was performed, equivalent to the Improved PHES Assumptions scenario with the additional maximum build limit constraints on flexible generation capacity. The differential evolution algorithm generated a broad range of solutions for these two optimisations which formed the Low Pumped Hydro region.

For the third optimisation, an additional minimum build limit was added to pumped hydro systems equal to the pumped hydro capacity in the global optimum of the 100% Renewables scenario that used the Fitted method in PLEXOS. An initial guess equal to the 100% Renewables PLEXOS solution was added to the population. The population of solutions evaluated by the differential evolution for this third optimisation formed the High Pumped Hydro region.

For each optimisation, the differential evolution algorithm iteratively mutated a large population of candidate solutions. For each iteration, the best performing (i.e., lowest cost) candidate solutions were retained in the population, progressively approaching a minimum solution. Every candidate solution tested during the three optimisations was saved.

The Low Pumped Hydro and High Pumped Hydro regions of the solution space were filtered to form near-optimal regions with build costs within 20% of the PLEXOS global optimum determined using the Fitted method. Within the near-optimal regions, candidate solutions were each converted to a corresponding build cost vector. The build cost vector ($\overrightarrow{BCV}$, $) was calculated from the element-wise product of the candidate solution vector ($\vec{\delta}$, kW or kWh), the annuity factor vector ($\overrightarrow{AF}$, unitless) and the capital cost vector ($\overrightarrow{CC}$, $/kW or $/kWh) as shown in Eq. (2). A description of the annuity factor is given in Supplementary Information A, Eq. (S3).

$$\overrightarrow{BCV} = \vec{\delta} \odot \overrightarrow{AF} \odot \overrightarrow{CC} \tag{2}$$

A mini-batch k-means algorithm was used to cluster near-optimal solutions in the Low Pumped Hydro and High Pumped Hydro regions separately. Distance for the purposes of clustering was based upon $\overrightarrow{BCV}$. The medoid $\vec{\delta}$ of each of the 20 clusters (10 per region), determined by finding the $\overrightarrow{BCV}$ closest to the cluster mean, was chosen as the cluster representative. By clustering according to $\overrightarrow{BCV}$, a variety of different grid configurations in the near-optimal regions could be sampled for polishing.

The new build power capacity and energy volumes of each asset were extracted from the representative $\vec{\delta}$ and used to define the units (solar PV and wind generators in REZs, as well as batteries), max build limits (OCGT generators), max power (pumped hydro) and capacity (pumped hydro) of scenarios in PLEXOS for the Low Pumped Hydro representative group. For the High

---

dispatched at similar times and powers to the FSO model. For simplicity, decision variables have been aggregated across the optimisation horizon before finding the difference, thereby focusing on long-term investment and energy supply outcomes.

Pumped Hydro group, $\vec{\delta}$ was used to define units (batteries), min build limits (solar PV and wind in REZs), max power (pumped hydro) and capacity (pumped hydro) in the corresponding PLEXOS scenarios. OCGT units were set to 0 for the High Pumped Hydro candidate solutions.

Each representative candidate was optimised using the Fitted method in PLEXOS to find the minimum OCGT capacity (High Pumped Hydro representatives) or additional solar PV and wind capacity (High Pumped Hydro representatives) required for a reliable grid configuration. These polished solutions were evaluated through the sensitivity analysis relative to real discount rate, gas generator costs, pumped hydro capital costs, and gas fuel costs. The full details of the sensitivity analysis data are available in an Excel workbook in Supplementary Information B.

# Data availability

Data inputs to the PLEXOS and FIRM models (solar traces, wind traces, demand traces, and hydro/gas annual generation constraints), data for figures and calculations, and raw output data from the models are provided in Supplementary Information C. Currency values in Supplementary Information C are in 2023 AUD and have been converted to 2025 USD using an exchange rate of 1 AUD to 0.65 USD and Australian All Groups CPI inflation.

Data for pumped hydro sites on the global atlases can be filtered using the Pumped Hydro Shortlisting Tool: https://re100.anu.edu.au/shortlisting/

# Code availability

The modified *2024 ISP Model* is provided in Supplementary Information C. The FIRM model developed for this analysis is available on GitHub: https://github.com/TimWeberRE100/FIRM_CE

# Supplementary Information

Supplementary Information B:
https://www.dropbox.com/scl/fo/6khb8uri25qt2w5q5m0gm/ADL9LuZy8yAj9VS3F27_7Co?rlkey=cc910eyg7kbigq22jqonfnqi8&st=6697i0jv&dl=0

Supplementary Information C:
https://www.dropbox.com/scl/fo/b962jbf2rpd211b82o2d0/AAnlr1Visvzz3nPtu33bCxw?rlkey=d3n9bvxldckzvv8nyyknj7h8a&st=me86uy6c&dl=0

# Acknowledgements

We would like to offer our thanks to Energy Exemplar for providing Timothy Weber with a PLEXOS student licence for this analysis. This research was supported by an Australian Government Research Training Program (RTP) Domestic Scholarship, an Australian Government Research Training Program (AGRTP) Fee Offset Scholarship, and the Australian Centre for Advanced Photovoltaics (ACAP). Service units for the Gadi supercomputer were awarded by the National Computational Infrastructure through the ANU Startup Scheme [project ce47].

## Declaration of competing interests

The authors declare the following financial interests/personal relationships which may be considered as potential competing interests: Timothy Weber, Cheng Cheng, Harry Thawley, Andrew Blakers, Bin Lu, and Anna Nadolny have previously received or currently receive funding from the Australian Renewable Energy Agency (ARENA) and Department of Foreign Affairs and Trade (DFAT). DFAT funding is related to separate projects supporting the uptake of solar PV and pumped hydro in Southeast Asia. ARENA and DFAT did not provide funding for this research article. Timothy Weber is a member of the Labor Environment Action Network (LEAN).

## References

[1] IRENA, "Renewable Capacity Highlights 2025," 26 March 2025. [Online]. Available: https://www.irena.org/-/media/Files/IRENA/Agency/Publication/2025/Mar/IRENA_DAT_RE_Capacity_Highlights_2025.pdf. [Accessed 7 May 2025].

[2] IRENA, "Renewable Capacity Statistics 2025," March 2025. [Online]. Available: https://www.irena.org/Publications/2025/Mar/Renewable-capacity-statistics-2025. [Accessed 7 May 2025].

[3] IAEA, "Power Reactor Information System: NPP Status Changes (2024)," 2025. [Online]. Available: https://pris.iaea.org/pris/. [Accessed 7 May 2025].

[4] M. Frysztacki, J. Hörsch, V. Hagenmeyer and T. Brown, "The strong effect of network resolution on electricity system models with high shares of wind and solar," *Applied Energy,* vol. 291, p. 116726, 2021.

[5] K. Poncelet, E. Delarue, D. Six, J. Duerinck and W. D'haeseleer, "Impact of the level of temporal and operational detail in energy-system planning models," *Applied Energy,* vol. 162, pp. 631-643, 2016.

[6] J. Merrick, "On representation of temporal variability in electricity capacity planning models," *Energy Economics,* vol. 59, pp. 261-271, 2016.

[7] J. Bistline, "The importance of temporal resolution in modeling deep decarbonization of the electric power sector," *Environmental Research Letters,* vol. 16, no. 8, p. 084005, 2021.

[8] A. Belderbos and E. Delarue, "Accounting for flexibility in power system planning with renewables," *International Journal of Electrical Power & Energy Systems,* vol. 71, pp. 33-41, 2015.

[9] C. Heuberger, I. Staffell, N. Shah and N. Dowell, "A systems approach to quantifying the value of power generation and energy storage technologies in future electricity networks," *Computers & Chemical Engineering,* vol. 107, pp. 247-256, 2017.

[10] D. Kirschen, J. Ma, V. Silva and R. Belhomme, "Optimizing the flexibility of a portfolio of generating plants to deal with wind generation," *2011 IEEE power and energy society general meeting,* pp. 1-7, 2011.

[11] M. Jafari, M. Korpås and A. Botterud, "Power system decarbonization: Impacts of energy storage duration and interannual renewables variability," *Renewable Energy,* vol. 156, pp. 1171-1185, 2020.

[12] L. Kotzur, L. Nolting, M. Hoffmann, T. Groß, A. Smolenko, J. Priesmann, H. Büsing, R. Beer, F. Kullmann, B. Singh and A. Praktiknjo, "A modeler's guide to handle complexity in energy systems optimization," *Advances in Applied Energy,* vol. 4, p. 1000063, 2021.

[13] S. Pfenninger, "Dealing with multiple decades of hourly wind and PV time series in energy models: A comparison of methods to reduce time resolution and the planning implications of inter-annual variability," *Applied Energy,* vol. 197, pp. 1-13, 2017.

[14] M. Prina, G. Manzolini, D. Moser, B. Nastasi and W. Sparber, "Classification and challenges of bottom-up energy system models - A review," *Renewable and Sustainable Energy Reviews,* vol. 129, p. 109917, 2020.

[15] C. Nweke, F. Leanez, G. Drayton and M. Kolhe, "Benefits of Chronological Optimization in Capacity Planning for Electricity Markets," *2012 IEEE international conference on power system technology (POWERCON),* pp. 1-6, 2012.

[16] F. Murphy and Y. Smeers, "Generation Capacity Expansion in Imperfectly Competitive Restructured Electricity Markets," *Operations Research,* vol. 53, no. 4, pp. 646-661, 2005.

[17] S. Kazempour, A. Conejo and C. Ruiz, "Strategic Generation Investment Using a Complementarity Approach," *IEEE Transactions on Power Systems,* vol. 26, no. 2, pp. 940-948, 2011.

[18] M. Caramanis, R. Tabors, K. Nochur and F. Schweppe, "The Introduction of NonDIispatchable Technologies a Decision Variables in Long-Term Generation Expansion Models," *IEEE Transactions on Power Apparatus and Systems,* Vols. PAS-101, no. 8, pp. 2658-2667, 1982.

[19] S. Wogrin, P. Duenas, A. Delgadillo and J. Reneses, "A new approach to model load levels in electric power systems with high renewable penetration," *IEEE Transactions on Power Systems,* vol. 29, no. 5, pp. 2210-2218, 2014.

[20] S. Wogrin, D. Galbally and J. Reneses, "Optimizing storage operations in medium-and long-term power system models," *IEEE Transactions on Power Systems,* vol. 31, no. 4, pp. 3129-3138, 2015.

[21] M. Hoffman, L. Kotzur, D. Stolten and M. Robinius, "A Review on Time Series Aggregation Methods for Energy System Models," *Energies,* vol. 13, no. 3, p. 641, 2020.

[22] P. de Guibert, B. Shirizadeh and P. Quirion, "Variable time-step: A method for improving computational tractability for energy system models with long-term storage," *Energy,* vol. 213, p. 119024, 2020.

[23] M. Hoffman, L. Kotzur and D. Stolten, "The Pareto-optimal temporal aggregation of energy system models," *Applied Energy,* vol. 315, p. 119029, 2022.

[24] B. Schyska, A. Kies, M. Schlott, L. von Bremen and W. Medjroubi, "The sensitivity of power system expansion models," *Joule,* vol. 5, no. 10, pp. 2606-2624, 2021.

[25] L. Kotzur, P. Markewitz, M. Robinius and D. Stolten, "Impact of different time series aggregation methods on optimal energy system design," *Renewable Energy,* vol. 117, pp. 474-487, 2018.

[26] W. Tso, C. Demirhan, C. Heuberger, J. Powell and E. Pistikopoulos, "A hierarchical clustering decomposition algorithm for optimizing renewable power systems with storage," *Applied Energy,* vol. 270, p. 115190, 2020.

[27] H. Teichgraeber and A. Brandt, "Time-series aggregation for the optimization of energy systems: Goals, challenges, approaches, and opportunities," *Renewable and Sustainable Energy Reviews,* vol. 157, p. 111984, 2022.

[28] Z. Li, L. Cong, J. Li, Q. Yang, X. Li and P. Wang, "Co-planning of transmission and energy storage by iteratively including extreme periods in time-series aggregation," *Energy Reports,* vol. 9, no. 7, pp. 1281-1291, 2023.

[29] P. A. Sánchez-Pérez, M. Staadecker, J. Szinai, S. Kurtz and P. Hidalgo-Gonzalez, "Effect of modeled time horizon on quantifying the need for long-duration storage," *Applied Energy,* vol. 317, p. 119022, 2022.

[30] P. Lund, J. Lindgren, J. Mikkola and J. Salpakari, "Review of energy system flexibility measures to enable high levels of variable renewable electricity," *Renewable and Sustainable Energy Reviews,* vol. 45, pp. 785-807, 2015.

[31] J. Twitchell, K. DeSomber and D. Bhatnagar, "Defining long duration energy storage," *Journal of Energy Storage,* vol. 60, p. 105787, 2023.

[32] CSIRO, "Renewable Energy Storage Roadmap," March 2023. [Online]. Available: https://www.csiro.au/en/work-with-us/services/consultancy-strategic-advice-services/csiro-futures/energy/renewable-energy-storage-roadmap. [Accessed 15 May 2025].

[33] S. Ludig, M. Haller, E. Schmid and N. Bauer, "Fluctuating renewables in a long-term climate change mitigation strategy," *Energy,* vol. 36, no. 11, pp. 6674-6685, 2011.

[34] R. Kannan and H. Turton, "A Long-Term Electricity Dispatch Model with the TIMES Framework," *Environmental Modeling & Assessment,* vol. 18, pp. 325-343, 2012.

[35] D. Mallapragada, D. Papageorgiou, A. Venkatesh, C. Lara and I. Grossman, "Impact of model resolution on scenario outcomes for electricity sector system expansion," *Energy,* vol. 163, pp. 1231-1244, 2018.

[36] P. Nahmmacher, E. Schmid, L. Hirth and B. Knopf, "Carpe diem: A novel approach to select representative days for long-term power system models with high shares of renewable energy sources," *USAEE Working Paper No. 14-194,* 2014.

[37] T. Levin, J. Bistline, R. Sioshansi, W. Cole, J. Kwon, S. Burger, G. Crabtree, J. Jenkins, R. O'Neil, M. Korpås and S. Wogrin, "Energy storage solutions to decarbonize electricity through enhanced capacity expansion modelling," *Nature Energy,* vol. 8, no. 11, pp. 1199-1208, 2023.

[38] J. Merrick, J. Bistline and G. Blanford, "On representation of energy storage in electricity planning models," *Energy Economics,* vol. 136, p. 107675, 2024.

[39] P. Gabrielli, M. Gazzani, E. Martelli and M. Mazzotti, "Optimal design of multi-energy systems with seasonal storage," *Applied Energy,* vol. 219, pp. 408-424, 2018.

[40] K. Poncelet, E. Delarue and W. D'haeseleer, "Unit commitment constraints in long-term planning models: Relevance, pitfalls and the role of assumptions on flexibility," *Applied Energy,* vol. 258, p. 113843, 2020.

[41] D. Tejada-Arango, M. Domeshek, S. Wogrin and E. Centeno, "Enhanced Representative Days and System States Modeling for Energy Storage Investment Analysis," *IEEE Transactions on Power Systems,* vol. 33, no. 6, pp. 6534-6544, 2018.

[42] L. Kotzur, P. Markewitz, M. Robinius and D. Stolten, "Time series aggregation for energy system design: Modeling seasonal storage," *Applied Energy,* vol. 213, pp. 123-135, 2018.

[43] R. Novo, P. Marocco, G. Giorgi, A. Lanzini, M. Santarelli and G. Mattiazzo, "Planning the decarbonisation of energy systems: The importance of applying time series clustering to long-term models," *Energy Conversion and Management: X,* vol. 15, p. 100274, 2022.

[44] M. Moradi-Sepahvand and S. Tindemans, "Capturing Chronology and Extreme Values of Representative Days for Planning of Transmission Lines and Long-Term Energy Storage Systems," *2023 IEEE Belgrade PowerTech,* pp. 1-6, 2023.

[45] J. Ma, N. Zhang, Q. Wen and Y. Wang, "An efficient local multi-energy systems planning method with long-term storage," *IET Renewable Power Generation,* vol. 18, no. 3, pp. 426-441, 2024.

[46] A. Hilbers, D. Brayshaw and A. Gandy, "Reducing climate risk in energy system planning: A posteriori time series aggregation for models with storage," *Applied Energy,* vol. 334, p. 120624, 2023.

[47] S. Gonzato, K. Bruninx and E. Delarue, "Long term storage in generation expansion planning models with a reduced temporal scope," *Applied Energy,* vol. 298, p. 117168, 2021.

[48] M. Staadecker, J. Szinai, P. Sánchez-Pérez, S. Kurtz and P. Hidalgo-Gonzalez, "The value of long-duration energy storage under various grid conditions in a zero-emissions future," *Nature Communications,* vol. 15, no. 1, p. 9501, 2024.

[49] N. Sepulveda, J. Jenkins, A. Edington, D. Mallapragada and R. Lester, "The design space for long-duration energy storage in decarbonized power systems," *Nature Energy,* vol. 6, pp. 506-516, 2021.

[50] M. Parzen, F. Neumann, A. Van Der Wijde, D. Friedrich and A. Kiprakis, "Beyond cost reduction: improving the value of energy storage in electricity systems," *Carbon Neutrality,* vol. 1, no. 26, 2022.

[51] J. Dowling, K. Rinaldi, T. Ruggles, S. Davis, M. Yuan, F. Tong, N. Lewis and K. Caldeira, "Role of Long-Duration Energy Storage in Variable Renewable Electricity Systems," *Joule,* vol. 4, no. 9, pp. 1907-1928, 2020.

[52] A. Blakers, B. Lu and M. Stocks, "100% renewable electricity in Australia," *Energy,* vol. 133, pp. 471-482, 2017.

[53] C. Cheng, A. Blakers, M. Stocks and B. Lu, "100% Renewable Energy in Japan," *Energy Conversion and Management,* vol. 255, no. 115299, 2022a.

[54] B. Lu, A. Blakers, M. Stocks and T. N. Do, "Low-cost, low-emission 100% renewable electricity in Southeast Asia supported by pumped hydro storage," *Energy,* vol. 236, 2021.

[55] A. Nadolny, C. Cheng, B. Lu, A. Blakers and M. Stocks, "Fully electrified land transport in 100% renewable electricity networks dominated by variable generation," *Renewable Energy,* vol. 182, pp. 562-577, 2022.

[56] D. Silalahi, A. Blakers and C. Cheng, "100% Renewable Electricity in Indonesia," *Energies,* vol. 17, no. 1, p. 3, 2023.

[57] T. Weber, A. Blakers, D. Silalahi, K. Catchpole and A. Nadolny, "Grids dominated by solar and pumped hydro in wind-constrained sunbelt countries," *Energy Conversion and Management,* vol. 308, p. 118354, 2024.

[58] B. Lu, A. Blakers, M. Stocks, C. Cheng and A. Nadolny, "A Zero-Carbon, Reliable and Affordable Energy Future in Australia," *Energy,* vol. 220, no. 119678, 2021.

[59] B. Elliston, I. MacGill and M. Diesendorf, "Least cost 100% renewable electricity scenarios in the Australian National Electricity Market," *Energy Policy,* vol. 59, pp. 270-282, August 2013.

[60] B. Lu, "Stabilising 100% Renewable Grids: The Integrated FIRM Strategy," *Net Zero,* vol. 1, no. 1, pp. 31-47, 2025.

[61] F. Lombardi, K. van Greevenbroek, A. Grochowicz, M. Lau, F. Neumann, N. Patankar and O. Vågerö, "Near-optimal energy planning strategies with modeling to generate alternatives to flexibly explore practically desirable options," *Joule,* p. 102144, 2025.

[62] USAID V-LEEP, Deloitte, "Technical Report: Integrating significant renewable energy in Vietnam's power sector: A PLEXOS based analysis of long-term power development planning," 18 March 2021. [Online]. Available:

https://www.sipet.org/images/document/14092022126861.pdf. [Accessed 20 May 2025].

[63]  AEMO, "2024 Integrated System Plan For the National Electricity Market," 26 June 2024. [Online]. Available: https://aemo.com.au/-/media/files/major-publications/isp/2024/2024-integrated-system-plan-isp.pdf?la=en. [Accessed 23 May 2025].

[64]  ASEAN Centre for Energy, "8th ASEAN Energy Outlook: 2023-2050," November 2024. [Online]. Available: https://aseanenergy.org/wp-content/uploads/2024/09/8th-ASEAN-Energy-Outlook.pdf. [Accessed 15 May 2025].

[65]  NREL, "2024 Standard Scenarios Report: A U.S. Electricity Sector Outlook," December 2024. [Online]. Available: https://docs.nrel.gov/docs/fy25osti/92256.pdf. [Accessed 15 May 2025].

[66]  US EIA, "Annual Energy Outlook 2025," 15 April 2025. [Online]. Available: https://www.eia.gov/outlooks/aeo/. [Accessed 23 May 2025].

[67]  Central Electricity Authority, "National Electricity Plan (Volume I) Generation," March 2023. [Online]. Available: https://mnre.gov.in/en/document/national-electricity-plan-volume-i-generation-by-cea/. [Accessed 16 May 2025].

[68]  IRENA, "Advancements in continental power system planning for Africa: Methodological framework of the African Continental Power Systems Masterplan's SPLAT-CMP model 2023," 2024. [Online]. Available: https://www.irena.org/-/media/Files/IRENA/Agency/Publication/2024/Jul/IRENA_Advancements_CMP_Africa_2024.pdf. [Accessed 16 May 2025].

[69]  Ceylon Electricity Board, "Long Term Generation Expansion Plan 2023-2042," February 2023. [Online]. Available: https://www.ceb.lk/publication-media/planing-documents/121/en. [Accessed 0 May 2025].

[70]  Ministerio para la Transición Ecológica y el Reto Demográfico, "Integrated National Energy and Climate Plan: Update 2023-2030," September 2024. [Online]. Available: https://commission.europa.eu/document/download/211d83b7-b6d9-4bb8-b084-4a3bfb4cad3e_en?filename=ES%20-%20FINAL%20UPDATED%20NECP%202021-2030%20%28English%29.pdf. [Accessed 20 May 2025].

[71]  NESO, "Future Energy Scenarios: ESO Pathways to Net Zero," July 2024. [Online]. Available: https://www.neso.energy/document/321041/download. [Accessed 15 May 2025].

[72]  RTE, "Futurs énergétiques 2050: Rapport complet," February 2022. [Online]. Available: https://rte-futursenergetiques2050.com/documents. [Accessed 15 May 2025].

[73]  Departement van Mineraalbronne en Energie, "Integrated Resource Plan, 2023," 4 January 2024. [Online]. Available: https://www.dmre.gov.za/Portals/0/Energy_Website/IRP/2023/IRP%20Government%20Gazzette%202023.pdf. [Accessed 23 May 2025].

[74]  Canada Energy Regulator, "Canada's Energy Future 2023: Energy Supply and Demand Projections to 2050 (EF2023)," 2023. [Online]. Available: https://www.cer-rec.gc.ca/en/data-analysis/canada-energy-future/2023/canada-energy-futures-2023.pdf. [Accessed 16 May 2025].

[75]  Philippine Department of Energy, Electric Power Industry Management Bureau, "Power Development Plan 2023-2050," 2023. [Online]. Available: https://legacy.doe.gov.ph/sites/default/files/pdf/electric_power/development_plans/Power%20Development%20Plan%202023-2050.pdf. [Accessed 16 May 2025].

[76]  F. Neuwahl, M. Wegener, R. Salvucci, M. Jaxa-Rozen, J. Gea Bermudez, P. Sikora and M. Rózsai, "Clean Energy Technology Observatory: POTEnCIA CETO 2024 Scenario – 2024 Energy System Modelling For Clean Energy Technology Scenarios," 2024. [Online]. Available: https://publications.jrc.ec.europa.eu/repository/handle/JRC139836. [Accessed 20 May 2025].

[77]  International Hydropower Association, "Pumped Storage Tracking Tool," 27 June 2022. [Online]. Available: https://www.hydropower.org/hydropower-pumped-storage-tool. [Accessed 31 October 2025].

[78]  International Energy Agency, "Batteries and Secure Energy Transitions," 25 April 2024. [Online]. Available: https://www.iea.org/reports/batteries-and-secure-energy-transitions. [Accessed 5 December 2025].

[79]  BloombergNEF, "Headwinds in Largest Energy Storage Markets Won't Deter Growth," 11 November 2024. [Online]. Available: https://about.bnef.com/insights/clean-energy/headwinds-in-largest-energy-storage-markets-wont-deter-growth/. [Accessed 5 December 2025].

[80]  M. Jafari, A. Botterud and A. Sakti, "Decarbonizing power systems: A critical review of the role of energy storage," *Renewable and Sustainable Energy Reviews,* vol. 158, p. 112077, 2022.

[81]  A. Blakers, T. Weber and D. Silalahi, "Pumped hydro energy storage to support 100% renewable energy," *Progress in Energy,* vol. 7, p. 022004, 2025.

[82]  T. Weber, R. Stocks, A. Blakers, A. Nadolny and C. Cheng, "A global atlas of pumped hydro systems that repurpose existing mining sites," *Renewable Energy,* vol. 224, p. 120113, 2024.

[83]  M. Stocks, R. Stocks, B. Lu, C. Cheng and A. Blakers, "Global Atlas of Closed-Loop Pumped Hydro Energy Storage," *Joule,* vol. 5, pp. 270-284, 2021.

[84]  National Transmission and Despatch Company , "Indicative Generation Capacity Expansion Plan (IGCEP) 2021-30," May 2021. [Online]. Available: https://fpcci.org.pk/wp-content/uploads/2021/11/IGCEP-2021.pdf. [Accessed 16 May 2025].

[85]  Snowy Hydro Limited, "Snowy 2.0 Updated Business Case," 24 May 2024. [Online]. Available: https://www.snowyhydro.com.au/wp-content/uploads/2024/05/Snowy-2.0-Updated-Business-Case.pdf. [Accessed 23 May 2025].

[86] AEMO, "2024 ISP Model," 26 June 2024. [Online]. Available: https://aemo.com.au/energy-systems/major-publications/integrated-system-plan-isp/2024-integrated-system-plan-isp. [Accessed 23 May 2025].

[87] AEMO, "ISP Methodology - June 2023," June 2023. [Online]. Available: https://aemo.com.au/-/media/files/stakeholder_consultation/consultations/nem-consultations/2023/isp-methodology-2023/isp-methodology_june-2023.pdf?la=en. [Accessed 15 May 2025].

[88] Departement van Mineraalbronne en Energie, "Department of Mineral Resources and Energy - Draft Integrated Resource Plan Stakeholder Workshops," November 2024. [Online]. Available: https://www.dmre.gov.za/Portals/0/Energy%20Resources/IRP/IRP%202023/Draft%20IRP%202024%20Outcomes%20Stakeholder%20Engagements.pdf?ver=4zU5hDlVy48zTCjoiJhmFA%3d%3d. [Accessed 16 May 2025].

[89] M. Stocks, R. Stocks, T. Weber, B. Lu, A. Nadolny, C. Cheng and A. Blakers, "ANU RE100 Map," 2025. [Online]. Available: https://re100.anu.edu.au/. [Accessed 12 December 2025].

[90] GHD, "Pumped Hydro Energy Storage: Parameter Review," 22 July 2025. [Online]. Available: https://www.aemo.com.au/-/media/files/stakeholder_consultation/consultations/nem-consultations/2024/2025-iasr-scenarios/final-docs/ghd-2025-pumped-hydro-energy-storage-cost-parameter-review.pdf?la=en. [Accessed 17 October 2025].

[91] Australian Office of Financial Management, "Treasury Bonds," 28 11 2025. [Online]. Available: https://www.aofm.gov.au/securities/treasury-bonds. [Accessed 4 12 2025].

[92] Australian Bureau of Statistics, "Consumer PRice Index, Australia," 26 November 2025. [Online]. Available: https://www.abs.gov.au/statistics/economy/price-indexes-and-inflation/consumer-price-index-australia/latest-release. [Accessed 4 December 2025].

[93] Australian Energy Regulator, "Rate of Return Instrument 2022," 24 February 2023. [Online]. Available: https://www.aer.gov.au/industry/registers/resources/guidelines/rate-return-instrument-2022. [Accessed 4 December 2025].

[94] AEMO, "2025 Inputs, Assumptions and Scenarios Report," August 2025. [Online]. Available: https://www.aemo.com.au/-/media/files/stakeholder_consultation/consultations/nem-consultations/2024/2025-iasr-scenarios/final-docs/2025-inputs-assumptions-and-scenarios-report.pdf?rev=63268acd3f044adb9f5f3a32b6880c27&sc_lang=en. [Accessed 9 November 2025].

[95] AEMC, "How the national energy objectives shape our decisions," 28 March 2025. [Online]. Available: https://www.aemc.gov.au/sites/default/files/2025-03/How%20the%20national%20energy%20objectives%20shape%20our%20decisions%20260325.pdf. [Accessed 9 November 2025].

[96]  S. Reynolds and I. f. E. E. a. F. Analysis, "Global gas turbine shortages add to LNG challenges in Vietnam and the Philippines," October 2025. [Online]. Available: https://ieefa.org/sites/default/files/2025-10/IEEFA%20Report_Global%20gas%20turbine%20shortages%20add%20to%20LNG%20challenges%20in%20Vietnam%20and%20the%20Philippines_October2025.pdf. [Accessed 10 October 2025].

[97]  J. Anderson, "US gas-fired turbine wait times as much as seven years; costs up sharply," S&P Global, 20 May 2025. [Online]. Available: https://www.spglobal.com/commodity-insights/en/news-research/latest-news/electric-power/052025-us-gas-fired-turbine-wait-times-as-much-as-seven-years-costs-up-sharply. [Accessed 31 October 2025].

[98]  L. Shen, D. Jacob, R. Gautam, M. Omara, T. Scarpelli, A. Lorente, D. Zavala-Araiza, X. Lu, Z. Chen and J. Lin, "National quantifications of methane emissions from fuel exploitation using high resolution inversions of satellite observations," *Nature Communications,* vol. 14, no. 1, p. 4948, 2023.

[99]  IEA, "Global Methane Tracker 2025," May 2025. [Online]. Available: https://iea.blob.core.windows.net/assets/b83c32dd-fc1b-4917-96e9-8cd918801cbf/GlobalMethaneTracker2025.pdf. [Accessed 11 July 2025].

[100]  IPCC, "IPCC Sixth Assessment Report," 2020. [Online]. Available: https://www.ipcc.ch/report/ar6/wg1/chapter/chapter-7/#7.6. [Accessed 31 October 2025].

[101]  International Energy Agency, "CCUS Projects Database," 30 April 2025. [Online]. Available: https://www.iea.org/data-and-statistics/data-product/ccus-projects-database. [Accessed 31 October 2025].

[102]  *Obligations of States in respect of Climate Change (Advisory Opinion),* 2025, ICJ, p. 122.

[103]  AEMO, "Quarterly Energy Dynamics Q3 2025," October 2025. [Online]. Available: https://www.aemo.com.au/-/media/files/major-publications/qed/2025/qed-q3-2025.pdf. [Accessed 12 December 2025].

[104]  *National Electricity (South Australia) Act 1996 (SA).*

[105]  Australian Government, "Australia's 2035 Nationally Determined Contribution," 2025. [Online]. Available: https://unfccc.int/sites/default/files/2025-09/Australias%20Second%20NDC.pdf. [Accessed 17 October 2025].

[106]  International Energy Agency, "WEO2025_AnnexA_Free_Dataset_World.csv," 12 November 2025. [Online]. Available: https://www.iea.org/data-and-statistics/data-product/world-energy-outlook-2025-free-dataset#data-files. [Accessed 5 December 2025].

[107]  AEMO, "2024 ISP Solar traces," 26 June 2024. [Online]. Available: https://aemo.com.au/-/media/files/major-publications/isp/2024/supporting-materials/2024-isp-solar-traces.zip?la=en. [Accessed 29 May 2025].

[108] AEMO, "2024 ISP Wind traces," 26 June 2024. [Online]. Available: https://aemo.com.au/-/media/files/major-publications/isp/2024/supporting-materials/2024-isp-wind-traces.zip?la=en. [Accessed 29 May 2025].

[109] AEMC, "National Electricity Rules version 231," 1 July 2025. [Online]. Available: https://energy-rules.aemc.gov.au/ner/659/612232#3.9.3C. [Accessed 2 July 2025].

[110] S. Lam, A. Pitrou and S. Seibert, "Numba: A llvm-based python jit compiler," *Proceedings of the Second Workshop on the LLVM Compiler Infrastructure in HPC,* pp. 1-6, 2015.

[111] A. Nelson, "scipy.optimize.differential_evolution," 2014. [Online]. Available: https://docs.scipy.org/doc/scipy/reference/generated/scipy.optimize.differential_evolution.html. [Accessed 30 January 2023].

[112] The pandas development team, "pandas-dev/pandas: Pandas," Zenodo, February 2020. [Online]. Available: https://doi.org/10.5281/zenodo.3509134.

[113] W. McKinney, "Data structures for statistical computing in Python," *Proceedings of the 9th Python in Science Conference,* pp. 56-61, 2010.

[114] C. Harris, K. Millman, S. Van Der Walt, R. Gommers, P. Virtanen, D. Cournapeau, E. Wieser, J. Taylor, S. Berg, N. Smith and R. Kern, "Array programming with NumPy," *Nature,* vol. 585, no. 7825, pp. 357-362, 2020.

# Supplementary Information A – Additional Details

*Table S1. Summary of Long-term Planning Model Assumptions that Impact Pumped Hydro (red represents very poor assumption with serious impact, yellow represents poor assumption that may impact results)*

| Organisation | Planning Software | Energy Plan | Temporal Aggregation | Coupled sample periods? | Temporal Resolution | Optimisation Step Time Horizon | Independent sizing of power and energy capacity? | PHES Overnight CAPEX (2025 USD) | Maximum PHES Duration | PHES Build Limit | Reference |
|---|---|---|---|---|---|---|---|---|---|---|---|
| Australian Energy Market Operator (Australia) | PLEXOS | ISP – Single-Stage Long-Term Model | Typical periods[1] | Yes | 30 min | 28 years | No | $47/kWh – $388/kWh | 48 h | 422 GWh[2] | [87, 115, 116, 63] |
|  | PLEXOS | ISP – Detailed Long-Term Model | Segmentation based on fitted step function | N/A | 5 – 8 fitted blocks per day | 7 years | No | $47/kWh – $388/kWh | 48 h | 422 GWh[3] | [87, 115, 116, 63] |
| ASEAN Centre for Energy (ASEAN Member States) | LEAP - NEMO | ASEAN Energy Outlook | Sampled 1 typical day per month | Yes | 1 hour | 28 years | No | $950/kW | Not specified | Not specified | [64, 117, 118] |
| National Renewable Energy Laboratory (United States of America) | ReEDS | Standard Scenarios Report: A U.S. Electricity Sector Outlook | Sampled 32 standard plus 9 extreme days per year | Yes | 3 hours | 25 years | No | $320/kWh – $485/kWh (Greenfield) $183/kWh – $350/kWh (Bluefield or Brownfield) | 8 – 12 h | 144,381 GWh | [65, 119, 120] |

---

[1] Specific clustering method in PLEXOS is proprietary. Specific sample settings used for the ISP - SSLT are not public, though [81] provides an example of two days per month sampling. Different sampling settings, including weeks per year, may be used across all of the ISP – SSLT scenarios.
[2] These build limits are expected to be relaxed in the 2026 ISP after AEMO requested a review of pumped hydro energy storage parameters by GHD [87].
[3] These build limits are expected to be relaxed in the 2026 ISP after AEMO requested a review of pumped hydro energy storage parameters by GHD [87].

| Organisation | Planning Software | Energy Plan | Temporal Aggregation | Coupled sample periods? | Temporal Resolution | Optimisation Step Time Horizon | Independent sizing of power and energy capacity? | PHES Overnight CAPEX (2025 USD) | Maximum PHES Duration | PHES Build Limit | Reference |
|---|---|---|---|---|---|---|---|---|---|---|---|
| National Energy System Operator (Great Britain) | PLEXOS | Future Energy Scenarios: Pathways to Net Zero | Not specified | Not specified | 1 hour | 28 years[4] | No | $465/kWh – $1241/kWh[4] | 4 h[4] | 6 GW[5] | [73, 121, 122] |
| Réseau de Transport d'Électricité (France) | Antares Simulator | Energy Pathways to 2050 | None | Yes | 1 hour | 1 week | N/A[6] | $1238/kW | Not specified | 3 GW | [72, 123] |
| Commonwealth Scientific and Industrial Research Organisation (Australia) | DIETER/ STABLE | Renewable Energy Storage Roadmap | None | N/A | 1 hour | 1 year | No[7] | $40/kWh – $328/kWh | 48 h | 398 GWh | [32, 124, 125, 126] |
| Departement van Mineraalbronne en Energie (South Africa) | PLEXOS | Integrated Resource Plan | Not specified | Not specified | 1 hour | 20 years | No | $1459/kW - $2330/kW | Not specified | Not specified | [127, 88, 128, 74] |
| US Energy Information Administration (United States of America) | NEMS – EMM and REStore | Annual Energy Outlook 2025 | Sampled 2 typical days (weekday and weekend) per month | Yes | 1 hour | 1 year | No | N/A | 12 h | 0 GWh[8] | [129, 130, 66] |
| Canada Energy Regulator (Canada) | Energy Futures | Canada's Energy Future | Not specified | Not specified | 1 hour | 5 years | No | $1673/kW | Not specified | 0 GWh | [75, 131] |

[4] Although the time horizon is 28 years, only weather data from 2013 is used for dispatch runs.
[5] Based upon [115]. Not explicitly referenced as a source of costs and build limits in [114], but NESO does state that technology costs were obtained from DESNZ Levelised Cost of Electricity generation work. The Mott MacDonald report is published on [135] alongside the relevant DESNZ reports.
[6] Analysis did not use a capacity expansion model. The Antares Simulator was used for unit commitment to evaluate scenarios.
[7] While the original version of DIETER allows for independent sizing of power and energy capacity, this does not appear to be the case in STABLE. Instead, additional storage capacity was estimated based upon the size and duration of energy shortfalls.
[8] Only new "diurnal storage" based upon battery costs is allowed within the model. Existing pumped hydro appears to be included.

| Organisation | Planning Software | Energy Plan | Temporal Aggregation | Coupled sample periods? | Temporal Resolution | Optimisation Step Time Horizon | Independent sizing of power and energy capacity? | PHES Overnight CAPEX (2025 USD) | Maximum PHES Duration | PHES Build Limit | Reference |
|---|---|---|---|---|---|---|---|---|---|---|---|
| | Modelling System – Electricity Supply Model (based on PyPSA-Can) | | | | | | | | | | |
| Central Electricity Authority (India) | ORDENA and PLEXOS | National Electricity Plan (Volume I) Generation | 2 typical days (peak and non-peak) for each season | Not specified | Unspecified number of fitted blocks per day | 5 years | No | $223/kW – $1116/kW | 6 – 11 h | 24.31 GW | [67] |
| Philippine Department of Energy (Philippines) | PLEXOS | Power Development Plan (PDP) 2023-2050 | Not specified | Not specified | 1 hour | 29 years | No | $212/kWh – $584/kWh | Not specified | Not specified | [76, 132] |
| National Transmission and Despatch Company (Pakistan) | PLEXOS | Indicative Generation Capacity Expansion Plan (IGCEP) 2021-30 | Not specified | Not specified | Not specified | 10 years | No | N/A | N/A | 0 GWh | [84] |
| African Union Development Agency / International Renewable Energy Agency (Africa) | SPLAT/ MESSAGE | African Continental Power Systems Masterplan (CMP) | 1 typical day per season (3 seasons per year), with peak demand included | Yes | 3 – 12 fitted blocks | 23 years | No | $214/kWh | Daily operation only (likely 6 h) | 0 GWh[9] | [133, 68, 134] |

---

[9] Pumped hydro is site specific based upon [126] which identified 36 potential sites in Africa with 86.8 GW capacity (restricted to 6 h storage duration), not allowed to build more during capacity expansion. Older long-term plans for the region, available from [136], only considered existing pumped hydro systems.

| Organisation | Planning Software | Energy Plan | Temporal Aggregation | Coupled sample periods? | Temporal Resolution | Optimisation Step Time Horizon | Independent sizing of power and energy capacity? | PHES Overnight CAPEX (2025 USD) | Maximum PHES Duration | PHES Build Limit | Reference |
|---|---|---|---|---|---|---|---|---|---|---|---|
| USAID Vietnam Low Emissions Energy Program (Vietnam) | PLEXOS | Technical Report: Integrating significant renewable energy in Vietnam's power sector: A PLEXOS based analysis of long-term power development planning[10] | Approximate LDC | N/A | 9 blocks per week, biased towards peak and off-peak | 26 years | No | $864/kW – $1110/kW | Not specified | 0 GWh[11] | [62] |
| Ceylon Electricity Board (Sri Lanka) | OPTGEN / SDDP | CEB LONG TERM GENERATION EXPANSION PLAN 2023-2042 | Cluster sampling of typical days within seasons[12] | Yes | 1 hour | 1 year rolling horizon | No | $1368/kW | Not specified | 2 GW | [69, 135] |
| Ministerio para la Transición Ecológica y el Reto Demográfico (Spain) | TIMES-SINERGIA | Integrated National Energy and Climate Plan: Update 2023-2030 | 1 typical day for each season (4 seasons per year) | Yes | 1 hour | 8 years | No | $4587/kW – $8901/kW[13] | Not specified | Not specified | [70, 136] |

---

[10] Authors of [57] note this report uses a similar model and assumptions to the draft Vietnam *Power Development Plan 8 Version 3*. The Revised *National Power Development Plan 8* did increase pumped hydro under the high scenario from 2.4 GW to 6 GW [137, 138].
[11] Only 6 pumped hydro systems included in the model, not allowed to build more during capacity expansion.
[12] Specific sample settings used by CEB are not specified.
[13] Based upon inputs to JRC POTEnCIA 2018 model, as noted in Table A.7 of [66]

| Organisation | Planning Software | Energy Plan | Temporal Aggregation | Coupled sample periods? | Temporal Resolution | Optimisation Step Time Horizon | Independent sizing of power and energy capacity? | PHES Overnight CAPEX (2025 USD) | Maximum PHES Duration | PHES Build Limit | Reference |
|---|---|---|---|---|---|---|---|---|---|---|---|
| Joint Research Centre – European Commission (Europe) | POTEnCIA | POTEnCIA CETO 2024 Scenario: Energy System Modelling for Clean Energy Technology Scenarios | None | N/A | 1 hour | 1 year | No | $1586/kW – $4530/kW[13] | Not specified | 6300 GWh[14] | [137, 71, 138] |

---

[14] Assuming theoretical energy storage and costs of pumped hydro from [131] was integrated into POTEnCIA, as implied by section 2.2.1 of [59].

# Supplementary Results

These supplementary results are intended to expand upon the discussion in the main article, support the reader in understanding the behaviour of the FIRM model, and provide additional details on the ANU parametric cost modelling for pumped hydro systems.

## Frequency Spectra of the State-of-Charge Profiles

The normalised magnitude of the state-of-charge frequency spectrum for each temporal aggregation method in the 100% Renewables PLEXOS scenario was plotted in Figure 1. The normalised magnitudes of the frequency spectra were calculated by taking the total stored energy time-series data for pumped hydro, performing a fast Fourier transform, multiplying frequencies by their complex conjugate, and normalising with respect to the largest magnitude. The DC offsets were excluded from the spectra.

A strong peak at 1 day$^{-1}$ means that a sinusoidal function with a frequency of 1 cycle per day contributed strongly to the overall stored energy time-series. A steep one-off, deep charge/discharge cycle would be approximately described by a Dirac function in the time domain, with a contribution from all frequencies in the frequency domain. This means that the frequencies cannot just be disaggregated from each other using simple window function filters to apportion the spectrum across different cycle durations (e.g., interannual, seasonal, weekly, and daily) because the noise across the entire spectrum might be important for describing key infrequent events in the time domain. Regardless, the dominant signals are still useful for understanding prominent cycling behaviour, even if the entire spectrum cannot be neatly apportioned between the storage duration categories.

The Sampled (days) method is dominated by a peak at the daily frequency, with no clear peaks in any lower frequencies. The Sampled (weeks) method also has a dominant daily frequency peak, as well as a very small weekly peak and interannual peak. This indicates that the typical period temporal aggregation methods in PLEXOS almost exclusively used energy storage systems for time-shifting of durations equal to or shorter than the length of the typical period. The small interannual peak for the Sampled (weeks) method is an artefact from using FY2052 demand traces for all 10 years in the modelling horizon, which led to identical clustering of typical weeks in each year.

The dominant peaks for the Fitted and Partial methods are at 1 year$^{-1}$. That is, deep discharging of the energy storage typically occurs only once per year. The Fitted method maintains a smaller peak at 1 day$^{-1}$, but the bulk of the contribution to the time-series is from seasonal frequencies. The absence of a visible peak at 1 day$^{-1}$ for the Partial method is likely due to the absence of chronology within each day when performing unit commitment according to the daily LDCs. Therefore, the only method that captured both shorter-duration overnight cycling and longer-duration seasonal and interannual cycling of the pumped hydro systems was the Fitted segmentation method.

## Time Domain of State-of-Charge Profiles

A two-week snapshot of the state-of-charge profiles (i.e., stored energy in the pumped hydro systems) for each temporal aggregation method in the 100% Renewables PLEXOS scenario was plotted in Figure 2. In Figure 2b, two typical days are highlighted within the snapshot based on the Sampled (days) method. The first typical day is repeated 12 times in a row, then the second typical day is linked, and then the first typical day ends the fortnight. Each month in the Sampled (days) method only has two typical days – the only example of PLEXOS sampled chronology used to describe the Single-Stage Long-Term (SSLT) model in the ISP also uses two representative days per month. It is the default setting for LT Plan sampled chronology in PLEXOS. It is possible that the ISP uses a range of sampled chronology settings for the SSLT over the iterative modelling process, though these are not described in the *ISP Methodology* [87]. Regardless, most long-term energy plans summarised in Tabel S1 that used typical period methods were explicit about using a very small number of typical days per year.

Figure 2c shows a typical week from the Sampled (weeks) method, repeated twice over the same period. Only one of the representative weeks is shown in the snapshot, though there are four for each year over the modelling horizon. Using four typical weeks per year is the default setting for PLEXOS LT Plan sampled chronology when swapping to use representative weeks. Since PLEXOS constrains the state-of-charge at the boundaries of the typical period to be equal, there is additional inter-daily flexibility in Sampled (weeks) compared to Sampled (days). This is why the frequency spectrum for the state-of-charge when using the Sampled (weeks) method contains prominent weekly frequencies as well as daily frequencies (refer Figure 1).

The Sampled (days) and Sampled (weeks) methods produced state-of-charge profiles that lack a resemblance to the FSO. In fact, the charging and discharging behaviour in the Sampled (weeks) method appears inverted relative to the FSO method. Within the snapshot, the lowest state-of-charge in the typical week coincidentally occurs on the same day as the maximum state-of-charge in the FSO. Since clustering of typical periods was performed exclusively on the demand profile, endogenous variables such as state-of-charge, solar and wind generation, and transmission flows were not considered by the clustering algorithm. PLEXOS does allow for additional variables to be considered for clustering through the use of load adjuster objects. Including *a posteriori* load adjusters in the clustering would require multiple optimisations to be iteratively performed [46], and the bias towards short-duration pumped hydro behaviour would still be imprinted on all solutions since the first optimisation informs the clustering of each subsequent optimisation. Using *a posteriori* clustering would still not allow for cycle durations longer than the length of the typical periods.

The state-of-charge profile for the Partial method, shown in Figure 2e, appears to have a very similar shape to the FSO. The energy volume constraint only appears to be breached for short periods of time. The default partial chronology setting for the LT Plan in PLEXOS fits an approximate LDC to each day. Chronology, and therefore state-of-charge constraints, are tracked between days. It is only within each day that the energy volume constraints (0 GWh minimum, energy capacity maximum) will be breached, not at the boundaries of the days. This is why daily cycling behaviour is lost within the Partial method, while longer-duration cycling is captured. Swapping to an approximate LDC for each month would result in loss of chronology within month-long periods instead. So, while the Sampled methods lose track of long-duration storage behaviour, the Partial method instead loses track of shorter-duration storage behaviour. Capturing overnight cycling behaviour is essential for pumped hydro because of the need to model overnight storage of solar, so losing track of this behaviour is unsuitable for modelling grids with a high penetration of renewables.

## Normalised L1-distance to the FSO

The normalised L1-distance between solution vectors for each temporal aggregation method and the FSO in the Improved PHES Assumptions and 100% Renewables scenarios is plotted in Figure S1.

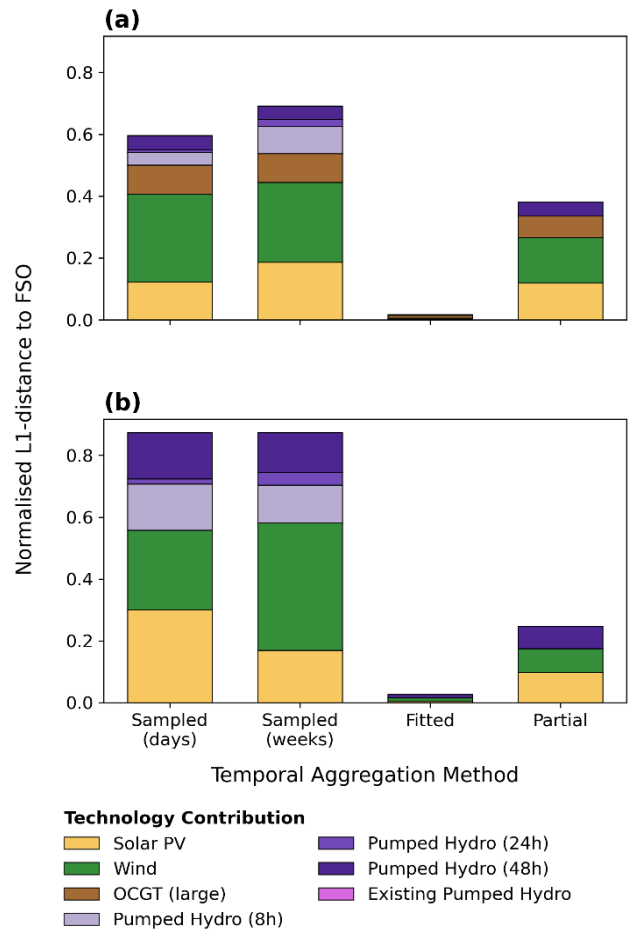

*Figure S1. Normalised L1-distance between solution vectors for each temporal aggregation method and the FSO in: (a) Improved PHES Assumptions scenario, and (b) 100% Renewables scenario*

The Sampled (days) and Sampled (weeks) temporal aggregation methods had a substantially higher L1-distance than the Fitted and Partial methods in both scenarios. Under the 100% Renewables scenario, the L1-distance of the Sampled (days) and Sampled (weeks) methods relative to the Improved PHES Assumptions scenario increased by 47% and 26% respectively. In fact, the L1-distance for the Sampled (days) and Sampled (weeks) methods in the 100% Renewables scenario was more than 87% of the total system costs of the FSO solution. Fixed costs for solar PV and wind generators were the largest driver of L1-distance for the typical period methods. Clustering into typical periods was based purely on the operational load profile, which is unlikely to be strongly correlated with wind velocity or cloud cover that influence wind and grid-scale solar PV generation. In both scenarios, the typical period methods preferred investment in short-duration 8-hour pumped hydro energy storage, while the FSO opted for longer-duration 48-hour pumped hydro systems. Since storage systems were constrained to cycles up to the length of the typical period, there was no incentive for these methods to develop longer-duration energy storage.

The Partial method had a normalised L1-distance from the FSO solution vector of 0.38 and 0.25 in the Improved PHES Assumptions and 100% Renewables scenarios respectively. The daily LDC unit commitment was able to capture long-term variation in solar and wind generation, but overestimated the value of solar PV due to the loss of chronological information within each day. For the 100% Renewables scenario, the Partial method over-invested in 86% of solar PV sites and under-invested in 99% of wind sites relative to the FSO. Every solar PV site was over-invested and every wind-site under-invested in the Improved PHES Assumptions scenario using the Partial method. Both the Partial method and the FSO exclusively invested in the 48-hour duration option for pumped hydro, but the Partial solution produced a smaller capacity, likely since the loss of chronological information within each day resulted in the shorter-term cycling being neglected (as indicated by the absence of daily peaks in Figure 1).

While the Fitted method had a lower temporal resolution compared to the FSO (8 variable-length blocks per day compared to 48 half-hour intervals), the investment costs and variable costs for every asset in the system were very similar. The loss of resolution in longer segments of the step function did not substantially influence the intra-day balancing behaviour of gas or energy storage systems.

The contribution to the L1-distance from OCGT in the Improved PHES Assumptions scenario was roughly comparable to the contribution from pumped hydro systems under each temporal aggregation method. Where the contribution from pumped hydro systems is driven by the choice to invest in longer or shorter duration storage, the OCGT contribution is primarily due to a bias towards dispatching the same asset for either short-term or long-term balancing. The difference in variable costs compared to the FSO accounted for 76% and 88% of the contribution to L1-distance by OCGT sites in the Sampled (days) and Sampled (weeks) methods respectively. That is, typical period methods in PLEXOS may invest in a similar amount of gas capacity as the FSO, but the reason for the investment is to balance short-term variability (on scales shorter than the typical period length) rather than seasonal variability.

Note that over the 1-year modelling horizon used to calculate the L1-distance of solution vectors relative to the FSO, none of the optimisations invested in 160-hour pumped hydro systems. Over the longer 10-year modelling horizon, the Fitted and Partial methods developed 610 GWh and 470 GWh of 160-hour pumped hydro systems respectively in the 100% Renewables scenario – this was the majority of new build pumped hydro capacity. That is, deep interannual storage can help manage extreme weather years at lower cost than smaller-scale storage on its own.

Negligible new build batteries were developed in any of the PLEXOS scenarios, though it is possible that a model with higher temporal resolution (<30 min) or power system considerations (e.g., frequency and voltage) would rely upon batteries for those short-duration services.

## Gas Capacity Compared to Generation in Each PLEXOS Scenario

When using the full 10-year modelling horizon, transitioning from the Original PHES Assumptions to the Improved PHES Assumptions scenario resulted in a reduction of up to 20% new build OCGT capacity and 15% gas generation, with the largest reduction observed in the Fitted scenario. There was no change in OCGT investment or gas generation in the Partial scenario through the introduction of improved pumped hydro assumptions, with the model opting to develop no additional pumped hydro systems in either scenario.

Sampled (days) had the largest OCGT capacity in each scenario, as well as the smallest decrease in OCGT capacity through the introduction of improved pumped hydro assumptions. Only 1.5 GW / 12 GWh of new pumped hydro systems were built in the Sampled (days) Improved PHES Assumptions scenario to displace OCGT capacity and generation in the least-cost solution. Since the Sampled (days) method was restricted to considering balancing for durations shorter than one day, it focused on investing in high-power OCGT. Since pumped hydro has a high power capacity cost, it is less competitive for short-duration balancing. Failing to capture long-duration balancing behaviour in the Sampled methods means that the lowest-cost options for pumped hydro (large-scale long-duration storage) are entirely overlooked.

*Figure S2. (a) New build OCGT capacity, and (b) OCGT total generation in each scenario*

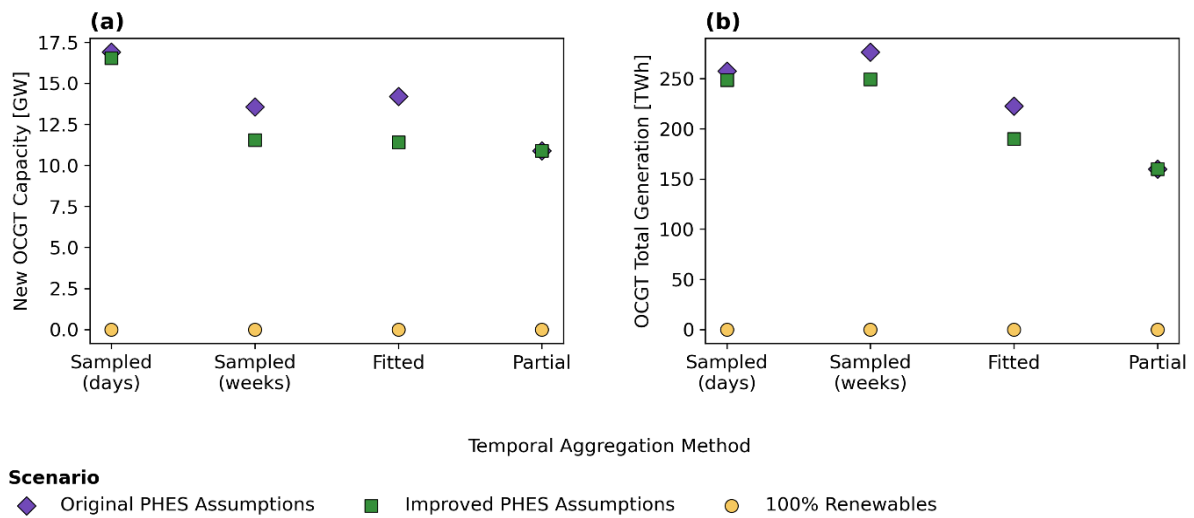

## Total Cost and Reliability of the PLEXOS Global Optima

The grid configurations for the global optimal solutions of each scenario were implemented in PLEXOS, and an FSO was performed across the 10-year modelling horizon in step sizes of 1 year. No additional new build capacity was allowed in the FSO evaluation relative to the least-cost grid configurations. The total unserved energy over the modelling period was calculated for each FSO optimisation (refer Figure S3b). Note that a hypothetical FSO with a step-size of 10 years would result in less unserved energy since the optimisation would capture interannual storage behaviour; however, this formulation was not possible to optimise using a reasonable amount of computing resources and time.

Over the 10-year horizon, the global optimal solutions for the Partial method resulted in 4300 GWh unserved energy in the scenarios containing gas and 24,000 GWh unserved energy for the 100% Renewables scenario when evaluated using the full time-series data – equivalent to between 0.16–0.94% of total electricity demand. Note that the NEM reliability standard requires 99.998% of annual demand to be met [109]. The global optimal grid configuration produced by the Partial method was not robust with respect to the full time-series data. The lack of chronology within each day caused the Partial method to underestimate the balancing requirements from both pumped hydro and gas, as well as under-build wind generators. Adding additional energy volume and power capacity reserves post-hoc could improve the reliability of solutions generated using the Partial method.

The global optimal solutions for the two scenarios containing gas were more reliable than the 100% Renewables scenario for all temporal aggregation methods. This is a structural error in the models, rather than an indication that gas-dependent systems are more reliable than storage-dependent systems. Energy storage systems are typically treated as having a finite amount of energy stored at any moment in time, while flexible generators, such as gas, are treated as having an effectively infinite reservoir of fuel immediately available over any length of time. When an optimisation is performed with a temporal aggregation method, the energy capacities of the storage systems are sized perfectly for the aggregated time-series in an attempt to minimise costs. When testing reliability of the grid configuration using the FSO, a small underestimate in energy volume requirements can lead to a period of unserved energy. The unserved energy in the Fitted grid configuration was accrued over one- or two-week periods in 3 of the 10 years. For gas generators, there is no energy volume that can be improperly sized, meaning that changes in gas generation between the aggregated time-series and FSO will never result in energy-constrained dispatch. All long-term energy plans evaluated in Table S1 featured this same structural error for gas generators. More robust 100% renewable energy systems could be modelled by post-hoc introducing an additional energy volume reserve to the global optima. More reasonable gas-based solutions ought to consider on-site gas storage capacity and flow rates of gas to replenish that storage.

Figure S3. Influence of temporal aggregation method in each scenario on: (a) total system cost, and (b) total unserved energy when global optimal grid configuration is dispatched according to FSO (10-year modelling horizon, 1-year optimisation steps)

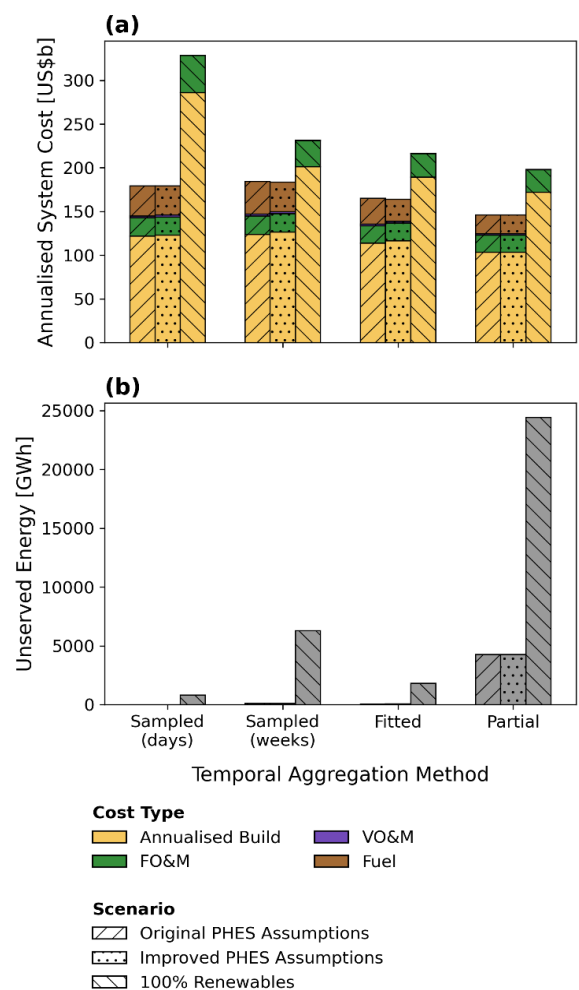

The costs of solutions for the Original PHES Assumptions and Improved PHES Assumptions scenarios were re-calculated using CCGT instead of large OCGT. While CCGTs have a lower heat rate (i.e., higher efficiency converting fuel to electricity) than OCGTs, they also have a much higher capital cost than OCGTs. This makes CCGTs more cost-competitive for mid-merit generation, where variable costs dominate, rather than peaking generation. A comparison of total system costs for both scenarios when assuming OCGT and CCGT assumptions is provided in Figure S4. Total system costs when using CCGT assumptions were within 6% of those when using OCGT (large) assumptions for all scenarios. Note that CCGT capital costs have increased even further than these assumptions (based on *2024 ISP Model*) due to turbine shortages recently.

*Figure S4. Difference in total system costs when assuming gas turbines are OCGT vs CCGT*

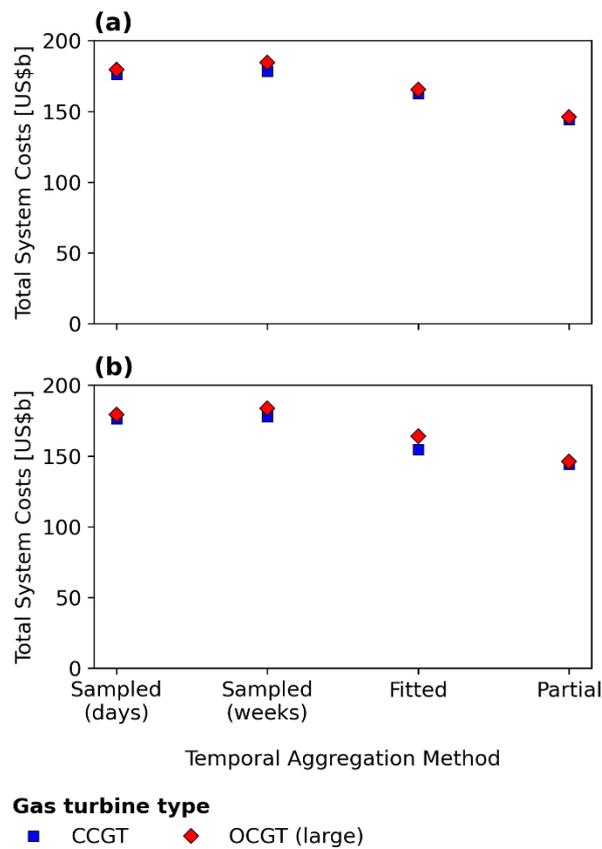

## FIRM Near Optimal Region

Three optimisations with FIRM were performed to explore the Low Pumped Hydro and High Pumped Hydro regions of the near-optimal solution space. Candidate solutions that were within 20% of the build costs of the Fitted PLEXOS global optimum were included in the near-optimal space. The total flexible capacity was plotted against the total new build pumped hydro capacity of the near-optimal solutions in Figure S5 and a boxplot of new build capacity in the full near-optimal space is provided in Figure S6.

The range of values depicted by the boxplots does not represent the full range of near-optimal solutions within the space; only a small subset that were actually explored by the differential evolution optimisation algorithm used by FIRM. The near-optimal solution space for the NEM model has over 100 dimensions which makes it infeasible to map the entire region, even with the rapid searching which is possible using a BR-LTP model, meaning that optimisation algorithms and the modeller are still required to make choices about which parts of the space to focus on.

A lack of population diversity, small mutation rate, or high recombination of the best performing candidates into the next iteration could lead to the optimiser becoming stuck in a local minimum. If the mutation rate was too high or recombination was too low, the optimiser would instead fail to converge [139, 111]. While a global optimum is not guaranteed by the metaheuristic optimiser, the ability to rapidly explore many candidate solutions for feasibility and approximate costs makes it suitable for evaluating regions in the near-optimal solution space that could achieve the objectives of the energy planner.

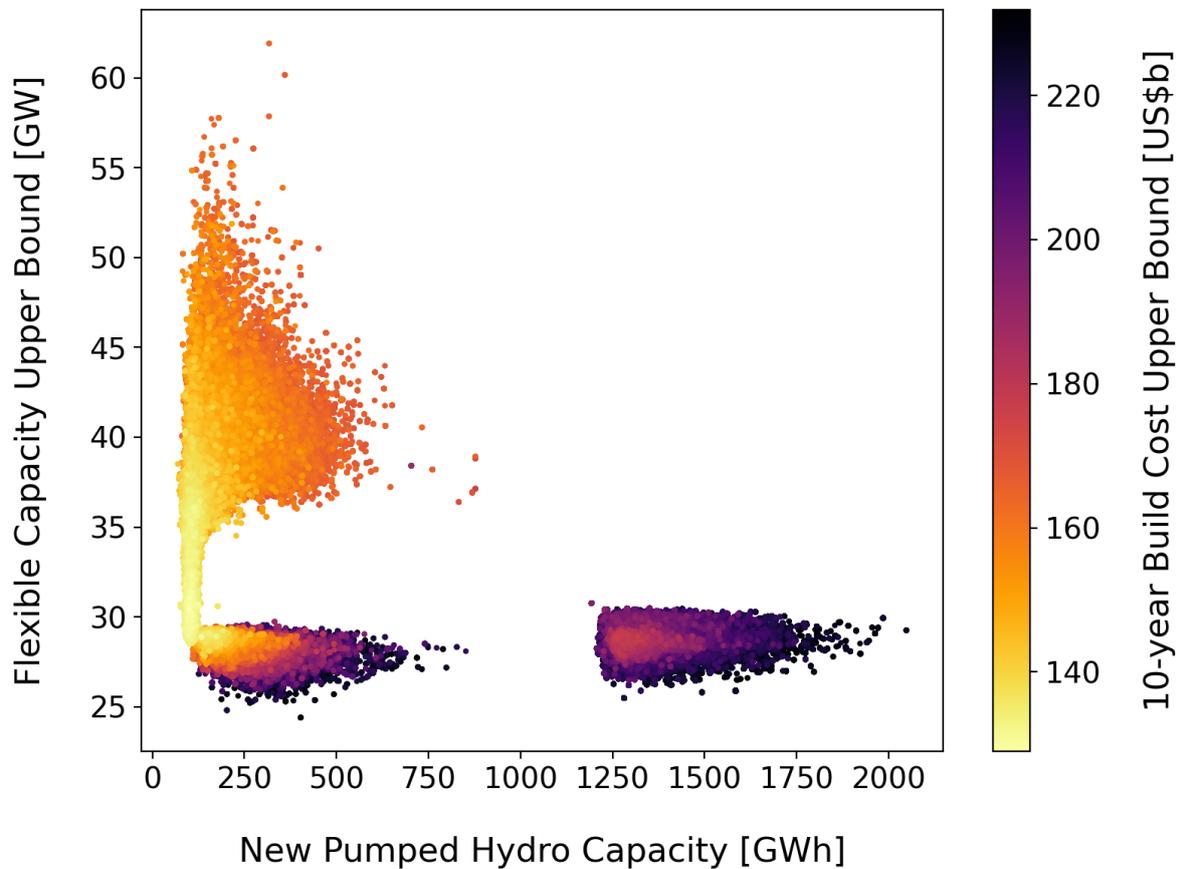

*Figure S5. Heatmap of total build costs for near-optimal candidate solutions (<120% of Fitted 100% Renewables build costs) based upon FIRM optimisations. Low Pumped Hydro (two candidate clusters on the left) and High Pumped Hydro (candidate cluster on the right) regions are depicted.*

The lower build cost of the Low Pumped Hydro region in Figure S5 is to be expected, since these solutions have higher costs related to gas fuel. The near-optimal region was based upon build costs, rather than total system costs, since BR-LTP models are expected to over-use capacity relative to optimal unit commitment formulations. The over-use of capacity means that gas generators are dispatched at a much higher capacity factor than necessary by FIRM. By polishing a candidate solution with a model that performs optimal unit commitment, the OCGT capacity and generation can be minimised to calculate more reasonable variable costs.

The first optimisation performed in FIRM was equivalent to the Improved PHES Assumptions scenario optimised in PLEXOS in § "The Bias towards Gas in Existing Long-term Energy Plan Formulations". In Figure S5, the evolution of the population for this first optimisation can be seen to start with a very large flexible capacity and a broad range of pumped hydro capacities. This population eventually narrows the pumped hydro capacity down to a tight range of values as flexible capacity is iteratively reduced. To capture some additional diversity in pumped hydro capacity, a second optimisation was performed by setting the flexible capacity max build limits equal to the final values in the least-cost solution of the first optimisation. This allowed the differential evolution to explore candidate solutions with a larger range of pumped hydro capacities in the near-optimal space, before these capacities narrowed down and converged. The third optimisation constrained pumped hydro minimum build limits to be equal to the values at the global optimum for the PLEXOS Fitted 100% Renewables scenario, allowing a High Pumped Hydro region of candidate solutions to be developed.

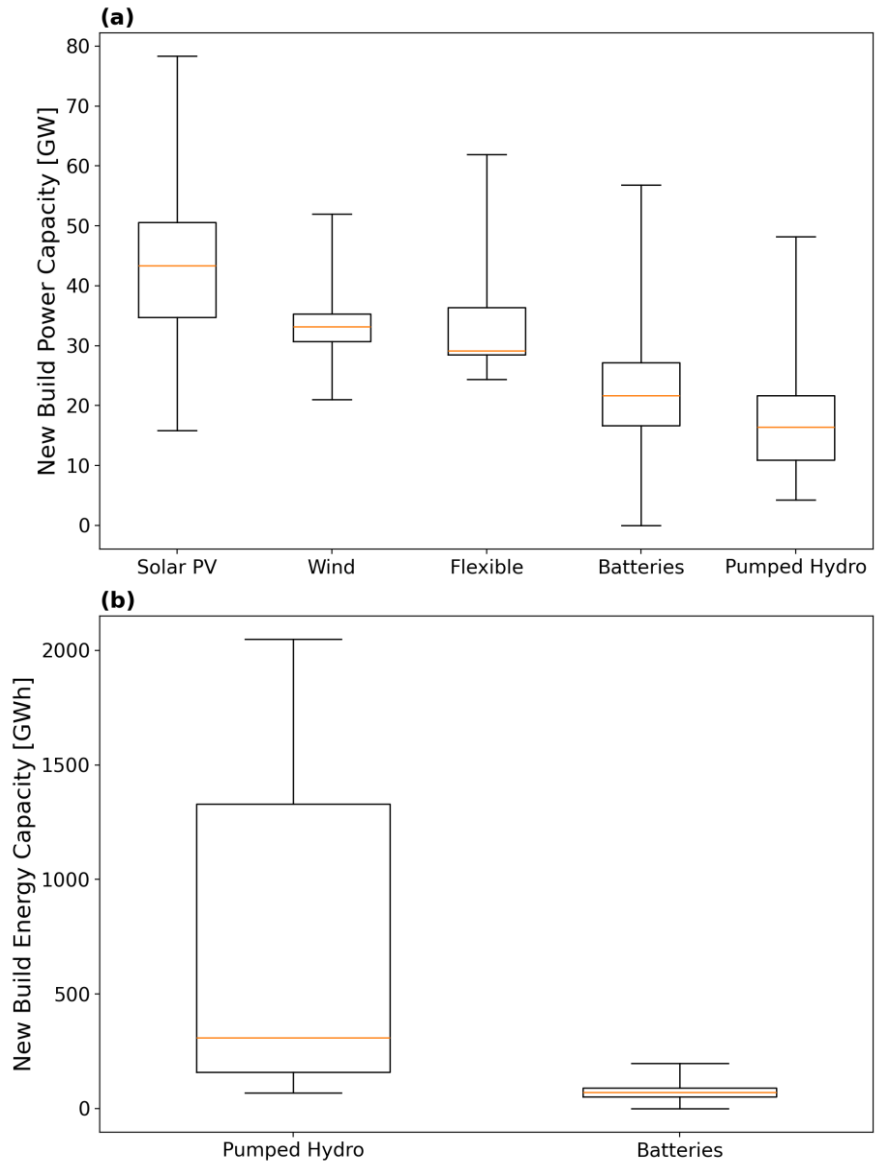

*Figure S6. Boxplot of the near-optimal space (<120% of Fitted 100% Renewables build costs) for: (a) new build power capacity of each technology, and (b) new build energy capacity of storage technologies. Whiskers extend to the minimum and maximum values.*

Solutions explored by the differential evolution algorithm are not to be considered a random sample or representative of the entire near-optimal solution space. These solutions are biased by the mutation and recombination process. Regardless, the algorithm still allows for a diverse range of feasible solutions to be explored and analysed. Future work should consider the development of optimisation algorithms for use with BR-LTP models with the explicit purpose of performing modelling to generate alternatives (MGA).

## Expanding the FIRM Near Optimal Region

Two additional optimisations were performed using the FIRM model to explore the solution space. One optimisation explicitly removed flexible capacity from the model, with the region of candidate solutions explored by the model depicted in the bottom-right corner of Figure S7. The second optimisation set the maximum build limits of flexible generators to the mid-point between the least-cost solution and 0 GW, with the region of candidate solutions shown in purple in the middle of Figure S7.

The cost cut-off was expanded in Figure S7 to include candidate solutions with build costs within 100% of the 100% Renewables optimum. In the absence of sufficient flexible capacity, build costs in the FIRM model were driven up beyond the original cut-off of the near-optimal region. The main driver of the increase in build costs was the substantial increase in storage volume. Load centres without pumped hydro options, such as the SNW and GG nodes, required a large battery capacity to replace the flexible generators that were otherwise built there. Larger pumped hydro energy volume was also required to ensure sufficient energy was stored to manage challenging winter weeks. The larger storage volume required in a 100% renewable energy scenario in FIRM compared to a linear programming model is a direct result of the inefficiencies associated with using business rules for unit commitment.

*Figure S7. Heatmap of total build costs for near-optimal candidate solutions (<200% of Fitted 100% Renewables build costs) based upon FIRM optimisations.*

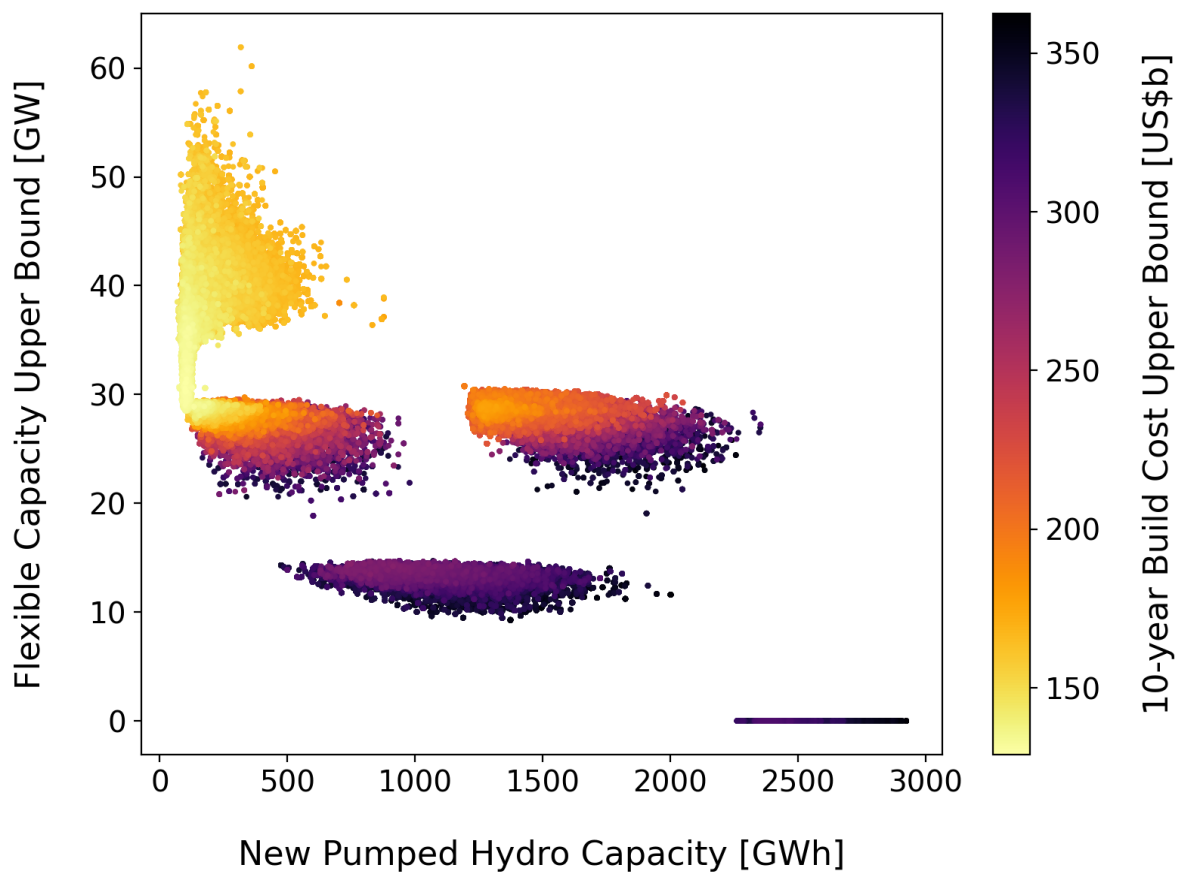

For energy to move from a pumped hydro system to meet peak demand in a load centre, it is bound by 3 different constraints: the power capacity of the pumped hydro system, the energy currently stored in the reservoir, and the remaining capacity available on the transmission lines. The best quality unit commitment algorithm would allow the pumped hydro to be dispatched at optimal times when transmission lines are not congested to recharge batteries at the load centres, ensuring the batteries and imports from transmission lines are available to meet peak load. The heuristics of BR-LTP models may struggle to thread this needle (something that would be optimised by real-world dispatch algorithms) since they rely on simplicity to minimise the time spent evaluating each candidate solution. The simple heuristics require an additional buffer of energy volume for batteries in load centres and pumped hydro in regional nodes to soak up the energy lost due to inefficient dispatch. By maintaining some flexible capacity in the FIRM system, over-build of storage energy volume can be minimised and interesting candidate solutions can be polished using a linear programming model with optimal unit commitment.

*Figure S8. Boxplot of the near-optimal space (<200% of Fitted 100% Renewables build costs) for: (a) new build power capacity of each technology, and (b) new build energy capacity of storage technologies. Whiskers extend to the minimum and maximum values.*

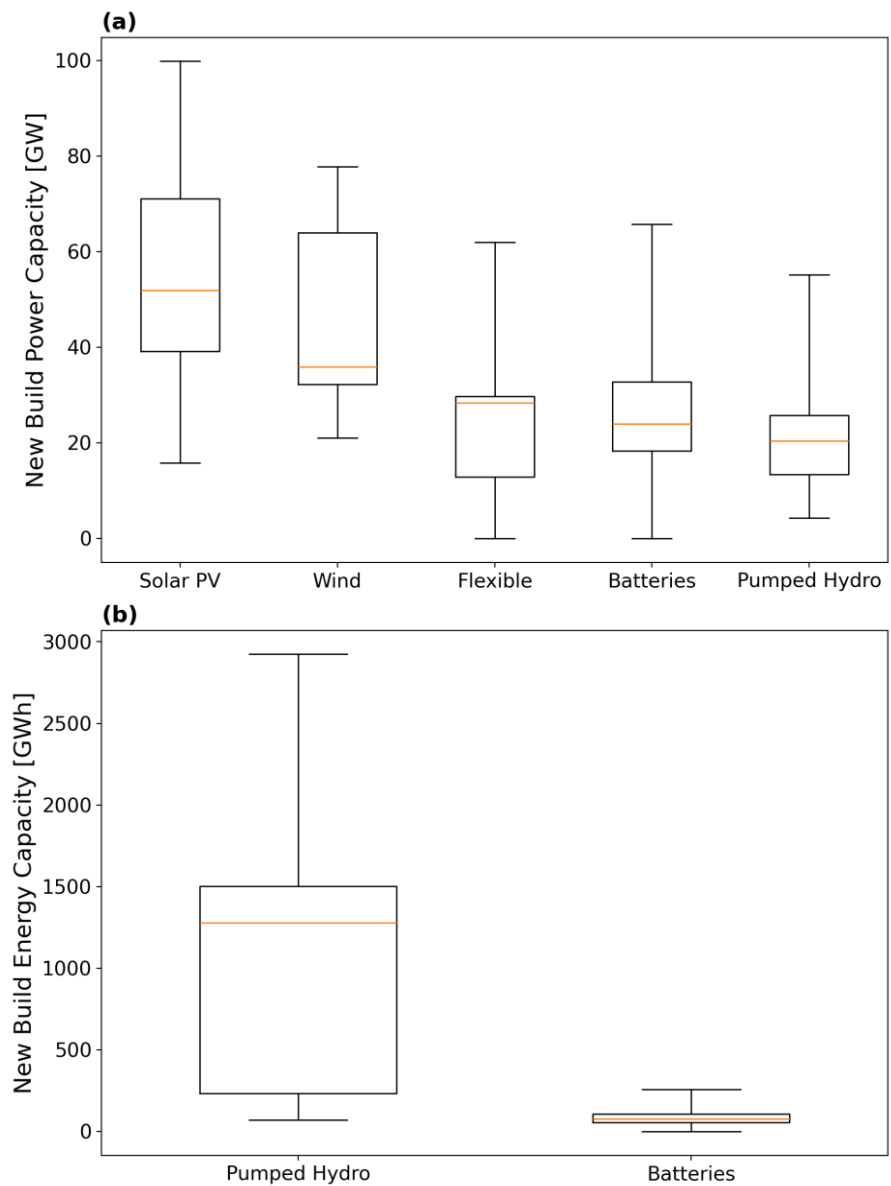

The over-build and over-use of capacity in a BR-LTP model implies that feasible candidate solutions represent an upper bound cost on the global optimum of a model with optimised unit commitment. When coupled with the fact that BR-LTP models do not require any temporal aggregation, an important characteristic of the candidate solutions is established: if a candidate solution from a BR-LTP model achieves the objectives of the energy planner, then it is guaranteed to achieve those goals under optimal unit commitment. To put it another way, a reliable, low-cost, low-emissions solution to a BR-LTP model will be even more reliable, lower cost, and lower emissions under optimal dispatch rules.

Still, it is useful to minimise over-build and over-use of capacity as much as possible in the BR-LTP model so that the modeller has a clearer understanding of the solution space before polishing any candidate solutions. Future BR-LTP models should attempt to refine the business rules used for unit commitment to represent optimal unit commitment behaviour more closely. One of the key drivers of pumped hydro energy volume is *dunkelflaute* in one or two weeks during winter each year. Therefore, improved business rules ought to focus on better management of energy storage reserves, and the efficient transmission of those reserves to load centre nodes.

## Additional Sensitivity Analysis Baselines

The sensitivity analysis was performed by varying parameters one-at-a-time from a baseline. Each baseline provides a different slice of the multi-parameter sensitivity analysis. In the main article, we presented the results for a baseline that used a real discount rate of 3% and a pumped hydro economic lifetime of 75 years, assuming that the pumped hydro was treated as a regulated monopoly (refer Figure 6). We used the upper bound on useful lifetime for civil infrastructure used by Snowy Hydro to define the 75-year economic lifetime of the pumped hydro systems [140]. The technical lifetime of pumped hydro is 150 years with periodic refurbishment.

A low discount rate and long economic lifetime is reasonable for large-scale pumped hydro energy storage, since the scale of these infrastructure projects makes it challenging to develop efficient competition. Additionally, both Borumba (48 GWh) and Snowy 2.0 (350 GWh) are being developed by government-owned corporations, with access to cheap capital. Public civil infrastructure typically has a longer economic lifetime than private assets.

However, AEMO is required to use a real discount rate suitable for private investment when developing the ISP in accordance with the Australian Energy Regulator's (AER) *Cost Benefit Analysis guidelines*. The cost of new build pumped hydro in the ISP is, therefore, based upon a real discount rate of 7% and economic lifetime of 40 years. Such an assumption would roughly double the annuity factor of pumped hydro capital costs compared to a 3% real discount rate with an economic lifetime of 75 years (refer Eq. (S3)). By treating a small number of large-scale pumped hydro systems the same as the dozens of competing solar farms, wind farms, utility-scale batteries, and gas generators, rather than treating them like a natural monopoly such as transmission, the cost analysis can become distorted (refer Figure S9).

In Figure S9, the High Pumped Hydro representative solutions have a much higher sensitivity to the energy volume cost and power capacity cost compared to Figure 6. Furthermore, increasing the economic lifetime for pumped hydro beyond 40 years does not substantially influence all-in LCOE relative to the baseline. At a 7% real discount rate, the asset has almost entirely depreciated at 40 years. Adjusting the pumped hydro economic lifetime ought to be done at the same time as adjusting the real discount rate.

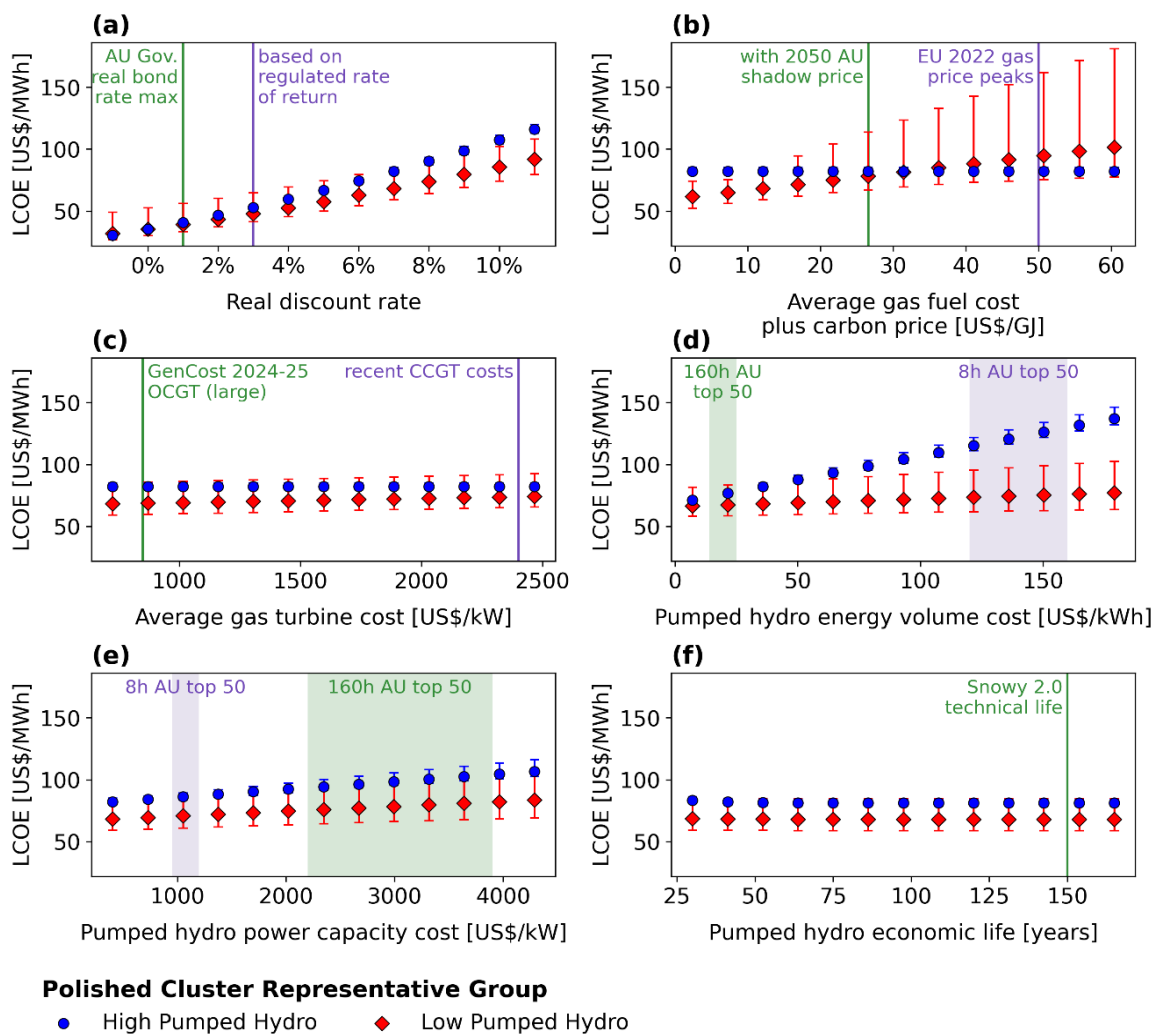

*Figure S9. Sensitivity of all-in LCOE for representative near-optimal solutions to: (a) real discount rate, (b) gas fuel costs, (c) gas turbine capital costs, (d) pumped hydro energy volume capital costs, (e) pumped hydro power capacity capital costs, and (f) pumped hydro economic life. Baseline values assume 7% real discount rate and 40-year economic life for pumped hydro.*

An additional baseline is provided in Figure S10 which uses a 3% real discount rate, 75-year economic lifetime for pumped hydro, and includes the Australian 2050 value of emissions reduction (i.e., carbon shadow price) in the cost of gas fuel. The value of emissions reduction is required to be used for evaluating elements of the National Electricity Objective related to achieving Australia's targets to reduce greenhouse gas emissions. The value of emissions reduction was not included in the baseline for Figure 6 presented in the main article since not all countries have assigned a price or shadow price to greenhouse gas emissions. However, the baseline in Figure S10 should be considered reasonable for the case of Australia, and countries with a price or a shadow price on carbon should incorporate those costs into their own calculations. Note that we only applied the value of emissions reduction to the scope 1 emissions associated with the combustion of gas and did not capture the cost of methane leakage along the gas supply chain. By considering the value of emissions reduction in Figure S10, the all-in LCOEs for the Low Pumped Hydro representative solutions were shifted up.

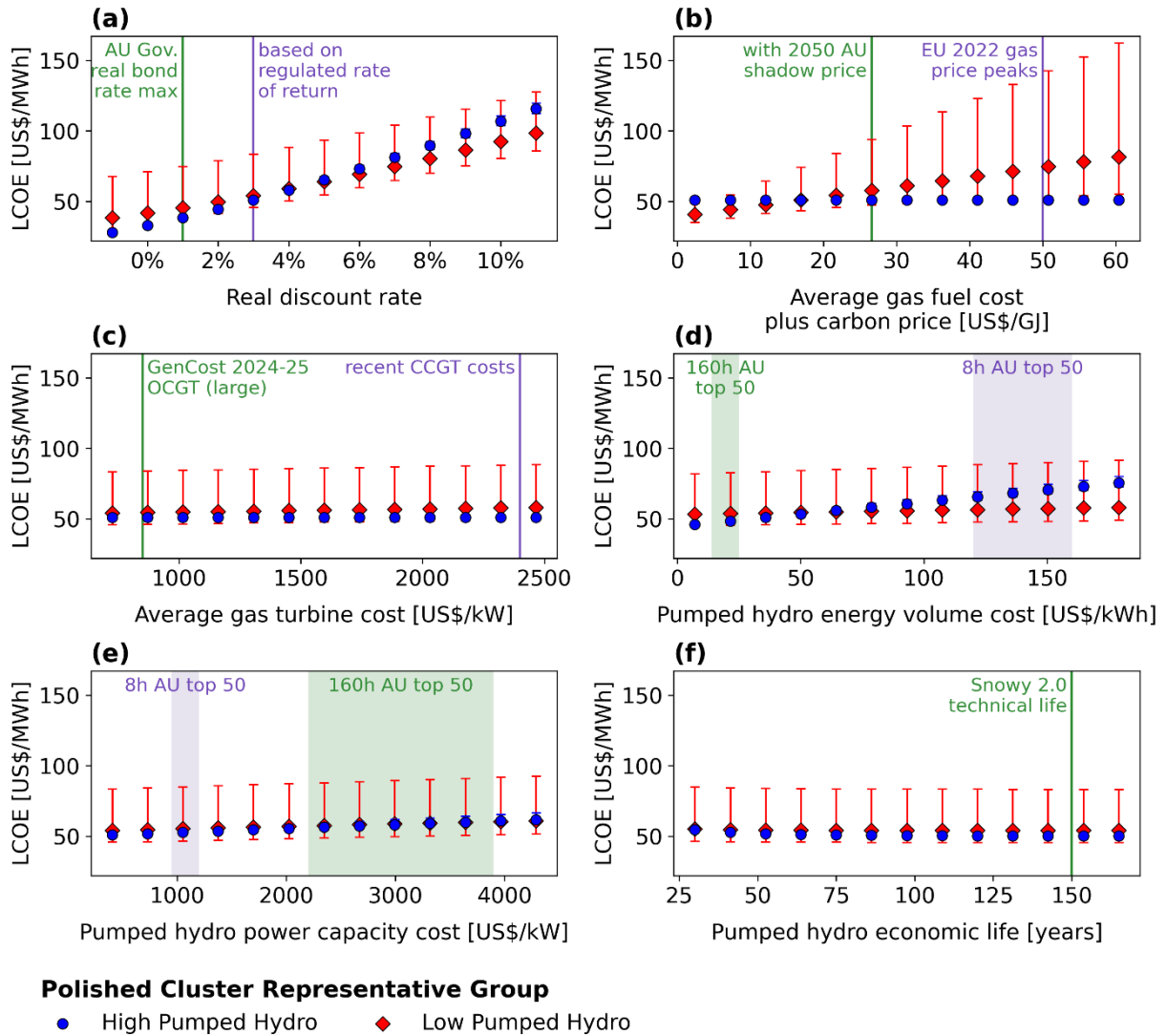

Figure S10. Sensitivity of all-in LCOE for representative near-optimal solutions to: (a) real discount rate, (b) gas fuel costs, (c) gas turbine capital costs, (d) pumped hydro energy volume capital costs, (e) pumped hydro power capacity capital costs, and (f) pumped hydro economic life. Baseline values assume 3% real discount rate, 75-year economic life for pumped hydro, and including the Australian 2050 value of emissions reduction (i.e., carbon shadow price) in the cost of gas fuel.

## Modelling the Cost of Pumped Hydro in the NEM

The ANU parametric cost model for pumped hydro systems was originally developed to determine cost classes for the global pumped hydro energy storage atlases [141]. The cost model estimates the costs of reservoirs, the powerhouse, switchyard and tunnels, but does not include land costs, water costs, or local taxes. It is intended as a comparative model to evaluate the quality of different sites relative to each other and should not be used as an authoritative source of real project costs. The same should be said of any top-down parametric cost model. Pumped hydro costs are highly dependent on the characteristics of specific sites, and may be influenced by factors such as geology, protected species, local heritage, land costs, social license, and complexity that can only be determined by actually visiting the site, engaging with local governments and communities, and digging for rock samples.

A shortlist of all 15–500 GWh pumped hydro options within 50 km of the NEM transmission network was generated using the Pumped Hydro Shortlisting Tool [142]. The ANU parametric cost model was used to estimate the power capacity and energy volume costs for each of the 20,500 shortlisted options. To account for some of the unknown costs, a 50% overhead add-on was included in the cost estimates. The costs of class AAA, AA, and A sites are shown in Figure S11. The intent of this figure is to demonstrate the broad range of site-specific costs available for pumped hydro systems, and that long-duration storage should be focused on sites that minimise energy volume cost.

The energy volume cost of each site was independent of the storage duration, since the duration was varied by adjusting the total power capacity of tunnels and pump-turbines at each location. Economies of scale are achieved for both energy volume (marginal increases in dam wall height raise the whole surface of the reservoir up, substantially increasing volume) and power capacity (costs associated with larger tunnels and powerhouses do not increase linearly with turbine size).

Long-duration storage requires a large energy capacity and a moderate power capacity. When coupled with batteries, pumped hydro power capacity within the grid can be reduced since the batteries can support the system to meet peak demand. During *dunkelflaute*, pumped hydro can trickle-charge the batteries when solar and wind generation are not available, ensuring that sufficient power capacity is still available for morning and evening peaks. If only 8-hour pumped hydro systems were built, then developing enough energy capacity for long-duration storage would result in a substantial over-build of the more expensive pumped hydro power capacity.

*Figure S11. Power capacity costs (powerhouse, switchyard, tunnels) and energy capacity costs (dam walls) for pumped hydro options within 50km of the NEM, assuming: (a) 8-hour duration, (b) 24-hour duration, (c) 48-hour duration, and (d) 160-hour duration. Includes an additional 50% overhead cost.*

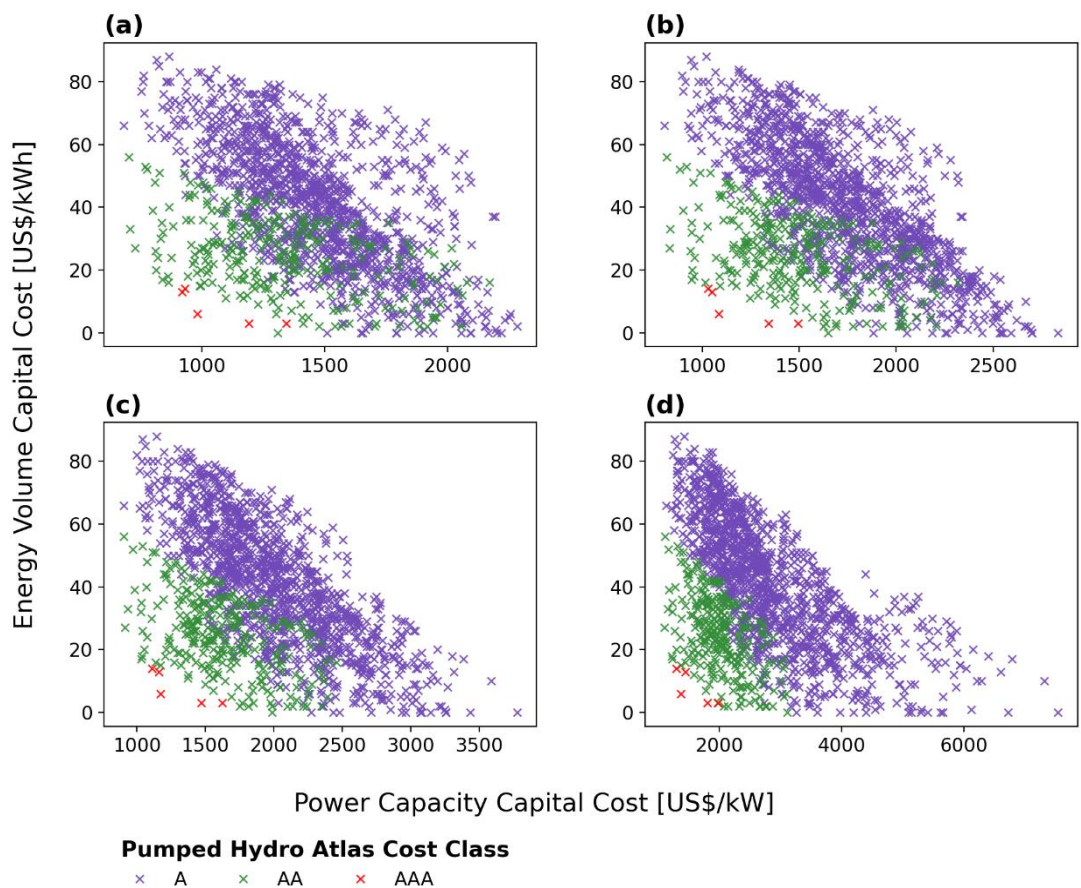

## Improving the Modelling of Long-duration Storage in the AEMO ISP

Currently, the AEMO ISP iteratively runs an SSLT model based upon typical periods and a DLT model based upon segmentation [87]. Even though the DLT model may capture long-duration storage behaviour, the SSLT model is likely to filter these solutions out through the iterative modelling process. The solution to the 100% Renewables scenario using the Sampled (days) typical period method was 83% more expensive than the solutions containing gas, while the Fitted method found a solution that was similar cost. The SSLT model is not fit-for-purpose when evaluating renewable-dominant systems and is expected to heavily bias solutions away from storage durations longer than the current 48-hour maximum and towards gas.

AEMO could instead implement MGA techniques [61] based on the DLT model to search the near-optimal space for solutions relying upon energy storage instead of gas. These changes ought to be coupled with more reasonable pumped hydro capital cost assumptions, much larger pumped hydro build limits, and new build pumped hydro storage durations longer than 48 hours. The 2026 version of the ISP is already seeing progress through updates to the pumped hydro build limits [90].

In the near-term, AEMO could include a 100% Renewables sensitivity scenario in the ISP. Similar sensitivity scenarios are already run to consider a small number of alternative pathways, such as the *Faster Coal Retirement* scenario in the 2026 ISP. By removing the SSLT model (which is not fit-for-purpose when modelling renewables-dominant grids) and running a *100% Renewables* sensitivity scenario, AEMO could demonstrate that the cost of a fully decarbonised electricity system supported by long-duration pumped hydro has a similar cost to the existing core scenarios. Once the method for the *100% Renewables* sensitivity scenario has been established and tested, it could take a more central role as a core scenario in future ISPs. A 100% Renewables scenario would demonstrate the most reliable pathway to achieving Australia's legislated target of net zero emissions by 2050.

The current pumped hydro cost model that is used for the ISP does not comport with the cost of Snowy 2.0. It also appears to substantially over-estimate the costs of the highest-quality large-scale sites across Australia (class A–AAA). Furthermore, the cost model provides a single average estimate for pumped hydro costs, adjusted by a locational cost factor in each NEM sub-region (node) to account for differences in labour costs, topography etc. In 2025, the locational cost factors were updated by GHD to reflect the mainland having similar quality sites to Tasmania [90]. However, these locational cost factors are relative values and were updated independent of the central cost model. Paradoxically, this led to pumped hydro in Tasmania suddenly appearing twice as expensive as it was two years ago, rather than the mainland sites being half as expensive. The pumped hydro cost model used by the ISP ought to reflect the best quality sites available in each NEM sub-region, since developers will target the highest quality sites to develop the very small number of large-scale long-duration storage required to decarbonise the grid. Storage durations longer than 48-hours, such as the 160-hour option included in our analysis, should also be costed and included in the ISP modelling. When using the segmentation method in the DLT, it is likely that the model will invest heavily in long-duration pumped hydro in the absence of gas.

Pumped hydro systems have technical lifetimes of a century or more, and the levelised cost is directly proportional to the discount rate. In contrast, gas generation is considerably less sensitive to discount rate because the system lifetime is 20–30 years and a substantial fraction of the cost is fuel and maintenance. "Discount rate" in this paper means "real discount rate". That is, the nominal discount rate is reduced by the inflation rate as measured by the All Groups

Consumer Price Index (CPI) published by the Australian Bureau of Statistics [92]. The wholesale price of electricity is expected to inflate at a similar rate when averaged over decades.

Within the Australian electricity system, transmission and distribution are natural monopolies regulated by the AER. The AER manages the rate of return for regulated monopolies in accordance with the Rate of Return Instrument [93]. At the time of publishing the *2025 Inputs, Assumptions, and Scenarios Report*, the latest determination by the AER reflected a real discount rate of 3% for regulated monopolies [94]. Given that only a small number of additional large-scale long-duration pumped hydro systems (roughly equivalent in scale to Snowy 2.0) would be required to support a 100% renewable electricity grid, it is possible that these systems could take on a monopolistic role in the market. The scale of these infrastructure projects also makes it challenging to develop efficient competition. It may be necessary for large-scale long-duration energy storage systems to be treated as a natural monopoly with a regulated rate of return to restrict monopolistic behaviour. As a regulated monopoly, they would have a much lower real discount rate than other riskier private investment.

For government-owned assets, the discount rate is even lower. The Commonwealth Government is currently offering bond lines with a coupon rate of between 1.00–4.75%,[15] while the Queensland Government is offering bonds with coupon rates of between 1.50–5.25% [91, 143]. At the current All Groups CPI inflation rate of 3.8% [92], this results in a real bond rate of –2.8% to 1.45% for these governments. Snowy Hydro (the developer of Snowy 2.0) and Queensland Hydro (the developer of Borumba) are Commonwealth and Queensland Government-owned corporations. Note that negative real discount rates are possible for government-owned assets.

For the upcoming 2026 Integrated System Plan, AEMO will assume an average 7% pre-tax real discount rate (appropriate for private enterprise investment), along with additional sensitivity scenarios at 3% (most recent AER rate of return determination) and 10% (Infrastructure Australia's guidelines for economic appraisal) [94]. AEMO is required to use a discount rate appropriate for private enterprise investment in the electricity grid across the NEM for the ISP, and to set a lower bound on the discount rate in accordance with the latest determination under the Rate of Return Instrument [144]. This requirement to use a real discount appropriate for private investors, as established by the AER's *Cost Benefit Analysis guidelines*, may not comport with reality when considering large-scale pumped hydro energy storage systems. Two large-scale systems already under development in the NEM, Snowy 2.0 and Borumba, are government-owned. Since the cost of capital-intensive projects is highly sensitive to discount rate, new build pumped hydro projects are likely unnecessarily penalised under the current *Cost Benefit Analysis guidelines* from the AER.

---

[15] Excluding the 0.5% AU0000106411 Treasury Bond line which matures on 21 September 2026.

*Table S2. LT Plan Settings for Each Temporal Simplification Scenario*

| Temporal Simplification | Scenario Code | Setting | Value |
|---|---|---|---|
| All temporal simplification methods | All scenarios | Step Size (years) | 10 |
| | | Overlap (years) | 0 |
| | | Discount Rate (%) | 7 |
| | | End Effects Method | None |
| | | Discount/Expansion Period | Year |
| | | Depreciation Method | None |
| | | Expansion Algorithm | Optimize |
| | | Expansion Decisions Integer Optimality | Linear |
| | | Integration Horizon (years) | -1 |
| | | Number of Solutions | 1 |
| | | Solution Quality (%) | 0 |
| | | Transmission | Nodal |
| | | Heat Rate | Simplest |
| | | Formulate Head Effects | Checked |
| | | Generation Pricing Method | Average |
| | | Bridge: Constraints | Checked |
| | | Bridge: Storage | Checked |
| Sample of typical time periods | Sampled (days) | Chronology | Sampled |
| | | Blocks in each Sample | 0 |
| | | Number of Years Sampled | 0 |
| | | Sample | 2 Days per Month |
| | Sampled (weeks) | Chronology | Sampled |
| | | Blocks in each Sample | 0 |
| | | Number of Years Sampled | 0 |
| | | Sample | 4 Weeks per Year |
| Approximate LDC | Partial | Chronology | Partial |
| | | One Duration Curve each | Day |
| | | Slicing Method | Weighted Least-squares Fit |
| | | Weight a (constant) | 0 |
| | | Weight b (linear) | 1 |
| | | Weight c (quadratic) | 0 |
| | | Weight d (cubic) | 0 |
| | | Pin Top | -1 |
| | | Pin Bottom | -1 |
| | | Blocks in each Duration Curve | 8 |
| Segmentation into adjacent intervals of similar demand | Fitted | Chronology | Fitted |
| | | Fit Step Function each | Day |
| | | Blocks in last curve in Horizon | 0 |
| | | Slicing Method | Weighted Least-squares Fit |
| | | Weight a (constant) | 0 |
| | | Weight b (linear) | 1 |
| | | Weight c (quadratic) | 0 |
| | | Weight d (cubic) | 0 |
| | | Pin Top | -1 |
| | | Pin Bottom | -1 |
| | | Blocks in each Day | 8 |

*Table S3. Modifications made to 2024 ISP Model*

| ID | Change | Explanation | Impact |
|---|---|---|---|
| 1.1 | Changed LT plan horizon to 10 years (FY 2042–2052) | Shorter time horizon makes fitted chronology (Fitted) and partial chronology (Partial) manageable in single step optimisation. Time horizon of 10 years still demonstrates multi-year unit commitment behaviour and improves robustness of solutions over a range of weather conditions. | Fewer weather years tested, which may reduce the robustness of solutions. |
| 1.2 | Used 2052 operational demand for every year within planning horizon | Grid configuration is evaluated for a point-in-time rather than considering true capacity expansion over the planning horizon. The 10-year horizon allows the grid configuration to be tested for reliable unit commitment over multiple years of weather data. | Demand growth over a period of time with full capacity expansion behaviour is not considered by the model. Lock-in effects, such as building gas generators quickly due to long lead times for large infrastructure projects (pumped hydro and transmission) are not taken into account. Retirement of existing assets is not considered. |
| 1.3 | Set Max Units Built in Year for all relevant generators to 0 after the first year | Forces all units to be built in the first year. This converts the model from a true capacity expansion model to a point-in-time model that finds a single grid configuration that can reliably solve the unit commitment problem. As a point-in-time model, the impact of temporal simplification on unit commitment can be evaluated without additional complexity related to development lead times, retirement costs, and annual demand growth. | All assets are built overnight to effectively find the least-cost grid configuration to reliably meet demand over the 10-year period. Lock-in effects, such as building gas quickly due to long lead times for large infrastructure projects (pumped hydro and transmission) are not taken into account. Retirement of existing assets is not considered. |
| 1.4 | Removed start and end dates for Generator and Battery Units that are anticipated or committed | Assumes that the anticipated and committed generators have been completed by FY2052. The only new build capacity is provided by new REZ solar, REZ wind, OCGT (large), batteries, and pumped hydro. | Anticipated and committed projects that may be delayed or cancelled are still assumed to be part of the grid in all solutions. |
| 2.1 | Removed coal, bioenergy, hydrogen, and liquid fuel generator objects | Solar PV and wind were 89% of total generation capacity installed globally in 2024. Based on these trends, bioenergy, hydrogen and liquid fuel are expected to be a negligible proportion of generation capacity compared to solar PV and wind. Coal is anticipated to be retired from the National Electricity Market before FY2052. | Some small amount of hydrogen and bioenergy could reduce the need for gas or energy storage. |
| 2.2 | CCGT and OCGT (small) generator objects were | The *2024 Integrated System Plan* forecasts 15 GW of gas, predominantly in the form of peaking OCGT (large), will be | Solutions that contain mid-merit CCGT and small OCGT gas are not considered by any of the models, even if those |
| | removed. Existing gas generator objects were removed. | required in 2050 [63]. Our analysis considers whether improvements to the modelling of long-duration pumped hydro in long-term energy planning can provide an alternative pathway to this flexible gas. | solutions could have slightly lower system costs. OCGT (large) generators are the main gas generators in 2050 within the ISP [63], so these effects are expected to be minimal. |
| 2.3 | Set OCGT (large) Max Units Built to 50 GW | A large enough build limit on new flexible gas in each node to make sure it is not a binding constraint. | Resource limits associated with the construction of new OCGT generators are not considered. |
| 2.4 | Removed Load Subtractor Generator objects | Load Subtractors are used to estimate residual load before the typical period clustering algorithm is executed (that is, the clustering method is an *a posteriori* method). The Load Subtractors are iteratively derived through multiple runs of the models. They have been removed to simplify the comparison between different temporal aggregation techniques for this paper. | For this paper, clustering was performed using *a priori* techniques which are expected to produce lower quality solutions than *a posteriori* typical period clustering. Despite *a posteriori* techniques producing better quality clusters [46], unit commitment within repetitions of the same typical period would still be identical. So, these methods still face issues with rapidly binding storage state-of-charge over many repetitions of the typical period. |
| 2.5 | Changed OCGT (large) Unit Size to 1 MW | Allows new build OCGT generators to be sized flexibly by the optimisation. | Requires the use of an average heat rate instead of the linear heat rate function. Otherwise, the heat rate base (GJ/h) would be applied to each 1 MW unit and substantially overestimate gas consumption. |
| 2.6 | Added Average Heat Rate attribute to OCGT (large) objects. Removed Heat Rate Base and Heat Rate Incr attributes. | Required due to the flexible sizing of OCGT (large) units in increments of 1 MW per unit. Average heat rates are based upon the AEMO *Inputs, Assumptions and Scenarios* report for the *2024 ISP* and correspond to those values used in the Single-Stage Long-Term model for that energy plan [115]. | The base heat rate is not considered independent of the incremental heat rate. In reality, an OCGT generator operating substantially below its nameplate capacity could be expected to be less efficient than one operating at its nameplate capacity. |
| 3.1 | Storage and Waterway objects were removed | All energy storage systems were defined as Battery objects and hydro power stations were assigned annual energy constraints. This removed extra complexity around water availability data which would be location-specific, while this analysis intends to make broader comment on global energy plans. | Water availability data could constrain the use of hydro power stations at certain point within a year. Therefore, hydro use is likely over-estimated within our model, which could bias away from new build gas or energy storage. |
| 3.2 | Hydroelectric power stations with relationships to Storage and Waterway | In the absence of water availability data, a generation constraint is required to limit the use of flexible hydro. These annual constraints were based upon the annual generation of | While hydro generation is constrained on an annual basis, it is not constrained on shorter timeframes. This could |
| | objects were modified to have an annual generation constraint | each power station calculated by the unmodified *2024 ISP Model*. | lead to over-use of the hydro assets compared to a real electricity system. |
| 3.3 | Existing, committed and anticipated pumped hydro systems were converted from Generator objects to Battery objects | Exogenously defined pumped hydro capacity is then treated the same as endogenously optimised new build pumped hydro. | Open-loop hydro systems which were converted to battery objects may be over-used compared to a real electricity system. This could drive down the need for new build pumped hydro in the model. |
| 4.1 | Initial SOC was defined to be 50% for all Battery objects | Assume that storage systems have some stored energy to manage challenging periods at the beginning of the planning horizon. | Different orders of reference weather years would influence the impact of this assumption. For example, starting an extreme weather year with a lower state-of-charge would drive the model to build more gas to cover early deficits. |
| 4.2 | Removed Max Cycles Day, Min SoC and Max SoC from Battery objects | Simplify battery objects to evaluate effects of temporal simplification in the absence of exogenously defined constraints that may vary between different battery operators or future battery technologies. | To minimise degradation under real conditions, batteries may avoid fully charging and discharging, as well as minimise their cycle frequency. This may mean that additional battery energy capacity is required to provide the storage necessary for a reliable grid. |
| 5.1 | Removed Forced Outage Rate, Outage Factor, and Mean Time to Repair from Generator and Battery objects | Simplification to evaluate the effects of temporal aggregation without stochastic maintenance events. | Maintenance events in a real electricity system will require additional storage or generation capacity to cover for the lost capacity during downtime. |
| 6.1 | Removed timeslice changes on Max Flow and Min Flow for Line objects. Instead, tested with only winter Max/Min Flows | Simplification assumes maximum transmission capacity is available. It is likely that transmission flows in a low-gas system are higher in winter as electricity is transmitted from large pumped hydro systems into load centres (where pumped hydro cannot be built), so overestimating summer transmission is not expected to substantially affect results. | Effects of climate change may change the frequency and timing of hot summer transmission line rating constraints. Slow transmission build-out may also constrain the use of transmission lines during unit commitment. Hot summer line ratings may be important when battery capacity at load centres is low. |
| 6.2 | Removed Loss Coefficient from Marinus lines | Simplification since transmission lines attributes are not the focus of this analysis. | Transmission may be over-used in the model compared to a real system, since line losses would drive the optimum towards supplying more electricity locally at each node. |
| 6.3 | Removed Marginal Loss Factor, Marginal Loss Factor Back, and Loss Allocation from Line objects | Only LT plan formulation in PLEXOS is relevant to this analysis, so variables related to the electricity market and prices were removed. Removal of line losses since transmission line attributes are not the focus of this analysis. | Transmission may be over-used in the model compared to a real system, since line losses would drive the optimum towards supplying more electricity locally at each node. |
| 6.4 | Start dates are removed from the anticipated and committed Line objects | It is assumed that the transmission lines that are forecast for development have been completed by 2052. | Transmission project delays or cancellations would reduce the availability of transmission compared to the model. |
| 7.1 | Removed Marginal Loss Factors from Generator and Battery objects. Removed Mark-up from Battery objects. | Only LT plan formulation in PLEXOS is relevant to this analysis, so variables related to the electricity market and prices were removed. | No impact on LT plan results. |
| 7.2 | Removed Max Capacity Factor, and Min Capacity Factor from Generator and Battery objects | Exogenously constrained capacity factors are not relevant to the analysis in this paper. | Generator and Battery objects are freely dispatched for unit commitment rather than constrained by exogenous capacity factor bounds. |
| 7.3 | Removed Aux Incr, Rating Factor, Rating, and Efficiency Incr from Generator objects | Detailed unit commitment attributes are removed to make the comparison between temporal aggregation techniques clearer. | Auxiliary power, generator efficiency, and changes to the rating throughout the year could slightly increase costs and drive the need for additional new build assets to provide reliable electricity supply. New build capacity may be slightly underestimated with the removal of these attributes. |
| 8.1 | Removed DSP Bid Price, DSP Bid Quantity, and Generator Settlement Model from Regions | Only LT plan formulation in PLEXOS is relevant to this analysis, so variables related to the electricity market and prices were removed. | No impact on LT plan results. |
| 8.2 | Removed Load Includes Losses, Load Metering Point, Load Settlement Model and VoLL from Regions | Only LT plan formulation in PLEXOS is relevant to this analysis, so variables related to the electricity market and prices were removed. | No impact on LT plan results. |
| 9.1 | Removed Constraint objects for REZ Generator objects. Added these constraints as Max Units Built to the REZ Generator objects. | The *2024 ISP Model* soft-constrains REZ deployment with an assumption that the cost of land will increase as more power stations are built within the REZ. These effects were excluded from our analysis to simplify the comparison between temporal aggregation techniques. | New build capacity in individual REZs may be overestimated, since the land acquisition cost would likely increase in reality. New build capacity in the model is likely more concentrated on a few locations with the best solar and wind resources. |
| 9.2 | Removed Constraint objects related to government policies (e.g., renewable energy targets and emission reduction targets) | Government policies are location-specific, while this analysis intended to make comments about global energy plan model formulations and assumptions. | Climate policies would likely drive faster deployment of renewable generators and energy storage systems, and constrain the use of gas. Since we developed a point-in-time model using 2052 demand, these impacts are likely minimal. |
| 10.1 | Removed Min Capacity Reserve from Zone objects | The effect of reserve constraints was beyond the scope of analysis in this paper. | OCGT and energy storage objects may be under-built in our model compared to a real electricity system that could be required to maintain reserve capacity. |
| 10.2 | Removed Firm Capacity from Generator and Battery objects | The effect of reserve constraints was beyond the scope of analysis in this paper. | OCGT and energy storage objects may be under-built in our model compared to a real electricity system that could be required to maintain reserve capacity. |
| 10.3 | Removed Min Capacity Reserves and Max Capacity Reserves from Line objects | The effect of reserve constraints was beyond the scope of analysis in this paper. | OCGT and energy storage objects may be under-built in our model compared to a real electricity system that could be required to maintain reserve capacity. |
| 11.1 | Added 160-hour pumped hydro option to CNSW, NNSW, SNSW, CQ, NQ, VIC, and TAS nodes with a build cost of $9435/kW (2023 AUD) for the Improved PHES Assumptions and 100% Renewables scenarios. | The NEM subregions to which the 160-hour pumped hydro option was added were found to have more than sufficient 150 GWh and 500 GWh options on the global pumped hydro energy storage atlases. The $9435/kW was a conservative estimate that would price a system the size of Snowy 2.0 at approximately AU$21 billion. | Finding near-optimal 100% renewable energy grids with a conservative cost estimate implies that good-quality pumped hydro site selection would make these systems even cheaper. It also allowed for contingency costs and other overheads to be comfortably captured in the cost, while still providing a cheaper option for deep storage compared to the 48-hour pumped hydro options. |

# FIRM Model Formulation

The equations and algorithms defining the FIRM BR-LTP formulation used for this analysis are provided in the subsections below.

## Objective Function and Penalty Functions

The objective function for the FIRM long-term planning model is defined in Eq. (S1). The system costs ($SC$, $/MWh) are minimised through a differential evolution. The total costs over the planning horizon are divided by the total demand, calculated from the load ($L$, MW) at each spatial node ($n$) in each time interval ($t$) in a year ($y$) at the temporal resolution of the data ($r$, hours), to provide a more manageable system-level levelised cost of energy (LCOE) for the user.

$$\min SC = \frac{FC+VC}{r \times \sum_y \sum_t \sum_n L_{n,y,t}} + PF^{\text{UE}} + PF^{\text{FC}} \tag{S1}$$

The fixed costs ($FC$, $) are calculated according to Eq. (S2) as the sum of new build power capacity ($\overline{\delta^{\text{np}}}$, kW) capital costs ($CP$, $/kW or $/kW-km), new build energy capacity ($\overline{\delta^{\text{ne}}}$, kWh) capital costs ($CE$, $/kWh), and fixed operation and maintenance costs ($F$, $/kW-year or $/kW-km-year) for the total power capacity ($\delta^{\text{tp}}$) of generators ($g \in \mathcal{G}$), storage systems ($s \in \mathcal{S}$), and transmission lines ($l \in \mathcal{L}$) of a given length ($d_l$, km) in each year of the planning horizon. Capital costs are annualised according to the annuity factor ($AF$). The $\overline{(\cdot)}$ indicates decision variables for the differential evolution.

$$FC = \sum_y \left( \begin{array}{c} \sum_g \left( \frac{CP_{g,y}^{\text{gen}} \times \overline{\delta_{g,y}^{\text{gen,np}}}}{AF_g} + F_{g,y}^{\text{gen}} \times \delta_{g,y}^{\text{gen,tp}} \right) + \\ \sum_s \left( \frac{CP_{s,y}^{\text{storage}} \times \overline{\delta_{s,y}^{\text{storage,np}}} + CE_{s,y} \times \overline{\delta_{s,y}^{\text{storage,ne}}}}{AF_s} + F_{s,y}^{\text{storage}} \times \delta_{s,y}^{\text{storage,tp}} \right) \\ + \sum_l d_l \times \left( \frac{CP_{l,y}^{\text{line}} \times \overline{\delta_{l,y}^{\text{line,np}}}}{AF_l} + F_{l,y}^{\text{line}} \times \delta_{l,y}^{\text{line,tp}} \right) \end{array} \right) \tag{S2}$$

The set of all assets $k \in \mathcal{K} = \mathcal{G} \cup \mathcal{S} \cup \mathcal{L}$. The annuity factor for asset $k$ is calculated according to its economic lifetime ($Y$, years) and discount rate ($dr$, %), as shown in Eq. (S3).

$$AF_k = \frac{1-(1+d_k)^{-Y_k}}{d_k} \tag{S3}$$

Total capacity of asset $k$ is the sum of new build capacity and existing capacity ($\delta^{\text{ep}}$, kW and $\delta^{\text{ee}}$, kWh), as per Eq. (S4) and (S5).

$$\delta_{k,y}^{\text{tp}} = \overline{\delta_{k,y}^{\text{np}}} + \delta_{k,y}^{\text{ep}} \tag{S4}$$

$$\delta_{k,y}^{\text{te}} = \overline{\delta_{k,y}^{\text{ne}}} + \delta_{k,y}^{\text{ee}} \tag{S5}$$

The variable costs ($VC$, $) are calculated according to Eq. (S6) from the variable operation and maintenance costs ($V$, $/MWh) based upon the generation from generators ($\varphi_{g,y,t}$, MW), power discharged from storage systems ($\max(\varphi_{s,y,t}, 0)$, MW), and magnitude of the power flowing through each transmission line ($|\varphi_{l,y,t}|$, MW) in each time interval. For generators, the average fuel costs ($U_g^{\text{avg.}}$, $/MWh) are also included in the calculation. Since the unit commitment problem is based upon deterministic business rules rather than an optimisation, there are no decision variables in Eq. (S6). Variables are calculated sequentially from business rules by

iterating through $t$ and $y$ (noting that business rules may sometimes move backwards through $t$ and $y$, such as when pre-charging storage systems).

$$VC = r \times \sum_y \sum_t \begin{pmatrix} \sum_g \left( \left( V_g^{\text{gen}} + U_g^{\text{avg.}} \right) \times \varphi_{g,y,t}^{\text{gen}} \right) + \\ \sum_s \left( V_s^{\text{storage}} \times \max\left( \varphi_{s,y,t}^{\text{storage}}, 0 \right) \right) + \\ \sum_l \left( V_l^{\text{line}} \times \left| \varphi_{l,y,t}^{\text{line}} \right| \right) \end{pmatrix} \quad (S6)$$

A penalty function for unserved energy ($PF^{\text{UE}}$, \$/MWh), calculated according to Eq. (S7), is used to apply a soft constraint that ensures the reliability standard ($RS$, %) of the grid is met. If the unserved energy ($UE$, MWh) breaches the reliability standard of the system (e.g., 99.998% of demand met for the National Electricity Market [109]), then it is multiplied by a large scalar to make the solution non-optimal. This penalty function is roughly analogous to assuming some large Value of Lost Load (VoLL). The penalty function is re-calculated at the end of each year $y$ and may end the unit commitment early if it has a value greater than 0 in order to reduce the optimisation time associated with unreliable solutions.

$$PF^{\text{UE}} = 10^6 \times r \times \max\left( \sum_t \sum_n (UE_{n,y,t}) - (1 - RS) \times \sum_t \sum_n (L_{n,y,t}), \ 0 \right) \quad (S7)$$

An additional fixed cost soft constraint may be applied using Eq. (S8) to reduce optimisation time by skipping unit commitment for high-cost solutions that are unlikely to meet the energy planning objectives. A fixed cost threshold ($FCT$, \$/MWh) can be exogenously defined by the user. Units of \$/MWh are used to define this fixed cost threshold as a proportion of system LCOE. Note that setting the $FCT$ too small carries the risk of the differential evolution getting stuck in a local minimum, where the fixed costs are just small enough for $PF^{\text{FC}} = 0$, but $PF^{\text{UE}} > 0$. Increasing the mutation rate of the differential evolution or increasing the $FCT$ can mitigate this risk.

$$PF^{\text{FC}} = \max\left( \frac{FC}{r \times \sum_y \sum_t \sum_n L_{n,y,t}} - FCT, \ 0 \right) \times 10^6 \quad (S8)$$

The objective function is formulated for each candidate solution $\left[ \overline{\delta_{g,y}^{\text{gen,np}}}, \overline{\delta_{s,y}^{\text{storage,np}}}, \overline{\delta_{s,y}^{\text{storage,ne}}}, \overline{\delta_{l,y}^{\text{line,np}}} \right]$ in the Scipy differential evolution population through **Algorithm 1**. The differential evolution is vectorised and a parallel range of solutions is tested simultaneously using Numba. This is an embarrassingly parallel process that could be run with any number of workers up to the size of the population.

---

**Algorithm 1** FIRM objective function

**Data:** $\left[ \overline{\delta_{g,y}^{\text{gen,np}}}, \overline{\delta_{s,y}^{\text{storage,np}}}, \overline{\delta_{s,y}^{\text{storage,ne}}}, \overline{\delta_{l,y}^{\text{line,np}}} \right]$ candidate solution from population. All other exogenous parameters.
**Result:** $SC$

Initialise solution object with all exogenous parameters.
Calculate $FC$ according to Eq. (S2).
Calculate fixed cost penalty $PF^{\text{FC}}$ according to Eq. (S8). If $PF^{\text{FC}} > 0$, set $SC = PF^{\text{FC}}$ and return objective function early.
Find $\varphi_{g,y,t}^{\text{gen}}$, $\max(\varphi_{s,y,t}^{\text{storage}}, 0)$, $\left| \varphi_{l,y,t}^{\text{line}} \right|$, and $UE_{n,y,t}$ and according to Algorithm 2. Within each year y evaluated by Algorithm 2, calculated unserved energy penalty $PF^{\text{UE}}$ according to Eq. (S7). If $PF^{\text{UE}} > 0$, set $SC = PF^{\text{UE}}$ and return objective function early.
Calculate totals $\sum_y \sum_t \varphi_{g,y,t}^{\text{gen}}$, $\sum_y \sum_t \max(\varphi_{s,y,t}^{\text{storage}}, 0)$, and $\sum_y \sum_t \left| \varphi_{l,y,t}^{\text{line}} \right|$

Calculate $VC$ according to Eq. (S6).
Calculate $SC$ according to Eq. (S1) and return objective function.

## Unit Commitment According to Business Rules

The unit commitment involves a series of business rules that work to balance the nodal loads in each time interval, along with defining storage charging and discharging behaviour. It involves a sequential iteration through each time interval, similar to a simulation model. Unlike a simulation, however, the model may iterate backwards through time to adjust previous dispatch decisions. The goal of the rules-based unit commitment formulation is to approximate the optimal solution that would be achieved through a perfect foresight optimisation, rather than to make specific assumptions about the forecasting capabilities or strategies of generator or storage system operators.

Within the model, there is a set of generator indices $g \in \mathcal{G} = \{\mathcal{G}^{\text{pv}} \cup \mathcal{G}^{\text{wind}} \cup \mathcal{G}^{\text{baseload}} \cup \mathcal{G}^{\text{flex}}\}$ where each subset consists of units labelled as either solar PV, wind, baseload or flexible generators. The $g$ index indicates an individual generator with $n(g)$ defining a map between generators and the nodes $n$ at which they are located. Similarly, $s$ indicates a specific storage system and $n(s)$ is a map between storage systems and their nodes.

The residual load ($RL_{n,y,t}$, GW) is calculated according to Eq. (S9).

$$RL_{n,y,t} = L_{n,y,t} - \left(G_{n,y,t}^{\text{pv}} + G_{n,y,t}^{\text{wind}} + G_{n,y,t}^{\text{baseload}}\right) \tag{S9}$$

The nodal solar PV ($G_{n,y,t}^{\text{pv}}$, GW), wind ($G_{n,y,t}^{\text{wind}}$, GW), and baseload ($G_{n,y,t}^{\text{baseload}}$, GW) dispatch powers are calculated according to equation (S10), (S11) and (S12) respectively. The availability of the generator ($A_{g,y,t}$, %) is imported from a data trace, with one trace per generator. The sets $\mathcal{G}^{\text{pv}}$, $\mathcal{G}^{\text{wind}}$ and $\mathcal{G}^{\text{baseload}}$ define the set of indices $g$ associated with each generator type. The sum is performed for generators at each node, as per the mapping $n(g) = n$. Note that curtailment is not explicitly determined for each generator in these equations. Rather, curtailment and spillage are captured in the same metric at the end of unit commitment.

$$G_{n,y,t}^{\text{pv}} = \sum_{n(g)} \left(\delta_{g,y}^{\text{pv,tp}} \times A_{g,y,t}\right) \tag{S10}$$

$$G_{n,y,t}^{\text{wind}} = \sum_{n(g)} \left(\delta_{g,y}^{\text{wind,tp}} \times A_{g,y,t}\right) \tag{S11}$$

$$G_{n,y,t}^{\text{baseload}} = \sum_{n(g)} \left(\delta_{g,y}^{\text{baseload,tp}} \times A_{g,y,t}\right) \tag{S12}$$

The initial energy stored in storage systems ($\gamma_{s,y=0,t=0}$, GWh) is calculated for all storage systems according to Eq. (S13).

$$\gamma_{s,y=0,t=0}^{\text{storage}} = 0.5 \times \delta_{s,y=0}^{\text{storage,te}} \tag{S13}$$

The energy balance within the unit commitment algorithm is performed according to **Algorithm 2**.

**Algorithm 2** Perform the energy balance over modelling horizon

**Data**: Solution object
**Result**: $\varphi_{g,y,t}^{\text{gen}}$, $\max(\varphi_{s,y,t}^{\text{storage}}, 0)$, $|\varphi_{l,y,t}^{\text{line}}|$, and $UE_{n,y,t}$ and for all $t$ and $y$. $PF^{\text{UE}}$ updated at the end of each $y$.

Initialise energy storage with Eq. (S13).
Calculate residual load $RL_{n,y,t}$ according to Eq. (S9).
Iterate through each $y$ to find $\varphi_{g,y,t}^{\text{gen}}$, $\varphi_{s,y,t}^{\text{storage}}$, $\varphi_{l,y,t}^{\text{line}}$, and $UE_{n,y,t}$ to balance $RL_{n,y,t}$ with Algorithm 3. At the end of each iteration, calculate $PF^{\text{UE}}$ and return early if penalty is greater than 0.
Calculate the magnitude of transmission flows $|\varphi_{l,y,t}^{\text{line}}|$
Calculate storage discharging powers $\max(\varphi_{s,y,t}^{\text{storage}}, 0)$.

---

Within a time interval, business rules progressively update unit commitment decisions and unit attributes (such as the energy stored in a storage system) in a sequence of steps. The constraint on storage discharging power ($S_{s,y,t}^{\text{dis}}$, GW) during a time interval is the minimum of power capacity and remaining storage, calculated from the stored energy, discharging efficiency ($\eta^{\text{dis}}$, %), and time interval resolution, as per Eq. (S14). The charging power is similarly constrained ($S_{s,y,t}^{\text{ch}}$, GW) by the minimum of power capacity and remaining storage, as per Eq. (S15). Charging efficiency is defined by $\eta^{\text{ch}}$ (%). The storage discharging ($S_{n,y,t}^{\text{dis}}$, GW) and charging ($S_{n,y,t}^{\text{ch}}$, GW) constraints at each node are calculated by Eq. (S16) and (S17).

When pre-charging the storage systems by moving backwards through time intervals, symmetry rules are applied to the charging/discharging behaviour. That is, a discharging action at interval $t$ in reverse time will add energy to the storage system at $t-1$. Similarly, charging in reverse time will remove energy from the system at $t-1$. Within the pre-charging period, $\gamma_{s,y,t+1}^{\text{storage}}$ is actually stored as a temporary value since it is overridden when resolving the discontinuity between stored energy in the forwards and reverse time directions once pre-charging has completed.

$$S_{s,y,t}^{\text{dis}} = \begin{cases} \min\left(\frac{\gamma_{s,y,t-1}^{\text{storage}} \times \eta^{\text{dis}}}{r}, \delta_{s,y}^{\text{storage,tp}}\right) & \text{for increasing } t \\ \min\left(\frac{\left(\delta_{s,y}^{\text{storage,te}} - \gamma_{s,y,t+1}^{\text{storage}}\right) \times \eta^{\text{dis}}}{r}, \delta_{s,y}^{\text{storage,tp}}\right) & \text{for decreasing } t \end{cases} \quad (S14)$$

$$S_{s,y,t}^{\text{ch}} = \begin{cases} \min\left(\frac{\delta_{s,y}^{\text{storage,te}} - \gamma_{s,y,t-1}^{\text{storage}}}{r \times \eta^{\text{ch}}}, \delta_{s,y}^{\text{storage,tp}}\right) & \text{for increasing } t \\ \min\left(\frac{\gamma_{s,y,t+1}^{\text{storage}}}{r \times \eta^{\text{ch}}}, \delta_{s,y}^{\text{storage,tp}}\right) & \text{for decreasing } t \end{cases} \quad (S15)$$

$$S_{n,y,t}^{\text{dis}} = \sum_{n(s)} S_{s,y,t}^{\text{dis}} \quad (S16)$$

$$S_{n,y,t}^{\text{ch}} = \sum_{n(s)} S_{s,y,t}^{\text{ch}} \quad (S17)$$

Flexible generators are treated in a similar way to storage systems, although they have no ability to charge and the available annual generation ($\gamma_{g,y,t}$, GWh) takes the place of the stored energy. The first time interval of each year initialises the available annual generation according to the pre-defined annual generation constraint ($\delta_{g,y}^{\text{flex,te}}$, GWh), as per Eq. (S18). The constraints on generation from each individual flexible generator ($S_{g,y,t}^{\text{flex}}$, GW) and all flexible generators at a node ($S_{n,y,t}^{\text{flex}}$, GW) are given by Eq. (S19) and (S20).

$$\gamma_{g,y,t=0}^{\text{flex}} = \delta_{g,y}^{\text{flex,te}} \quad \forall g \in \mathcal{G}^{\text{flex}} \tag{S18}$$

$$S_{g,y,t}^{\text{flex}} = \begin{cases} \min\left(\frac{\gamma_{g,y,t}^{\text{flex}}}{r}, \delta_{g,y}^{\text{flex,tp}}\right) & \forall g \in \mathcal{G}^{\text{flex}} \text{ for increasing } t \\ \min\left(\frac{\delta_{g,y}^{\text{flex,te}} - \gamma_{g,y,t}^{\text{flex}}}{r}, \delta_{g,y}^{\text{flex,tp}}\right) & \forall g \in \mathcal{G}^{\text{flex}} \text{ for decreasing } t \end{cases} \tag{S19}$$

$$S_{n,y,t}^{\text{flex}} = \sum_{n(g)} S_{g,y,t}^{\text{flex}} \quad \forall g \in \mathcal{G}^{\text{flex}} \tag{S20}$$

We use the term "residual load" to refer to the load that is not balanced by solar PV, wind, and baseload generation, while "net load" refers to the outstanding residual load at a particular node in a given time interval throughout the unit commitment process. Residual load is calculated once, net load is periodically updated as transmission occurs, and storage systems and flexible generators are dispatched. Surplus generation ($SG$, MW) occurs whenever the net load ($NL$, GW) is negative. Surplus generation is calculated for each node in each time interval according to Eq. (S21).

$$SG_{n,y,t} = -1 \times \min(NL_{n,y,t}, 0) \tag{S21}$$

When balancing unserved energy using storage systems located at the same node as the unserved energy, storage dispatch power is calculated according to Eq. (S22). It is calculated as the sum of discharging (positive GW) and charging (negative GW) power, constrained by the nodal storage discharge and charging limits. Net load is updated after transmission to include the energy imported and exported from each node, so storage systems can be charged and discharged to meet those transmission requirements.

$$\varphi_{n,y,t}^{\text{storage}} = \max(\min(NL_{n,y,t}, S_{n,y,t}^{\text{dis}}), 0) + \min(\max(NL_{n,y,t}, S_{n,y,t}^{\text{ch}}), 0) \tag{S22}$$

The nodal storage dispatch is apportioned across individual storage systems according to the merit order at each node, as per Eq. (S23). Storage systems are dispatched in order of shortest to longest storage duration $\left(\frac{\delta_{s,y}^{\text{storage,te}}}{\delta_{s,y}^{\text{storage,tp}}}, \text{hours}\right)$, with the index $o$ indicating their position in that order. The map $o(s, n)$ determines the location of a specific storage system $s$ node $n$ in the order.

$$\varphi_{o(s,n)=O,y,t}^{\text{storage}} = \begin{cases} \min\left(\varphi_{n,y,t}^{\text{storage}} - \sum_{o=0}^{O-1} \varphi_{o(s,n),y,t}^{\text{storage}}, S_{s,y,t}^{\text{dis}}\right) & \text{if } \varphi_{n,y,t}^{\text{storage}} > 0 \\ \max\left(\varphi_{n,y,t}^{\text{storage}} - \sum_{o=0}^{O-1} \varphi_{o(s,n),y,t}^{\text{storage}}, -S_{s,y,t}^{\text{ch}}\right) & \text{if } \varphi_{n,y,t}^{\text{storage}} < 0 \end{cases} \tag{S23}$$

After storage systems have been dispatched, flexible generators are dispatched at each node to balance remaining netload according to Eq. (S24). Flexible generator dispatch power is calculated in a similar manner to storage systems, although there is no charging power component.

$$\varphi_{n,y,t}^{\text{flex}} = \min(\max(NL_{n,y,t} - \varphi_{n,y,t}^{\text{storage}}, 0), S_{n,y,t}^{\text{flex}}) \tag{S24}$$

Flexible generators are ordered according to short-run marginal cost. The map $o(g, n)$ determines the location of a specific flexible generator $g$ at node $n$ in the order. The dispatch of individual flexible generators at each node is determined by Eq. (S25).

$$\varphi_{o(g,n)=O,y,t}^{\text{flex}} = \min\left(\varphi_{n,y,t}^{\text{flex}} - \sum_{o=0}^{O-1} \varphi_{o(g,n),y,t}^{\text{flex}}, S_{g,y,t}^{\text{flex}}\right) \tag{S25}$$

The energy stored in storage systems and remaining annual flexible generation is then calculated for each asset according to Eq. (S26) and (S27).

$$\gamma_{s,y,t}^{\text{storage}} = \gamma_{s,y,t-1}^{\text{storage}} - r \times \left(\frac{\max\left(\varphi_{s,y,t}^{\text{storage}},0\right)}{\eta^{\text{dis}}} + \eta^{\text{ch}} \times \min\left(\varphi_{s,y,t}^{\text{storage}},0\right)\right) \tag{S26}$$

$$\gamma_{g,y,t}^{\text{flex}} = \gamma_{g,y,t-1}^{\text{flex}} - r \times \varphi_{g,y,t}^{\text{flex}} \tag{S27}$$

**Algorithm 3** iterates through each time interval and attempts to balance the residual load first with storage systems, then with flexible generators. Surplus generation is used to charge storage systems. Electricity is transmitted along lines between nodes to either balance residual load or charge storage systems as required. If the residual load cannot be balanced according to these business rules, then storage systems are pre-charged in an attempt to resolve remaining residual deficits. The final time interval in a year is indexed by $t = T_y$.

---

**Algorithm 3** Balance residual demand in each time interval for a given year

    **Data**: Solution object
    **Result**: $\varphi_{g,y,t}^{\text{gen}}$, $\max(\varphi_{s,y,t}^{\text{storage}},0)$, $\varphi_{l,y,t}^{\text{line}}$, and $UE_{n,y,t}$, for all $t$ in given $y$. $PF^{\text{UE}}$ updated for $y$.

    Set storage pre-charging decision $\rho = \text{False}$
    **for** $t = 1, T_y$ **do**
        Initialise storage and flexible limits with Eq. (S14) – (S20).
        Initialise net load $NL_{n,y,t}$ according to corresponding $RL_{n,y,t}$
        **if** $\max(NL_{n,y,t}, 0) > 0$ for any $n$ **then**
            Calculate surplus generation $SG_{n,y,t}$ at each node according to Eq. (S21).
            Transmit surplus generation to balance unserved energy with Algorithm 4.
            Update nodal net load $NL_{n,y,t}$ based upon $\varphi_{n,y,t}^{\text{imp}}$, and $\varphi_{n,y,t}^{\text{exp}}$
        **end if**

        Calculate $\varphi_{n,y,t}^{\text{storage}}$ and $\varphi_{s,y,t}^{\text{storage}}$ using storage local to each node with Eq. (S22) and (S23).
        Update nodal net load $NL_{n,y,t}$ based upon storage dispatch $\varphi_{n,y,t}^{\text{storage}}$

        **if** $\max(NL_{n,y,t}, 0) > 0$ for any $n$ **then**
            Transmit available storage discharging power to balance deficits with Algorithm 4.
            Update nodal net load $NL_{n,y,t}$ based upon $\varphi_{n,y,t}^{\text{imp}}$, and $\varphi_{n,y,t}^{\text{exp}}$
            Apportion net load to storage systems with Eq. (S22) and (S23).
            Apportion remaining net load to local flexible generators with Eq. (S24) and (S25).
            Update nodal net load $NL_{n,y,t}$ based upon $\varphi_{n,y,t}^{\text{storage}}$ and $\varphi_{n,y,t}^{\text{flex}}$

        **end if**

---

**if** $\max(NL_{n,y,t}, 0) > 0$ for any $n$ **then**
    Transmit available flexible power to balance deficits with Algorithm 4.
    Update nodal net load $NL_{n,y,t}$ based upon $\varphi_{n,y,t}^{\text{imp}}$, and $\varphi_{n,y,t}^{\text{exp}}$
    Apportion net load to flexible generators with Eq. (S24) and (S25).
**end if**

Calculate surplus generation $SG_{n,y,t}$ according to Eq. (S21).

**if** $SG_{n,y,t} > 0$ for any $n$ **then**
    Transmit surplus generation to charge storages with Algorithm 4.
    Update nodal net load $NL_{n,y,t}$ based upon $\varphi_{n,y,t}^{\text{imp}}$, and $\varphi_{n,y,t}^{\text{exp}}$
    Apportion net load to storage systems with Eq. (S22) and (S23).
**end if**

Update the $\gamma_{s,y,t}^{\text{storage}}$ and $\gamma_{g,y,t}^{\text{flex}}$ using Eq. (S26) and (S27).

**if** not $\rho$ and $(\sum_n \max(NL_{n,y,t}, 0) > 0)$ **then**
    Update storage pre-charging decision $\rho = \text{True}$.
**end if**

**if** $\rho$ and $(\sum_n \max(NL_{n,y,t}, 0) < 0)$ **then**
    Pre-charge storage systems according to Algorithm 5.
    Update storage pre-charging decision $\rho = \text{False}$.
**end if**
**end for**

---

Power is transmitted throughout the network according to the pre-defined topology, based upon the start and end nodes assigned to each line. Ordered sequences of lines form routes ($\mathcal{R}_{n^{\text{fill}},p,c}$) through the topology. Routes are indexed by a "fill node", a length or leg ($p$), and a path to a final node ($c$). The final node in a route is called the "surplus node". Two routes may have the same length, fill node, and surplus node, but take a different path of lines to get between those end points.

Fill values ($TF$, GW) define the power that each node is attempting to import, while surplus values ($TS$, GW) define the power available for export at each node in a given time interval. During transmission, routes are iterated through (starting at the shortest $p$) in attempt to transmit surplus power from surplus nodes to balance fill power at fill nodes. The maximum power that can flow through a route ($MF_{\mathcal{R},y,t}$, GW) is based upon the smallest maximum flow of the lines ($MF_{l,y,t,p}$, GW) that form that route, as per Eq. (S28) and (S29). There is a mapping $n = n^{\text{surplus}}(\mathcal{R})$ that defines a relationship between routes and their surplus nodes.

$$MF_{l,y,t,p} = \min\left(\delta_{l,y}^{\text{line,tp}} - \varphi_{l,y,t}\right) \tag{S28}$$

$$MF_{\mathcal{R}} = \min\left(\min(MF_{l,y,t,p}), TS_{n^{\text{surplus}}(\mathcal{R})}\right) \quad \forall l \in \mathcal{R} \tag{S29}$$

The available imports for a fill node ($\varphi^{\text{imp,max}}_{n(l),y,t,p}$, GW) is the sum of the maximum route flows into that node for a given leg. This is provided by Eq. (S30).

$$\varphi^{\text{imp,max}}_{n=n^{\text{fill}},y,t,p} = \sum_c MF_{\mathcal{R}_{n^{\text{fill}},p,c}} \tag{S30}$$

Note that when iterating through paths $c$ for a given leg $p$, any given line may belong to multiple routes and any given node may be the surplus node for multiple routes. This means that temporary line flows and assigned surpluses must be used to avoid double-counting commitments made on previous paths. These temporary values have been excluded from the equations shown above for simplicity, though explanations can be found in the `firm_ce.fast_methods.network_m` module of the codebase.

The imports across all routes on a given leg are rescaled if they exceed the amount of energy sought by the fill node. The rescaled flow updates ($FU_\mathcal{R}$, GW) are calculated from Eq. (S31).

$$FU_\mathcal{R} = \min\left(1, TF_{n^{\text{fill}},y,t}/\varphi^{\text{imp,max}}_{n=n^{\text{fill}},y,t,p}\right) \times MF_\mathcal{R} \tag{S31}$$

Finally, the total imports to the fill node are updated based on the route flows to that node according to Eq. (S32). Similarly, line flows along the route and exports from the surplus nodes ($\varphi^{\text{exp}}_{n^{\text{surplus}},y,t}$, GW) are updated according to the route flow update. Imports are positive values and exports are negative values.

$$\varphi^{\text{imp}}_{n=n^{\text{fill}},y,t} \mathrel{+}= \sum_c FU_{\mathcal{R}_{n^{\text{fill}},p,c}} \tag{S32}$$

Equations (S28) – (S32) are intended to provide a description of the key calculations involved in determining transmission throughout the network, but are not a comprehensive description of the process. Different transmission cases (e.g., transmitting surpluses, discharging from neighbouring storage systems, discharging from neighbouring flexible generators etc.) are assigned different node fill and surplus values. Readers should refer to Algorithm 4 or the `firm_ce.fast_methods.network_m` module of the Python code for details.

---

**Algorithm 4** Perform transmission to fill each node using surplus power

---

**Data**: $TF_{n,y,t}$, $TS_{n,y,t}$, $\delta^{\text{line,tp}}_{l,y}$, $\max(p)$, $\varphi^{\text{line}}_{l,y,t}$, $\varphi^{\text{imp}}_{n,y,t}$, and $\varphi^{\text{exp}}_{n,y,t}$

**Result**: $\varphi^{\text{imp}}_{n,y,t}$, $\varphi^{\text{exp}}_{n,y,t}$, and $\varphi^{\text{line}}_{l,y,t}$

Set $FU_\mathcal{R} = 0$ for all routes.
**if** $\sum_n \sum_y \sum_t TS_{n,y,t} = 0$ **or** $\sum_n \sum_y \sum_t TF_{n,y,t} = 0$ **then**
    Return with no updates to $\varphi^{\text{imp}}_{n,y,t}$, $\varphi^{\text{exp}}_{n,y,t}$, and $\varphi_{l,y,t}$
**end if**

**for** $p = 1, \max(p)$ **do**
    **for** $n^{\text{fill}} = 1, N$ **do**
        **if** $TF_{n^{\text{fill}},y,t} == 0$ **then**
            continue
        **end if**
        **if** $TS_{n,y,t} == 0$ for all $n = n^{\text{surplus}}\left(\mathcal{R}_{n^{\text{fill}},p,c}\right)$ **then**
            continue
        **end if**

---

                Reset line temporary flows for leg $p$.
                Reset node temporary assigned surpluses for leg $p$.
                Determine maximum imports available for $n^{\text{fill}}$ according to Eq. (S30).
                Re-scale the imports to find the route flow updates $FU_{\mathcal{R}}$ using Eq. (S31).
                Update node imports $\varphi_{n,y,t}^{\text{imp}}$ with Eq. (S32), node exports $\varphi_{n,y,t}^{\text{exp}}$ and line flows $\varphi_{l,y,t}^{\text{line}}$ based on route flow updates.
                Update the fill $TF_{n,y,t}$ and surplus $TS_{n,y,t}$ values remaining for each node.
          **end for**
    **end for**

Upon reaching the end of a contiguous block of $t$ containing unserved energy (i.e., a deficit block), the pre-charging process is initiated. The first step is to attempt inter-storage transfers from storage systems with surplus stored energy (trickle-chargers) to empty storage systems (pre-chargers). Once no more inter-storage transfers are possible, flexible generators are dispatched to pre-charge the storage systems. Algorithm 5 provides the high-level function that manages storage system pre-charging.

**Algorithm 5** Perform inter-storage transfers and pre-charging of storage with flexible generators
        **Data**: Solution object and final $t$ of the deficit block
        **Result**: $\varphi_{g,y,t}^{\text{gen}}$, $\varphi_{s,y,t}^{\text{storage}}$, $\varphi_{l,y,t}^{\text{line}}$, and $UE_{n,y,t}$ for $t$ and $y$ during pre-charging period and deficit block.

        Determine energy that storage systems must be pre-charged with using Algorithm 6
        Update $\varphi_{s,y,t}^{\text{storage}}$ and $\varphi_{g,y,t}^{\text{flex}}$ required to achieve pre-charge energies with Algorithm 7
        Update $\gamma_{s,y,t}^{\text{storage}}$ and $\gamma_{g,y,t}^{\text{flex}}$ based upon feasible pre-charging with Algorithm 8

The first step, outlined in Algorithm 6, is to determine the trickle-charging energy available and the amount of energy that must be pre-charged for each storage system. The process involves iterating backwards through $t$ to perform the energy balance, using the symmetry rules for charging and discharging in reverse time. That is, charging a storage system in $t$ will remove energy from $t - 1$ (and vice versa for discharging).

The energy that must be pre-charged for each storage system ($PE_{s,y,t}$, GWh) is constrained based upon the difference in maximum ($\gamma_{s,y,t}^{\text{storage,max\_db}}$, GWh) and minimum ($\gamma_{s,y,t}^{\text{storage,min\_db}}$, GWh) stored energy within the deficit block. These minimum and maximum stored energy values are calculated according to Eq. (S33) and (S34). The amount of pre-charge energy required for each storage system is calculated from the difference in the stored energy calculated moving forwards in time compared to backwards in time in the first interval of the deficit block (refer Eq. (S35)). The + and – superscripts are used to distinguish between variables calculated when moving forwards or backwards in time respectively.

Energy available for trickle charging ($TE_{s,y,t}$, GWh) is also constrained according to the minimum and maximum stored energy values, according to Eq. (S36). The amount of flexible generation available for pre-charging ($FE_{g,y,t}$, GWh) is also constrained by ensuring sufficient $\gamma_{g,y,t}^{\text{flex}}$ remains to dispatch during the deficit block.

$$\gamma_{s,y,t}^{storage,max\_db} = \max(\gamma_{s,y,t+1}^{storage,max\_db}, \gamma_{s,y,t}^{storage,-}) \tag{S33}$$

$$\gamma_{s,y,t}^{storage,min\_db} = \min(\gamma_{s,y,t+1}^{storage,min\_db}, \gamma_{s,y,t}^{storage,-}) \tag{S34}$$

$$PE_{s,y,t} = \begin{cases} \gamma_{s,y,t}^{storage,-} - \gamma_{s,y,t}^{storage,+}, \text{if } \gamma_{s,y,t}^{storage,+} > \gamma_{s,y,t}^{storage,max\_db} - \gamma_{s,y,t}^{storage,min\_db} \\ 0, \text{ if } \gamma_{s,y,t}^{storage,+} < \gamma_{s,y,t}^{storage,max\_db} - \gamma_{s,y,t}^{storage,min\_db} \end{cases} \tag{S35}$$

$$TE_{s,y,t} = \gamma_{s,y,t}^{storage,max\_db} - \gamma_{s,y,t}^{storage,min\_db} \tag{S36}$$

---

**Algorithm 6** Determine energy that must be pre-charged for each storage system (deficit block)

**Data**: Solution object and first $t$ immediately following deficit block

**Result**: First $t$ in deficit block, $PE_{s,y,t}$, $TE_{s,y,t}$ $FE_{g,y,t}$, and updated $\varphi_{s,y,t}^{storage}$, $\varphi_{g,y,t}^{flex}$, $\varphi_{n,y,t}^{imp}$, $\varphi_{n,y,t}^{exp}$, and $\varphi_{l,y,t}^{line}$ for $t$ within the deficit block.

**while** True **do**

    Go back to previous time interval $t = t - 1$
    Initialise storage and flexible limits with Eq. (S14) – (S20)
    Reset transmission for the time interval with $\varphi_{l,y,t}^{line}, \varphi_{n,y,t}^{imp}, \varphi_{n,y,t}^{exp} = 0$
    Reset storage and flexible generator dispatch with $\varphi_{s,y,t}^{storage}, \varphi_{g,y,t}^{flex} = 0$

    Initialise storage and flexible limits with Eq. (S14) – (S20).
    Initialise net load $NL_{n,y,t}$ according to corresponding $RL_{n,y,t}$

    **if** $\max(NL_{n,y,t}, 0) > 0$ for any $n$ **then**
        Calculate surplus generation $SG_{n,y,t}$ at each node according to Eq. (S21).
        Transmit surplus generation to balance unserved energy with Algorithm 4.
        Update nodal net load $NL_{n,y,t}$ based upon transmission $\varphi_{n(l),y,t}$
    **end if**

    Calculate $\varphi_{n,y,t}^{storage}$ and $\varphi_{s,y,t}^{storage}$ using storage local to each node with Eq. (S22) and (S23).
    Update nodal net load $NL_{n,y,t}$ based upon storage dispatch $\varphi_{n,y,t}^{storage}$

    **if** $\max(NL_{n,y,t}, 0) > 0$ for any $n$ **then**
        Transmit available storage discharging power to balance deficits with Algorithm 4.
        Update nodal net load $NL_{n,y,t}$ based upon transmission $\varphi_{n(l),y,t}$
        Apportion net load to storage systems with Eq. (S22) and (S23).
        Apportion remaining net load to local flexible generators with Eq. (S24) and (S25).
        Update nodal net load $NL_{n,y,t}$ based upon $\varphi_{n,y,t}^{storage}$ and $\varphi_{n,y,t}^{flex}$
    **end if**

**if** $\max(NL_{n,y,t}, 0) > 0$ for any $n$ **then**
    Transmit available flexible power to balance deficits with Algorithm 4.
    Update nodal net load $NL_{n,y,t}$ based upon transmission $\varphi_{n(l),y,t}$
    Apportion net load to flexible generators with Eq. (S24) and (S25).
**end if**

Calculate surplus generation $SG_{n,y,t}$ according to Eq. (S21).

**if** $SG_{n,y,t} > 0$ for any $n$ **then**
    Transmit surplus generation to charge storages with Algorithm 4.
    Update nodal net load $NL_{n,y,t}$ based upon transmission $\varphi_{n(l),y,t}$
    Apportion net load to storage systems with Eq. (S22) and (S23).
**end if**

Update the $\gamma_{s,y,t}^{\text{storage},-}$ and $\gamma_{g,y,t}^{\text{flex},-}$ using Eq. (S26) and (S27)

Update $\gamma_{s,y,t}^{\text{storage,max\_db}}$ with Eq. (S33) using $\gamma_{s,y,t}^{\text{storage},-}$
Update $\gamma_{s,y,t}^{\text{storage,min\_db}}$ with Eq. (S34) using $\gamma_{s,y,t}^{\text{storage},-}$

**if** ($\max(NL_{n,y,t-1}, 0) == 0$ for all $n$) or $t == 0$ **then**
    Calculate $\gamma_{s,y,t}^{\text{storage},+}$ based on rules for moving forward in time
    Update $\gamma_{s,y,t}^{\text{storage,max\_db}}$ with Eq. (S33) using $\gamma_{s,y,t}^{\text{storage},+}$
    Update $\gamma_{s,y,t}^{\text{storage,min\_db}}$ with Eq. (S34) using $\gamma_{s,y,t}^{\text{storage},+}$

    Determine energies that must be pre-charged $PE_{s,y,t}$ with Eq. (S35)
    Calculate energies available for trickle charging $TE_{s,y,t}$ with Eq. (S36)
    Determine energies available for flexible pre-charging $FE_{g,y,t}$
    Return current value of $t$
    **break**
**end if**
**end while**

After **Algorithm 6**, a discontinuity exists between $\gamma_{s,y,t}^{\text{storage},+}$ and $\gamma_{s,y,t}^{\text{storage},-}$ for the time interval at the start of the deficit block (also for $\gamma_{g,y,t}^{\text{flex},+}$ and $\gamma_{g,y,t}^{\text{flex},-}$). To resolve that discontinuity, inter-storage transfers and pre-charging from flexible generators are required in the pre-charging period leading up to the deficit block using **Algorithm 7**. Continuing backwards in reverse time, trickle chargers are dispatched to fill pre-chargers by adjusting the previously determined dispatch powers. Once no more trickle chargers remain, the generation from flexible generators is adjusted to fill the pre-chargers.

**Algorithm 8** then iterates forwards in time from the start of the pre-charging period to the end of the deficit block, updating $\gamma_{s,y,t}^{\text{storage}}$ and $\gamma_{g,y,t}^{\text{flex}}$ based upon the adjusted dispatch powers. If any update to the stored energy or remaining energy is found to be infeasible, the dispatch powers are adjusted to a feasible value. The infeasibility indicates that the pre-charging business rules were unable to balance all of the unserved energy within the deficit block.

**Algorithm 7** Adjust dispatch powers required to achieve pre-charge energy (pre-charging period)

**Data**: Solution object and first t of the deficit block

**Result:** First $t$ of the pre-charging period, adjusted values of $\varphi_{s,y,t}^{\text{storage}}$, $\varphi_{g,y,t}^{\text{flex}}$, $\varphi_{n,y,t}^{\text{imp}}$, $\varphi_{n,y,t}^{\text{exp}}$, and $\varphi_{l,y,t}^{\text{line}}$ for $t$ within the pre-charging period.

Assign pre-charging and trickle charging flags to storage systems and flexible generators.
**while** True **do**
    Go back to previous time interval $t = t - 1$

    **if** $t < 0$ **then**
        Return $t = 0$ as first $t$ in pre-charging period
        **break**
    Initialise net load $NL_{n,y,t}$ according to $RL_{n,y,t}$, $\varphi_{n,y,t}^{\text{imp}}$, and $\varphi_{n,y,t}^{\text{exp}}$
    Initialise storage and flexible limits with Eq. (S14) – (S20).
    Set $TF_{n,y,t}$ and $TS_{n,y,t}$ at each node based on sum of $PE_{s,y,t}$ and $TE_{s,y,t}$ at each node.

    Calculate surplus generation $SG_{n,y,t}$ at each node according to Eq. (S21).
    Apportion $SG_{n,y,t}$ to balance $PE_{s,y,t}$ based on $TF_{n,y,t}$ at local nodes. Adjust $\varphi_{s,y,t}^{\text{storage}}$ for pre-chargers as required.
    Transmit $SG_{n,y,t}$ to balance $PE_{s,y,t}$ based on $TF_{n,y,t}$ at other nodes. Adjust $\varphi_{s,y,t}^{\text{storage}}$ for pre-chargers and $\varphi_{n,y,t}^{\text{imp}}$, $\varphi_{n,y,t}^{\text{exp}}$, and $\varphi_{l,y,t}^{\text{line}}$ as required.

    Apportion $TE_{s,y,t}$ to balance $PE_{s,y,t}$ based on $TF_{n,y,t}$ and $TS_{n,y,t}$ at local nodes. Adjust $\varphi_{s,y,t}^{\text{storage}}$ for pre-chargers and trickle chargers as required.
    Transmit $TE_{s,y,t}$ to balance $PE_{s,y,t}$ based on $TF_{n,y,t}$ and $TS_{n,y,t}$ at other nodes. Adjust $\varphi_{s,y,t}^{\text{storage}}$ for pre-chargers and trickle chargers and $\varphi_{n,y,t}^{\text{imp}}$, $\varphi_{n,y,t}^{\text{exp}}$, and $\varphi_{l,y,t}^{\text{line}}$ as required.

    **if** there are any pre-chargers remaining **then**
        Set $TF_{n,y,t}$ and $TS_{n,y,t}$ at each node based on sum of $PE_{s,y,t}$ and $FE_{g,y,t}$ at each node.

        Apportion $FE_{g,y,t}$ to balance $PE_{s,y,t}$ based on $TF_{n,y,t}$ and $TS_{n,y,t}$ at local nodes. Adjust $\varphi_{s,y,t}^{\text{storage}}$ for pre-chargers and $\varphi_{g,y,t}^{\text{flex}}$ for trickle chargers as required.
        Transmit $FE_{g,y,t}$ to balance $PE_{s,y,t}$ based on $TF_{n,y,t}$ and $TS_{n,y,t}$ at other nodes. Adjust $\varphi_{s,y,t}^{\text{storage}}$ for pre-chargers and $\varphi_{g,y,t}^{\text{flex}}$ for trickle chargers and $\varphi_{n,y,t}^{\text{imp}}$, $\varphi_{n,y,t}^{\text{exp}}$, and $\varphi_{l,y,t}^{\text{line}}$ as required.
    **end if**

    **if** there are no pre-chargers or no trickle chargers remaining **then**
        Return current $t$
        **break**
    **end if**
**end while**

**Algorithm 8** Update stored energy and remaining energy according to feasible dispatch behaviour

**Data**: First $t$ in pre-charging period, first $t$ after deficit block, solution object
**Result**: Updated $\gamma_{s,y,t}^{\text{storage}}$ and $\gamma_{g,y,t}^{\text{flex}}$. If an infeasibility is found, adjusted $\varphi_{s,y,t}^{\text{storage}}$, $\varphi_{g,y,t}^{\text{flex}}$, $\varphi_{n,y,t}^{\text{imp}}$, $\varphi_{n,y,t}^{\text{exp}}$, and $\varphi_{l,y,t}^{\text{line}}$

**for** $t$ = precharging period start, $t$ = interval after deficit block **do**
    Initialise net load $NL_{n,y,t}$ according to $RL_{n,y,t}$, $\varphi_{n,y,t}^{\text{imp}}$, and $\varphi_{n,y,t}^{\text{exp}}$
    Initialise storage and flexible limits with Eq. (S14) – (S20).

    **if** $\varphi_{s,y,t}^{\text{storage}}$ is infeasible for any $s$ **then**
        Reset $\varphi_{l,y,t}^{\text{line}}$, $\varphi_{n,y,t}^{\text{flex}}$ and $\varphi_{g,y,t}^{\text{flex}}$ to 0.
        Transmit surplus to balance remaining storage power with Algorithm 4.
        Apportion net load to storage systems with Eq. (S22) and (S23).
        Apportion remaining net load to local flexible generators with Eq. (S24) and (S25).
        Update nodal net load $NL_{n,y,t}$ based upon $\varphi_{n,y,t}^{\text{imp}}$, and $\varphi_{n,y,t}^{\text{exp}}$

        **if** $\max(NL_{n,y,t}, 0) > 0$ for any $n$ **then**
            Transmit available flexible power to balance deficits with Algorithm 4.
            Update nodal net load $NL_{n,y,t}$ based upon transmission $\varphi_{n,y,t}^{\text{imp}}$, and $\varphi_{n,y,t}^{\text{exp}}$
            Apportion net load to flexible generators with Eq. (S24) and (S25).
        **end if**
    **else**
        **if** $\varphi_{g,y,t}^{\text{flex}}$ is infeasible for any $s$ **then**
            Reset $\varphi_{l,y,t}^{\text{line}}$ to 0.
            Determine $\varphi_{s,y,t}^{\text{storage}}$ actually available for dispatch.
            Transmit surplus to balance net load with Algorithm 4.
            Transmit surplus to balance remaining storage power with Algorithm 4.

            Adjust unbalanced $\varphi_{s,y,t}^{\text{storage}}$ to feasible charging values.
        **else**
            Update nodal net load $NL_{n,y,t}$ based upon transmission $\varphi_{n,y,t}^{\text{imp}}$, and $\varphi_{n,y,t}^{\text{exp}}$
        **end if**
    **end if**

    Calculate spillage and unserved energy from net load $NL_{n,y,t}$
    Update the $\gamma_{s,y,t}^{\text{storage}}$ and $\gamma_{g,y,t}^{\text{flex}}$ using Eq. (S26) and (S27).
**end for**

# References

[1] AEMO, "ISP Methodology - June 2023," June 2023. [Online]. Available: https://aemo.com.au/-/media/files/stakeholder_consultation/consultations/nem-consultations/2023/isp-methodology-2023/isp-methodology_june-2023.pdf?la=en. [Accessed 15 May 2025].

[2] AEMO, "2023 Inputs, Assumptions and Scenarios Report," July 2023. [Online]. Available: https://aemo.com.au/-/media/files/major-publications/isp/2023/2023-inputs-assumptions-and-scenarios-report.pdf?la=en. [Accessed 15 May 2025].

[3] AEMO, "2023 IASR Assumptions Workbook," 8 September 2023. [Online]. Available: https://aemo.com.au/energy-systems/major-publications/integrated-system-plan-isp/2024-integrated-system-plan-isp/current-inputs-assumptions-and-scenarios. [Accessed 15 May 2025].

[4] AEMO, "2024 Integrated System Plan For the National Electricity Market," 26 June 2024. [Online]. Available: https://aemo.com.au/-/media/files/major-publications/isp/2024/2024-integrated-system-plan-isp.pdf?la=en. [Accessed 23 May 2025].

[5] ASEAN Centre for Energy, "8th ASEAN Energy Outlook: 2023-2050," November 2024. [Online]. Available: https://aseanenergy.org/wp-content/uploads/2024/09/8th-ASEAN-Energy-Outlook.pdf. [Accessed 15 May 2025].

[6] ASEAN Centre for Energy, "An Energy Sector Roadmap to Net Zero Emissions for Lao PDR," 2025. [Online]. Available: https://aseanenergy.org/wp-content/uploads/2025/03/An-Energy-Sector-Roadmap-to-Net-Zero-Emissions-for-Lao-PDR_Report.pdf. [Accessed 15 May 2025].

[7] Stockholm Energy Institute, "Time Slicing - NEMO," 2025. [Online]. Available: https://sei-international.github.io/NemoMod.jl/stable/time_slicing/. [Accessed 15 May 2025].

[8] NREL, "2024 Standard Scenarios Report: A U.S. Electricity Sector Outlook," December 2024. [Online]. Available: https://docs.nrel.gov/docs/fy25osti/92256.pdf. [Accessed 15 May 2025].

[9] NREL, "Electricity Annual Technology Baseline (ATB) Data," 2024. [Online]. Available: https://atb.nrel.gov/electricity/2024/data. [Accessed 15 May 2025].

[10] NREL, "Model Documentation - ReEDS 2.0," 2024. [Online]. Available: https://nrel.github.io/ReEDS-2.0/model_documentation.html#storage-technologies. [Accessed 15 May 2025].

[11] NESO, "Future Energy Scenarios: ESO Pathways to Net Zero," July 2024. [Online]. Available: https://www.neso.energy/document/321041/download. [Accessed 15 May 2025].

[12] NESO, "FES 2024 Levers and Assumptions V001," 21 August 2024. [Online]. Available: https://www.neso.energy/document/321416/download. [Accessed 15 May 2025].

[13] Mott MacDonald, "Storage cost and technical assumptions for BEIS: Summary document," 8 August 2018. [Online]. Available: https://assets.publishing.service.gov.uk/media/5f3cf6c9d3bf7f1b0fa7a165/storage-costs-technical-assumptions-2018.pdf. [Accessed 15 May 2025].

[14] RTE, "Futurs énergétiques 2050: Rapport complet," February 2022. [Online]. Available: https://rte-futursenergetiques2050.com/documents. [Accessed 15 May 2025].

[15] RTE, "Antares Simulator Documentation: Overview," 2024. [Online]. Available: https://antares-simulator.readthedocs.io/en/latest/user-guide/01-overview/. [Accessed 15 May 2025].

[16] CSIRO, "Renewable Energy Storage Roadmap," March 2023. [Online]. Available: https://www.csiro.au/en/work-with-us/services/consultancy-strategic-advice-services/csiro-futures/energy/renewable-energy-storage-roadmap. [Accessed 15 May 2025].

[17] AEMO, "Inputs, assumptions and scenarios workbook," 30 June 2022. [Online]. Available: https://aemo.com.au/energy-systems/major-publications/integrated-system-plan-isp/2022-integrated-system-plan-isp/2022-isp-inputs-assumptions-and-scenarios. [Accessed 15 May 2025].

[18] CSIRO, "Dieter.jl - Model Formulation," 2023. [Online]. Available: https://csiro-energy-systems.github.io/Dieter.jl/dev/tutorials/optmodel/. [Accessed 15 May 2025].

[19] A. Zerrahn and W. Schill, "Long-run power storage requirements for high shares of renewables: review and a new model," *Renewable and Sustainable Energy Reviews,* vol. 79, pp. 1518-1534, 2017.

[20] Departement van Mineraalbronne en Energie, "Updated New Tech Assumptions IRP 2023," 4 January 2024. [Online]. Available: https://www.dmre.gov.za/mining-minerals-energy-policy-development/integrated-resource-plan/irp-2023. [Accessed 16 May 2025].

[21] Departement van Mineraalbronne en Energie, "Department of Mineral Resources and Energy - Draft Integrated Resource Plan Stakeholder Workshops," November 2024. [Online]. Available: https://www.dmre.gov.za/Portals/0/Energy%20Resources/IRP/IRP%202023/Draft%20IRP%202024%20Outcomes%20Stakeholder%20Engagements.pdf?ver=4zU5hDlVy48zTCjoiJhmFA%3d%3d. [Accessed 16 May 2025].

[22] B. Merven and Sanedi, "Demand Projection Model in Support of IRP Update 2023," 3 November 2023. [Online]. Available: https://www.dmre.gov.za/Portals/0/Energy_Website/IRP/2023/Electricity%20demand%20projection%20model%20ESRG%20SANEDI%20Report%20Rev%201.pdf. [Accessed 16 May 2025].

[23] Departement van Mineraalbronne en Energie, "Integrated Resource Plan, 2023," 4 January 2024. [Online]. Available: https://www.dmre.gov.za/Portals/0/Energy_Website/IRP/2023/IRP%20Government%20Gazzette%202023.pdf. [Accessed 23 May 2025].

[24] US EIA, "Electricity Market Module of the National Energy Modeling System: Model Documentation 2022," September 2022. [Online]. [Accessed 16 May 2025].

[25] US EIA, "Assumptions to the Annual Energy Outlook 2025: Electricity Market Module," April 2025. [Online]. Available: https://www.eia.gov/outlooks/aeo/assumptions/pdf/EMM_Assumptions.pdf. [Accessed 16 May 2025].

[26] US EIA, "Annual Energy Outlook 2025," 15 April 2025. [Online]. Available: https://www.eia.gov/outlooks/aeo/. [Accessed 23 May 2025].

[27] Canada Energy Regulator, "Canada's Energy Future 2023: Energy Supply and Demand Projections to 2050 (EF2023)," 2023. [Online]. Available: https://www.cer-rec.gc.ca/en/data-analysis/canada-energy-future/2023/canada-energy-futures-2023.pdf. [Accessed 16 May 2025].

[28] T. Brown, J. Hörsch, F. Hofmann, F. Neumann, L. Zeyen, M. Frysztacki, P. Glaum, M. Parzen, D. Schlachberger, L. Trippe and P. Developers, "PyPSA - Suer Guide - System Optimization," 2025. [Online]. Available: https://pypsa.readthedocs.io/en/latest/user-guide/optimal-power-flow.html. [Accessed 16 May 2025].

[29] Central Electricity Authority, "National Electricity Plan (Volume I) Generation," March 2023. [Online]. Available: https://mnre.gov.in/en/document/national-electricity-plan-volume-i-generation-by-cea/. [Accessed 16 May 2025].

[30] Philippine Department of Energy, Electric Power Industry Management Bureau, "Power Development Plan 2023-2050," 2023. [Online]. Available: https://legacy.doe.gov.ph/sites/default/files/pdf/electric_power/development_plans/Power%20Development%20Plan%202023-2050.pdf. [Accessed 16 May 2025].

[31] NREL, "Pumped Storage Hydropower | Electricity | 2022 | ATB | NREL," 2022. [Online]. Available: https://atb.nrel.gov/electricity/2022/pumped_storage_hydropower. [Accessed 16 May 2025].

[32] National Transmission and Despatch Company , "Indicative Generation Capacity Expansion Plan (IGCEP) 2021-30," May 2021. [Online]. Available: https://fpcci.org.pk/wp-content/uploads/2021/11/IGCEP-2021.pdf. [Accessed 16 May 2025].

[33] S. Viennet and Stantec, "The European Union Global Technical Assistance Facility on Sustainable Energy: Africa - Specific Support Study on hydro reservoir and pumped storage plants. Assignment number: GT088," 17 February 2023. [Online]. Available: https://cmpmwanga.nepad.org/PSH_del6. [Accessed 16 May 2025].

[34] IRENA, "Advancements in continental power system planning for Africa: Methodological framework of the African Continental Power Systems Masterplan's SPLAT-CMP model 2023," 2024. [Online]. Available: https://www.irena.org/-/media/Files/IRENA/Agency/Publication/2024/Jul/IRENA_Advancements_CMP_Africa_2024.pdf. [Accessed 16 May 2025].

[35] IRENA, "Planning and prospects for renewable power: Central Africa," 2024. [Online]. Available: https://www.irena.org/-/media/Files/IRENA/Agency/Publication/2025/Jan/IRENA_Planning_Prospects_Central_Africa_2025.pdf. [Accessed 20 May 2025].

[36] USAID V-LEEP, Deloitte, "Technical Report: Integrating significant renewable energy in Vietnam's power sector: A PLEXOS based analysis of long-term power development planning," 18 March 2021. [Online]. Available: https://www.sipet.org/images/document/14092022126861.pdf. [Accessed 20 May 2025].

[37] Ceylon Electricity Board, "Long Term Generation Expansion Plan 2023-2042," February 2023. [Online]. Available: https://www.ceb.lk/publication-media/planing-documents/121/en. [Accessed 0 May 2025].

[38] A. Soares, R. Perez and F. Thome, "Optimal generation and transmission expansion planning addressing short-term constraints with co-optimization of energy and reserves," *arXiv preprint arXiv:1910.00446,* 2019.

[39] Ministerio para la Transición Ecológica y el Reto Demográfico, "Integrated National Energy and Climate Plan: Update 2023-2030," September 2024. [Online]. Available: https://commission.europa.eu/document/download/211d83b7-b6d9-4bb8-b084-4a3bfb4cad3e_en?filename=ES%20-%20FINAL%20UPDATED%20NECP%202021-2030%20%28English%29.pdf. [Accessed 20 May 2025].

[40] Joint Research Centre - European Commission, "Central_2018_ES: Power generation technology assumptions - Developed to serve as input to the Central_2018 scenario," October 2019. [Online]. Available: https://jeodpp.jrc.ec.europa.eu/ftp/jrc-opendata/POTEnCIA/Central_2018/. [Accessed 20 May 2025].

[41] L. Mantzos, N. Matei, M. Rózsai, P. Russ and A. Ramirez, "POTEnCIA: A new EU-wide energy sector model," *2017 14th international conference on the European Energy Market (EEM),* pp. 1-5, 2017.

[42] F. Neuwahl, M. Wegener, R. Salvucci, M. Jaxa-Rozen, J. Gea Bermudez, P. Sikora and M. Rózsai, "Clean Energy Technology Observatory: POTEnCIA CETO 2024 Scenario – 2024 Energy System Modelling For Clean Energy Technology Scenarios," 2024. [Online]. Available: https://publications.jrc.ec.europa.eu/repository/handle/JRC139836. [Accessed 20 May 2025].

[43] E. Quaranta, A. Georgakaki, S. Letout, A. Mountraki, E. Ince and J. Gea Bermudez, " Clean Energy Technology Observatory: Hydropower and Pumped Storage Hydropower in the European Union - 2024 Status Report on Technology Development, Trends, Value Chains and Markets," 2024. [Online]. Available:

https://publications.jrc.ec.europa.eu/repository/handle/JRC139225. [Accessed 20 May 2025].

[44] A. Hilbers, D. Brayshaw and A. Gandy, "Reducing climate risk in energy system planning: A posteriori time series aggregation for models with storage," *Applied Energy,* vol. 334, p. 120624, 2023.

[45] AEMC, "National Electricity Rules version 231," 1 July 2025. [Online]. Available: https://energy-rules.aemc.gov.au/ner/659/612232#3.9.3C. [Accessed 2 July 2025].

[46] R. Storn and K. Price, "Differential evolution–a simple and efficient heuristic for global optimization over continuous spaces," *Journal of global optimization,* vol. 11, no. 4, pp. 341-359, 1997.

[47] A. Nelson, "scipy.optimize.differential_evolution," 2014. [Online]. Available: https://docs.scipy.org/doc/scipy/reference/generated/scipy.optimize.differential_evolution.html. [Accessed 30 January 2023].

[48] Snowy Hydro, "Annual Report For the Financial Year ended 30 June 2025," 2025. [Online]. Available: https://www.snowyhydro.com.au/wp-content/uploads/2025/10/Snowy-Hydro-Annual-Report-2024-25.pdf. [Accessed 22 December 2025].

[49] M. Stocks, R. Stocks, B. Lu, C. Cheng and A. Blakers, "Global Atlas of Closed-Loop Pumped Hydro Energy Storage - Supplemental Information," *Joule,* vol. 5, no. 1, pp. 270-284, 2021.

[50] T. Weber, "Pumped Hydro Shortlisting Tool," 2025. [Online]. Available: https://re100.anu.edu.au/shortlisting/. [Accessed 9 November 2025].

[51] F. Lombardi, K. van Greevenbroek, A. Grochowicz, M. Lau, F. Neumann, N. Patankar and O. Vågerö, "Near-optimal energy planning strategies with modeling to generate alternatives to flexibly explore practically desirable options," *Joule,* p. 102144, 2025.

[52] GHD, "Pumped Hydro Energy Storage: Parameter Review," 22 July 2025. [Online]. Available: https://www.aemo.com.au/-/media/files/stakeholder_consultation/consultations/nem-consultations/2024/2025-iasr-scenarios/final-docs/ghd-2025-pumped-hydro-energy-storage-cost-parameter-review.pdf?la=en. [Accessed 17 October 2025].

[53] Australian Bureau of Statistics, "Consumer PRice Index, Australia," 26 November 2025. [Online]. Available: https://www.abs.gov.au/statistics/economy/price-indexes-and-inflation/consumer-price-index-australia/latest-release. [Accessed 4 December 2025].

[54] Australian Energy Regulator, "Rate of Return Instrument 2022," 24 February 2023. [Online]. Available: https://www.aer.gov.au/industry/registers/resources/guidelines/rate-return-instrument-2022. [Accessed 4 December 2025].

[55] AEMO, "2025 Inputs, Assumptions and Scenarios Report," August 2025. [Online]. Available: https://www.aemo.com.au/-/media/files/stakeholder_consultation/consultations/nem-consultations/2024/2025-iasr-scenarios/final-docs/2025-inputs-assumptions-and-scenarios-report.pdf?rev=63268acd3f044adb9f5f3a32b6880c27&sc_lang=en. [Accessed 9 November 2025].

[56] Australian Office of Financial Management, "Treasury Bonds," 28 11 2025. [Online]. Available: https://www.aofm.gov.au/securities/treasury-bonds. [Accessed 4 12 2025].

[57] Queensland Treasury Corporation, "Benchmark bonds," 11 December 2025. [Online]. Available: https://www.qtc.com.au/institutional-investors/aud-benchmark-bonds/. [Accessed 13 December 2025].

[58] Australian Energy Regulator, "Cost Benefit Analysis guidelines," November 2024. [Online]. Available: https://www.aer.gov.au/industry/registers/resources/guidelines/cost-benefit-analysis-guidelines. [Accessed 4 December 2025].

[59] IRENA, "Renewable Capacity Highlights 2025," 26 March 2025. [Online]. Available: https://www.irena.org/-/media/Files/IRENA/Agency/Publication/2025/Mar/IRENA_DAT_RE_Capacity_Highlights_2025.pdf. [Accessed 7 May 2025].

[60] IRENA, "Renewable Capacity Statistics 2025," March 2025. [Online]. Available: https://www.irena.org/Publications/2025/Mar/Renewable-capacity-statistics-2025. [Accessed 7 May 2025].

[61] IAEA, "Power Reactor Information System: NPP Status Changes (2024)," 2025. [Online]. Available: https://pris.iaea.org/pris/. [Accessed 7 May 2025].

[62] M. Frysztacki, J. Hörsch, V. Hagenmeyer and T. Brown, "The strong effect of network resolution on electricity system models with high shares of wind and solar," *Applied Energy,* vol. 291, p. 116726, 2021.

[63] K. Poncelet, E. Delarue, D. Six, J. Duerinck and W. D'haeseleer, "Impact of the level of temporal and operational detail in energy-system planning models," *Applied Energy,* vol. 162, pp. 631-643, 2016.

[64] J. Merrick, "On representation of temporal variability in electricity capacity planning models," *Energy Economics,* vol. 59, pp. 261-271, 2016.

[65] J. Bistline, "The importance of temporal resolution in modeling deep decarbonization of the electric power sector," *Environmental Research Letters,* vol. 16, no. 8, p. 084005, 2021.

[66] A. Belderbos and E. Delarue, "Accounting for flexibility in power system planning with renewables," *International Journal of Electrical Power & Energy Systems,* vol. 71, pp. 33-41, 2015.

[67]   C. Heuberger, I. Staffell, N. Shah and N. Dowell, "A systems approach to quantifying the value of power generation and energy storage technologies in future electricity networks," *Computers & Chemical Engineering,* vol. 107, pp. 247-256, 2017.

[68]   D. Kirschen, J. Ma, V. Silva and R. Belhomme, "Optimizing the flexibility of a portfolio of generating plants to deal with wind generation," *2011 IEEE power and energy society general meeting,* pp. 1-7, 2011.

[69]   M. Jafari, M. Korpås and A. Botterud, "Power system decarbonization: Impacts of energy storage duration and interannual renewables variability," *Renewable Energy,* vol. 156, pp. 1171-1185, 2020.

[70]   L. Kotzur, L. Nolting, M. Hoffmann, T. Groß, A. Smolenko, J. Priesmann, H. Büsing, R. Beer, F. Kullmann, B. Singh and A. Praktiknjo, "A modeler's guide to handle complexity in energy systems optimization," *Advances in Applied Energy,* vol. 4, p. 1000063, 2021.

[71]   S. Pfenninger, "Dealing with multiple decades of hourly wind and PV time series in energy models: A comparison of methods to reduce time resolution and the planning implications of inter-annual variability," *Applied Energy,* vol. 197, pp. 1-13, 2017.

[72]   M. Prina, G. Manzolini, D. Moser, B. Nastasi and W. Sparber, "Classification and challenges of bottom-up energy system models - A review," *Renewable and Sustainable Energy Reviews,* vol. 129, p. 109917, 2020.

[73]   C. Nweke, F. Leanez, G. Drayton and M. Kolhe, "Benefits of Chronological Optimization in Capacity Planning for Electricity Markets," *2012 IEEE international conference on power system technology (POWERCON),* pp. 1-6, 2012.

[74]   F. Murphy and Y. Smeers, "Generation Capacity Expansion in Imperfectly Competitive Restructured Electricity Markets," *Operations Research,* vol. 53, no. 4, pp. 646-661, 2005.

[75]   S. Kazempour, A. Conejo and C. Ruiz, "Strategic Generation Investment Using a Complementarity Approach," *IEEE Transactions on Power Systems,* vol. 26, no. 2, pp. 940-948, 2011.

[76]   M. Caramanis, R. Tabors, K. Nochur and F. Schweppe, "The Introduction of NonDIispatchable Technologies a Decision Variables in Long-Term Generation Expansion Models," *IEEE Transactions on Power Apparatus and Systems,* Vols. PAS-101, no. 8, pp. 2658-2667, 1982.

[77]   S. Wogrin, P. Duenas, A. Delgadillo and J. Reneses, "A new approach to model load levels in electric power systems with high renewable penetration," *IEEE Transactions on Power Systems,* vol. 29, no. 5, pp. 2210-2218, 2014.

[78]   S. Wogrin, D. Galbally and J. Reneses, "Optimizing storage operations in medium-and long-term power system models," *IEEE Transactions on Power Systems,* vol. 31, no. 4, pp. 3129-3138, 2015.

[79] M. Hoffman, L. Kotzur, D. Stolten and M. Robinius, "A Review on Time Series Aggregation Methods for Energy System Models," *Energies,* vol. 13, no. 3, p. 641, 2020.

[80] P. de Guibert, B. Shirizadeh and P. Quirion, "Variable time-step: A method for improving computational tractability for energy system models with long-term storage," *Energy,* vol. 213, p. 119024, 2020.

[81] M. Hoffman, L. Kotzur and D. Stolten, "The Pareto-optimal temporal aggregation of energy system models," *Applied Energy,* vol. 315, p. 119029, 2022.

[82] B. Schyska, A. Kies, M. Schlott, L. von Bremen and W. Medjroubi, "The sensitivity of power system expansion models," *Joule,* vol. 5, no. 10, pp. 2606-2624, 2021.

[83] L. Kotzur, P. Markewitz, M. Robinius and D. Stolten, "Impact of different time series aggregation methods on optimal energy system design," *Renewable Energy,* vol. 117, pp. 474-487, 2018.

[84] W. Tso, C. Demirhan, C. Heuberger, J. Powell and E. Pistikopoulos, "A hierarchical clustering decomposition algorithm for optimizing renewable power systems with storage," *Applied Energy,* vol. 270, p. 115190, 2020.

[85] H. Teichgraeber and A. Brandt, "Time-series aggregation for the optimization of energy systems: Goals, challenges, approaches, and opportunities," *Renewable and Sustainable Energy Reviews,* vol. 157, p. 111984, 2022.

[86] Z. Li, L. Cong, J. Li, Q. Yang, X. Li and P. Wang, "Co-planning of transmission and energy storage by iteratively including extreme periods in time-series aggregation," *Energy Reports,* vol. 9, no. 7, pp. 1281-1291, 2023.

[87] P. A. Sánchez-Pérez, M. Staadecker, J. Szinai, S. Kurtz and P. Hidalgo-Gonzalez, "Effect of modeled time horizon on quantifying the need for long-duration storage," *Applied Energy,* vol. 317, p. 119022, 2022.

[88] P. Lund, J. Lindgren, J. Mikkola and J. Salpakari, "Review of energy system flexibility measures to enable high levels of variable renewable electricity," *Renewable and Sustainable Energy Reviews,* vol. 45, pp. 785-807, 2015.

[89] J. Twitchell, K. DeSomber and D. Bhatnagar, "Defining long duration energy storage," *Journal of Energy Storage,* vol. 60, p. 105787, 2023.

[90] S. Ludig, M. Haller, E. Schmid and N. Bauer, "Fluctuating renewables in a long-term climate change mitigation strategy," *Energy,* vol. 36, no. 11, pp. 6674-6685, 2011.

[91] R. Kannan and H. Turton, "A Long-Term Electricity Dispatch Model with the TIMES Framework," *Environmental Modeling & Assessment,* vol. 18, pp. 325-343, 2012.

[92] D. Mallapragada, D. Papageorgiou, A. Venkatesh, C. Lara and I. Grossman, "Impact of model resolution on scenario outcomes for electricity sector system expansion," *Energy,* vol. 163, pp. 1231-1244, 2018.

[93] P. Nahmmacher, E. Schmid, L. Hirth and B. Knopf, "Carpe diem: A novel approach to select representative days for long-term power system models with high shares of renewable energy sources," *USAEE Working Paper No. 14-194,* 2014.

[94] T. Levin, J. Bistline, R. Sioshansi, W. Cole, J. Kwon, S. Burger, G. Crabtree, J. Jenkins, R. O'Neil, M. Korpås and S. Wogrin, "Energy storage solutions to decarbonize electricity through enhanced capacity expansion modelling," *Nature Energy,* vol. 8, no. 11, pp. 1199-1208, 2023.

[95] J. Merrick, J. Bistline and G. Blanford, "On representation of energy storage in electricity planning models," *Energy Economics,* vol. 136, p. 107675, 2024.

[96] P. Gabrielli, M. Gazzani, E. Martelli and M. Mazzotti, "Optimal design of multi-energy systems with seasonal storage," *Applied Energy,* vol. 219, pp. 408-424, 2018.

[97] K. Poncelet, E. Delarue and W. D'haeseleer, "Unit commitment constraints in long-term planning models: Relevance, pitfalls and the role of assumptions on flexibility," *Applied Energy,* vol. 258, p. 113843, 2020.

[98] D. Tejada-Arango, M. Domeshek, S. Wogrin and E. Centeno, "Enhanced Representative Days and System States Modeling for Energy Storage Investment Analysis," *IEEE Transactions on Power Systems,* vol. 33, no. 6, pp. 6534-6544, 2018.

[99] L. Kotzur, P. Markewitz, M. Robinius and D. Stolten, "Time series aggregation for energy system design: Modeling seasonal storage," *Applied Energy,* vol. 213, pp. 123-135, 2018.

[100] R. Novo, P. Marocco, G. Giorgi, A. Lanzini, M. Santarelli and G. Mattiazzo, "Planning the decarbonisation of energy systems: The importance of applying time series clustering to long-term models," *Energy Conversion and Management: X,* vol. 15, p. 100274, 2022.

[101] M. Moradi-Sepahvand and S. Tindemans, "Capturing Chronology and Extreme Values of Representative Days for Planning of Transmission Lines and Long-Term Energy Storage Systems," *2023 IEEE Belgrade PowerTech,* pp. 1-6, 2023.

[102] J. Ma, N. Zhang, Q. Wen and Y. Wang, "An efficient local multi-energy systems planning method with long-term storage," *IET Renewable Power Generation,* vol. 18, no. 3, pp. 426-441, 2024.

[103] S. Gonzato, K. Bruninx and E. Delarue, "Long term storage in generation expansion planning models with a reduced temporal scope," *Applied Energy,* vol. 298, p. 117168, 2021.

[104] M. Staadecker, J. Szinai, P. Sánchez-Pérez, S. Kurtz and P. Hidalgo-Gonzalez, "The value of long-duration energy storage under various grid conditions in a zero-emissions future," *Nature Communications,* vol. 15, no. 1, p. 9501, 2024.

[105] N. Sepulveda, J. Jenkins, A. Edington, D. Mallapragada and R. Lester, "The design space for long-duration energy storage in decarbonized power systems," *Nature Energy,* vol. 6, pp. 506-516, 2021.

[106] M. Parzen, F. Neumann, A. Van Der Wijde, D. Friedrich and A. Kiprakis, "Beyond cost reduction: improving the value of energy storage in electricity systems," *Carbon Neutrality,* vol. 1, no. 26, 2022.

[107] J. Dowling, K. Rinaldi, T. Ruggles, S. Davis, M. Yuan, F. Tong, N. Lewis and K. Caldeira, "Role of Long-Duration Energy Storage in Variable Renewable Electricity Systems," *Joule,* vol. 4, no. 9, pp. 1907-1928, 2020.

[108] A. Blakers, B. Lu and M. Stocks, "100% renewable electricity in Australia," *Energy,* vol. 133, pp. 471-482, 2017.

[109] C. Cheng, A. Blakers, M. Stocks and B. Lu, "100% Renewable Energy in Japan," *Energy Conversion and Management,* vol. 255, no. 115299, 2022a.

[110] B. Lu, A. Blakers, M. Stocks and T. N. Do, "Low-cost, low-emission 100% renewable electricity in Southeast Asia supported by pumped hydro storage," *Energy,* vol. 236, 2021.

[111] A. Nadolny, C. Cheng, B. Lu, A. Blakers and M. Stocks, "Fully electrified land transport in 100% renewable electricity networks dominated by variable generation," *Renewable Energy,* vol. 182, pp. 562-577, 2022.

[112] D. Silalahi, A. Blakers and C. Cheng, "100% Renewable Electricity in Indonesia," *Energies,* vol. 17, no. 1, p. 3, 2023.

[113] T. Weber, A. Blakers, D. Silalahi, K. Catchpole and A. Nadolny, "Grids dominated by solar and pumped hydro in wind-constrained sunbelt countries," *Energy Conversion and Management,* vol. 308, p. 118354, 2024.

[114] B. Lu, A. Blakers, M. Stocks, C. Cheng and A. Nadolny, "A Zero-Carbon, Reliable and Affordable Energy Future in Australia," *Energy,* vol. 220, no. 119678, 2021.

[115] B. Elliston, I. MacGill and M. Diesendorf, "Least cost 100% renewable electricity scenarios in the Australian National Electricity Market," *Energy Policy,* vol. 59, pp. 270-282, August 2013.

[116] B. Lu, "Stabilising 100% Renewable Grids: The Integrated FIRM Strategy," *Net Zero,* vol. 1, no. 1, pp. 31-47, 2025.

[117] International Hydropower Association, "Pumped Storage Tracking Tool," 27 June 2022. [Online]. Available: https://www.hydropower.org/hydropower-pumped-storage-tool. [Accessed 31 October 2025].

[118] International Energy Agency, "Batteries and Secure Energy Transitions," 25 April 2024. [Online]. Available: https://www.iea.org/reports/batteries-and-secure-energy-transitions. [Accessed 5 December 2025].

[119] BloombergNEF, "Headwinds in Largest Energy Storage Markets Won't Deter Growth," 11 November 2024. [Online]. Available: https://about.bnef.com/insights/clean-

energy/headwinds-in-largest-energy-storage-markets-wont-deter-growth/. [Accessed 5 December 2025].

[120]  M. Jafari, A. Botterud and A. Sakti, "Decarbonizing power systems: A critical review of the role of energy storage," *Renewable and Sustainable Energy Reviews,* vol. 158, p. 112077, 2022.

[121]  A. Blakers, T. Weber and D. Silalahi, "Pumped hydro energy storage to support 100% renewable energy," *Progress in Energy,* vol. 7, p. 022004, 2025.

[122]  T. Weber, R. Stocks, A. Blakers, A. Nadolny and C. Cheng, "A global atlas of pumped hydro systems that repurpose existing mining sites," *Renewable Energy,* vol. 224, p. 120113, 2024.

[123]  M. Stocks, R. Stocks, B. Lu, C. Cheng and A. Blakers, "Global Atlas of Closed-Loop Pumped Hydro Energy Storage," *Joule,* vol. 5, pp. 270-284, 2021.

[124]  Snowy Hydro Limited, "Snowy 2.0 Updated Business Case," 24 May 2024. [Online]. Available: https://www.snowyhydro.com.au/wp-content/uploads/2024/05/Snowy-2.0-Updated-Business-Case.pdf. [Accessed 23 May 2025].

[125]  AEMO, "2024 ISP Model," 26 June 2024. [Online]. Available: https://aemo.com.au/energy-systems/major-publications/integrated-system-plan-isp/2024-integrated-system-plan-isp. [Accessed 23 May 2025].

[126]  M. Stocks, R. Stocks, T. Weber, B. Lu, A. Nadolny, C. Cheng and A. Blakers, "ANU RE100 Map," 2025. [Online]. Available: https://re100.anu.edu.au/. [Accessed 12 December 2025].

[127]  AEMC, "How the national energy objectives shape our decisions," 28 March 2025. [Online]. Available: https://www.aemc.gov.au/sites/default/files/2025-03/How%20the%20national%20energy%20objectives%20shape%20our%20decisions%20260325.pdf. [Accessed 9 November 2025].

[128]  S. Reynolds and I. f. E. E. a. F. Analysis, "Global gas turbine shortages add to LNG challenges in Vietnam and the Philippines," October 2025. [Online]. Available: https://ieefa.org/sites/default/files/2025-10/IEEFA%20Report_Global%20gas%20turbine%20shortages%20add%20to%20LNG%20%20challenges%20in%20Vietnam%20and%20the%20Philippines_October2025.pdf. [Accessed 10 October 2025].

[129]  J. Anderson, "US gas-fired turbine wait times as much as seven years; costs up sharply," S&P Global, 20 May 2025. [Online]. Available: https://www.spglobal.com/commodity-insights/en/news-research/latest-news/electric-power/052025-us-gas-fired-turbine-wait-times-as-much-as-seven-years-costs-up-sharply. [Accessed 31 October 2025].

[130]  L. Shen, D. Jacob, R. Gautam, M. Omara, T. Scarpelli, A. Lorente, D. Zavala-Araiza, X. Lu, Z. Chen and J. Lin, "National quantifications of methane emissions from fuel exploitation using high resolution inversions of satellite observations," *Nature Communications,* vol. 14, no. 1, p. 4948, 2023.

[131] IEA, "Global Methane Tracker 2025," May 2025. [Online]. Available: https://iea.blob.core.windows.net/assets/b83c32dd-fc1b-4917-96e9-8cd918801cbf/GlobalMethaneTracker2025.pdf. [Accessed 11 July 2025].

[132] IPCC, "IPCC Sixth Assessment Report," 2020. [Online]. Available: https://www.ipcc.ch/report/ar6/wg1/chapter/chapter-7/#7.6. [Accessed 31 October 2025].

[133] International Energy Agency, "CCUS Projects Database," 30 April 2025. [Online]. Available: https://www.iea.org/data-and-statistics/data-product/ccus-projects-database. [Accessed 31 October 2025].

[134] *Obligations of States in respect of Climate Change (Advisory Opinion),* 2025, ICJ, p. 122.

[135] AEMO, "Quarterly Energy Dynamics Q3 2025," October 2025. [Online]. Available: https://www.aemo.com.au/-/media/files/major-publications/qed/2025/qed-q3-2025.pdf. [Accessed 12 December 2025].

[136] *National Electricity (South Australia) Act 1996 (SA).*

[137] Australian Government, "Australia's 2035 Nationally Determined Contribution," 2025. [Online]. Available: https://unfccc.int/sites/default/files/2025-09/Australias%20Second%20NDC.pdf. [Accessed 17 October 2025].

[138] International Energy Agency, "WEO2025_AnnexA_Free_Dataset_World.csv," 12 November 2025. [Online]. Available: https://www.iea.org/data-and-statistics/data-product/world-energy-outlook-2025-free-dataset#data-files. [Accessed 5 December 2025].

[139] AEMO, "2024 ISP Solar traces," 26 June 2024. [Online]. Available: https://aemo.com.au/-/media/files/major-publications/isp/2024/supporting-materials/2024-isp-solar-traces.zip?la=en. [Accessed 29 May 2025].

[140] AEMO, "2024 ISP Wind traces," 26 June 2024. [Online]. Available: https://aemo.com.au/-/media/files/major-publications/isp/2024/supporting-materials/2024-isp-wind-traces.zip?la=en. [Accessed 29 May 2025].

[141] S. Lam, A. Pitrou and S. Seibert, "Numba: A llvm-based python jit compiler," *Proceedings of the Second Workshop on the LLVM Compiler Infrastructure in HPC,* pp. 1-6, 2015.

[142] The pandas development team, "pandas-dev/pandas: Pandas," Zenodo, February 2020. [Online]. Available: https://doi.org/10.5281/zenodo.3509134.

[143] W. McKinney, "Data structures for statistical computing in Python," *Proceedings of the 9th Python in Science Conference,* pp. 56-61, 2010.

[144] C. Harris, K. Millman, S. Van Der Walt, R. Gommers, P. Virtanen, D. Cournapeau, E. Wieser, J. Taylor, S. Berg, N. Smith and R. Kern, "Array programming with NumPy," *Nature,* vol. 585, no. 7825, pp. 357-362, 2020.